\documentclass[11pt,english]{article}
\usepackage{graphicx}
\usepackage{xcolor}
\usepackage{amstext}
\usepackage{caption}
\usepackage{etoolbox}
\usepackage{makeidx}
\include{epsf}
\usepackage{amsfonts}
\usepackage{authblk} 
\usepackage{sectsty}
\usepackage{amsmath,amssymb,epsfig}
\usepackage{amscd}
\usepackage{amsthm}
\usepackage{mathrsfs}
\usepackage{dsfont}
\usepackage[applemac]{inputenc}
\usepackage[english]{babel}
\usepackage{enumitem} 
\usepackage[]{latexsym}
\usepackage{caption}
\usepackage{subfigure}
\usepackage{hyperref}
\usepackage{blindtext}
\usepackage{verbatim}
\usepackage{cancel}
\usepackage{cases}

\hypersetup{
pdftitle={},%
pdfauthor={},%
pdfsubject={},%
pdfkeywords={},%
colorlinks=true,%
linkcolor=blue,%
citecolor=crimson,%
linktocpage=true,%
hyperfootnotes=true,%
pageanchor=true
}

\definecolor{crimson}{rgb}{0.7, 0.08, 0.24}

\newcommand*{\affaddr}[1]{#1}
\newcommand*{\affmark}[1][*]{\textsuperscript{#1}}

\newtheorem*{proof*}{Proof}

\newcommand{\be}{\begin{equation}}

\newcommand{\ee}{\end{equation}}
\def\beqa{\begin{eqnarray}}
\def\eeqa{\end{eqnarray}}
\def\bean{\begin{eqnarray*}}
\def\eean{\end{eqnarray*}}

\newcommand{\del}{\partial}
\newcommand{\dd}{\mathrm{d}}

\renewenvironment{thebibliography}[1]
         {\section*{References}\frenchspacing\small
          \begin{list}{[\arabic{enumi}]}
         {\usecounter{enumi}\parsep=2pt\topsep 0pt
         \settowidth{\labelwidth}{[#1]}
         \leftmargin=\labelwidth\advance\leftmargin\labelsep
         \rightmargin=0pt\itemsep=1pt\sloppy}}{\end{list}}

 \numberwithin{equation}{section}

\input xy
\xyoption{all}

\usepackage[normalem]{ulem}

\newcommand{\ket}[1]{\left| #1 \right\rangle}

\newcommand{\braket}[2]{\left\langle \vphantom {#1 #2} #1 \hphantom{|} \right| \left. \vphantom {#1 #2} #2 \right\rangle}

\title{\textbf{\textsf{Effective Quantum Extended Spacetime of Polymer Schwarzschild Black Hole}}\vspace{0.35cm}}

\author{
\textsf{Norbert Bodendorfer\affmark[1]\footnote{\texttt{norbert.bodendorfer@physik.uni-r.de}}, Fabio M. Mele\affmark[1]\footnote{\texttt{fabio.mele@physik.uni-r.de}}, and Johannes M\"unch\affmark[1]\footnote{\texttt{johannes.muench@physik.uni-r.de}}}\\
\affaddr{\affmark[1]\textsf{Institute for Theoretical Physics, University of Regensburg,}}\\
\affaddr{\textsf{93040 Regensburg, Germany}}\vspace{-0.5cm}
}

\begin{document}

\maketitle

\begin{abstract}
\textsf{The physical interpretation and eventual fate of gravitational singularities in a theory surpassing classical general relativity are puzzling questions that have generated a great deal of interest among various quantum gravity approaches. In the context of loop quantum gravity (LQG), one of the major candidates for a non-perturbative background-independent quantisation of general relativity, considerable effort has been devoted to construct effective models in which these questions can be studied. In these models, classical singularities are replaced by a ``bounce'' induced by quantum geometry corrections. Undesirable features may arise however depending on the details of the model. In this paper, we focus on Schwarzschild black holes and propose a new effective quantum theory based on polymerisation of new canonical phase space variables inspired by those successful in loop quantum cosmology. The quantum corrected spacetime resulting from the solutions of the effective dynamics is characterised by infinitely many pairs of trapped and anti-trapped regions connected via a space-like transition surface replacing the central singularity. Quantum effects become relevant at a unique mass independent curvature scale, while they become negligible in the low curvature region near the horizon. The effective quantum metric describes also the exterior regions and asymptotically classical Schwarzschild geometry is recovered. We however find that physically acceptable solutions require us to select a certain subset of initial conditions, corresponding to a specific mass (de-)amplification after the bounce. We also sketch the corresponding quantum theory and explicitly compute the kernel of the Hamiltonian constraint operator.}
\end{abstract}

\section{Introduction}

The study of Einstein's field equations has revealed that under generic conditions space-like singularities arise in solutions of general relativity \cite{PenroseGravitationalCollapseAnd,HawkingPropertiesOfExpanding}. The two paradigmatic physical situations in which such gravitational singularities appear are the Big Bang or the Big Crunch singularities in cosmological scenarios, and in the interior region of black holes. The occurrence of gravitational singularities in solutions of Einstein's field equations signals that general relativity breaks down once spacetime curvature reaches the Planck regime and hence its predictions cannot be trusted at such scales where quantum gravitational effects are expected to be relevant. It is commonly believed that once a complete quantum theory of gravity is employed, the classical singularities will be resolved, see e.g. \cite{BojowaldSingularitiesAndQuantum, NatsuumeTheSingularityProblem} for an overview. The understanding of the fate of gravitational singularities and their physical interpretation in a theory surpassing classical general relativity are puzzling questions that have generated a great deal of interest among various quantum gravity approaches, most notably loop quantum gravity (LQG) \cite{BojowaldSingularitiesAndQuantum,BojowaldAbsenceOfSingularity,AshtekarLoopQuantumCosmology,WilsonEwingTheloopquantum,OlmedoFromblackholesto}, string theory \cite{NatsuumeTheSingularityProblem,BalasubramanianAspace-timeorbifold,CornalbaANewCosmological,CornalbaAresolutionof,BerkoozCommentsoncosmological}, AdS/CFT \cite{CrapsQuantumResolutionOf,FidkowskiTheBlackHole,MaldacenaEternalBlackHoles,HanadaHolographicdescriptionof,EngelhardtHolographicConsequencesof,EngelhardtNewInsightsinto}, as well as non-commutative geometry \cite{MadoreOntheresolution,MacedaOntheResolution,MacedaCannoncommutativityresolve,GorjiSpacetimesingularityresolution} and related contexts \cite{ChamseddineResolvingCosmologicalSingularities, ChamseddineNonsingularBlackHole,NicoliniNoncommutativegeometryinspired,BenAchourNon-singularblackholes}. However, there is still no agreement on whether and how spacetime singularities are resolved in quantum gravity. For instance, in the context of the gauge/gravity correspondence, it was argued in \cite{EngelhardtHolographicConsequencesof,EngelhardtNewInsightsinto} that not all singularities may be resolved by quantum gravity effects.

As a complete theory of quantum gravity is still lacking nowadays, it becomes important to construct effective models in which such issues can be investigated and eventually to try also to extract from them useful lessons for the full theory. Within loop quantum gravity and related formalisms, the simplest example is provided by homogeneous and isotropic FLRW cosmological spacetimes where much progress has been made \cite{AshtekarLoopQuantumCosmology,OritiBouncingCosmologiesFrom,AgulloLoopQuantumCosmology,AshtekarQuantumNatureOf,AshtekarLoopQuantumCosmologyFrom} (see also \cite{DienerNumericalSimulationsOf,AshtekarLoopQuantumCosmologyBianchiI} for results in non-isotropic cosmology). In these models, quantum geometry effects provide a Planck scale cutoff for spacetime curvature invariants which in turn induces a critical finite maximal value of the matter energy density, thus naturally resolving the initial singularity. Quantum effects lead then to an effective spacetime where the ``Big Bang'' is replaced by a ``Big Bounce'', i.e. a quantum regime which interpolates between a contracting and an expanding branch. The heart of the construction of the effective quantum theory and the source of the resulting bounce mechanism solving the classical singularity rely on a phase space regularisation usually called \textit{polymerisation} in the LQG literature, see e.g. \cite{CorichiPolymerQuantumMechanics}. The basic idea behind this procedure is the following: starting from the canonically conjugate phase space variables $(q,p)$ describing the geometry of the minisuperspace model under consideration (e.g., the volume $v$ and its conjugate momentum $b$ for FLRW cosmology), the passage to the effective quantum theory is achieved by replacing the momenta $p$ with their regularised version $\sin(\lambda p)/\lambda$, where $\lambda$ is a parameter (called ``polymerisation scale'') controlling the onset of quantum effects. The choice of $\lambda$ may be inspired by heuristic considerations of about the cosmological sector of full loop quantum gravity as e.g. in \cite{AshtekarQuantumNatureOf}, or by physical considerations of when quantum effects are supposed to become relevant, usually when the involved curvatures reach the Planck curvature. The structure of the modification is inspired by similar ones in loop quantum gravity that are closely related to lattice gauge theory supplemented with quantum geometry considerations, which suggest to take $\lambda$ at the Planck scale instead of taking the limit $\lambda \rightarrow 0$, see e.g. \cite{AshtekarQuantumNatureOf}.

The resulting phase space is then described by the configuration variables $q$ and the exponentiated momenta $e^{\pm i\lambda p}$ whose canonical Poisson bracket algebra corresponds to an adaptation to the symmetry reduced framework of the holonomy-flux algebra used in LQG. Remarkably, in the context of loop quantum cosmology (LQC) it was shown that the effective dynamics generated by the polymerised Hamiltonian agrees with the full quantum dynamics projected on a finite-dimensional submanifold spanned by properly constructed semiclassical states \cite{TaverasCorrectionstothe,DienerNumericalSimulationsOf,RovelliWhyAreThe}. The effective polymerised theory is thus capturing quantum geometry corrections descending from the loop quantised cosmological theory. 

In the light of the promising results obtained in the cosmological setting, the following question then naturally arises: are black hole singularities also resolved by LQG quantum geometry effects? The prototype spacetime geometry for addressing this question is provided by the Schwarzschild solution which as such has gained considerable attention over the last twenty years \cite{AshtekarQuantumGeometryand,ModestoLoopquantumblack,GambiniQuantumblackholes,BianchiWhiteHolesasRemnants,CorichiLoopquantizationof,ModestoSemiclassicalLoopQuantum,BoehmerLoopquantumdynamics,ChiouPhenomenologicalloopquantum,OlmedoFromblackholesto,AshtekarQuantumTransfigurarationof,AshtekarQuantumExtensionof,OritiBlackHolesas,BenAchourPolymerSchwarzschildBlack,BojowaldEffectivelineelements,LoboRainbow-likeBlackHole,Morales-TecotlEffectivedynamicsof}. The starting point of LQG-inspired analyses is the observation that the Schwarzschild interior region is isometric to the vacuum Kantowski-Sachs cosmological model. Techniques from homogeneous and anisotropic LQC can thus be imported to construct a Hamiltonian framework for the effective quantum theory according to the polymerisation procedure mentioned above. A common feature of these investigations is that in the quantum corrected Schwarzschild spacetime resulting from the effective equations, the central singularity is replaced by a transition surface between a trapped and an anti-trapped region respectively interpreted as black hole and white hole interior regions. However, although the qualitative picture of the quantum-extended interior regions derived in these effective models seems to agree, subtle differences and undesirable physical predictions come out in the previous proposals depending on whether the polymerization scales are considered to be purely constant or phase space dependent functions. According to this methodological distinction, previous LQG investigations can be divided in two main classes. In the so-called $\mu_o$-\textit{type schemes} \cite{AshtekarQuantumGeometryand,ModestoLoopquantumblack,CampigliaLoopquantizationof,ModestoSemiclassicalLoopQuantum}, the quantum parameters are assumed to be constant. These approaches however turn out to have drawbacks such as the final outcome fails to be independent of the fiducial structures introduced in the construction of the classical phase space and large quantum effects may survive even in the low-curvature regime. In the so-called $\bar\mu$-\textit{type schemes} \cite{ChiouPhenomenologicaldynamicsof,JoeKantowski-Sachsspacetimein,BoehmerLoopquantumdynamics,ChiouPhenomenologicalloopquantum} instead the quantum parameters are selected to be functions of the classical phase space. Although the dependence on fiducial structures is removed in these approaches, large quantum corrections near the horizon still survive. More recently, a generalisation of the $\mu_o$-scheme has been proposed in \cite{CorichiLoopquantizationof,OlmedoFromblackholesto,AshtekarQuantumTransfigurarationof,AshtekarQuantumExtensionof}. In these models, a mass dependence is introduced in the quantum parameters which then become Dirac observables, i.e. constant only along the trajectories solving the effective dynamics. These choices remarkably lead to effective models where both the two problems mentioned above are removed. In \cite{CorichiLoopquantizationof}, for instance, a mass dependence in one of the quantum parameters is introduced by the identification of the radius of the fiducial sphere with a physical length scale settled to be the classical Schwarzschild radius. Although the fiducial cell dependence is thus cured in \cite{CorichiLoopquantizationof}, the curvature scale at which quantum effects become dominant depends on the black hole mass and furthermore there is a huge mass amplification in the transition from the black to the white hole side. The \textit{generalised $\mu_o$-scheme} introduced in \cite{OlmedoFromblackholesto} has been recently improved in \cite{AshtekarQuantumTransfigurarationof,AshtekarQuantumExtensionof} by introducing a recursively defined effective Hamiltonian in which the quantum parameters are functions of the Hamiltonian itself.  The mass dependence of the quantum parameters, which was phenomenologically introduced in \cite{OlmedoFromblackholesto}, is then determined by means of quantum geometry arguments based on rewriting the curvature in terms of the holonomies of the gravitational connection along suitably chosen plaquettes enclosing the minimal area at the transition surface. Among the other desired features, this leads to a symmetric bounce for macroscopic black holes with no mass amplification with a universal upper bound on spacetime curvature invariants at which quantum effects get relevant. However, the equations of motion used in \cite{AshtekarQuantumTransfigurarationof,AshtekarQuantumExtensionof} to derive these results are qualitatively different from those following from the effective Hamiltonian in that paper due to a technical error \cite{BodendorferANoteOnTheHamiltonian}, which obfuscates the relation to LQG type models.

In this paper, we take a different route to construct an effective quantum theory of Schwarzschild black holes. Instead of using the standard connection variables on which all previous LQG investigations are based, we introduce a new classical phase space description based on canonical variables inspired by physical considerations about the onset of quantum effects. In particular, this allows to relate the on-shell momenta to the curvature and inverse area of the 2-spheres. The effective dynamics is then obtained via polymerisation with a constant scale. The resulting effective theory is characterised by a remarkably simple form of the polymerised Hamiltonian and all the desirable mentioned features of the resulting quantum corrected spacetime can be obtained, except for the absence of mass (de-)amplification (that may however not be ruled out by general arguments). 
Moreover, the simple form of the Hamiltonian allows us to explicitly construct the quantum theory and already perform some steps in solving it explicitly.
There are however some important differences between our approach and previous ones that will be discussed in more detail throughout the paper, such as a restriction of the possible initial conditions for physical viability.

The paper is organized as follows. In Sec. \ref{sec:Classical} we briefly recall the spacetime description of the classical Schwarzschild solution, its Hamiltonian formulation, and introduce then our new canonical variables.  In Sec. \ref{sec:effectivequantumtheory}, the polymerisation of the model is discussed and the corresponding effective equations of motion are solved. Sec. \ref{sec:curvinvariants} focusses on the physical consequences of the polymerisation scheme adopted, especially the curvature scale at which quantum effects become relevant. The structure of the quantum corrected effective spacetime is analysed and the Penrose diagram is construed in Sec. \ref{effectivestruc}. In Sec. \ref{sec:comparison}, we report a detailed comparison with previous investigations. The paper closes with a brief sketch of a possible quantisation of our model in Sec. \ref{quantumtheory}, while some concluding remarks and future directions are reported in Sec. \ref{sec:conclusions}.

\section{Classical theory}\label{sec:Classical}

\subsection{Spacetime description of classical Schwarzschild solution}

Let us start by recalling the main aspects of the classical Schwarzschild solution \cite{HawkingTheLargeScale,MisnerGravitation}. As this is intended just to be compared with the effective spacetime description resulting from the polymerised model, we will not enter the details and only report those aspects which are relevant for the purposes of the paper.

Spherically symmetric solutions of Einstein field equations are locally isometric to the Schwarzschild metric whose line element is given by

\be\label{clmetric}
\dd s^2=-\left(1-\frac{2M}{r}\right)\dd t^2+\left(1-\frac{2M}{r}\right)^{-1}\dd r^2+r^2\dd\Omega_2^2\;,
\ee

\noindent
where $\dd\Omega_2^2=\dd\theta^2+\sin^2\theta\,\dd\phi^2$ is the round metric on the unit sphere and we are using natural units in which $G=c=1$. The radial coordinate $r\in(0,+\infty)$ is defined by the requirement that $4\pi r^2$ be the area of the 2-spheres identified by $t=const$, $r=const$ which are the transitivity surfaces of the $\textsf{SO(3)}$ isometry group. The spacetime described by the metric \eqref{clmetric} is asymptotically flat since as $r\to\infty$ it reduces to the Minkowski metric in polar coordinates. The vector field $\partial/\partial t$ orthogonal to the hypersurfaces $t=const$ is a Killing vector field of the metric \eqref{clmetric} so that in the region $r>2M$ spacetime is static. The metric has a curvature singularity at $r=0$ as can be seen from the Kretschmann scalar
\be\label{clkretschmann}
\mathcal K=R_{\alpha\beta\gamma\delta}R^{\alpha\beta\gamma\delta}=\frac{48M^2}{r^6}\;.
\ee

\noindent
For $M=0$ the metric \eqref{clmetric} describes a flat spacetime. For $M>0$ it describes black holes, while for $M<0$ it describes naked singularities. In this work, we consider black hole solutions for which the constant $M$ can be interpreted as the black hole mass. The metric becomes singular also at $r=r_s=2M$ but this is just a coordinate singularity as it can be removed by changing the coordinate system. The null hypersurface $r=2M$ separates regions of spacetime where $r=const$ hypersurfaces  are time-like hypersurfaces ($r>2M$) from regions where these are space-like hypersurfaces ($r<2M$). The null hypersurface $r=2M$ is called horizon as objects crossing it from $r > 2 M$ can never come back. The maximal analytic extension of \eqref{clmetric} is obtained by introducing the so-called Kruskal-Szekeres coordinates as follows \cite{KruskalMaximalExtensionof,SzekeresOnTheSingularities}. First, we change coordinates from $(t,r,\theta,\phi)$ to $(u,v,\theta,\phi)$ with
\be
u=t-r_*\qquad,\qquad v=t+r_*
\ee

\noindent
where $r_*$ is the so-called tortoise coordinate defined by

\be
r_*=r+2M\log\left(\frac{r-2M}{2M}\right)\;.
\ee

\noindent
Lines of constant $v$ and $u$ respectively correspond to ingoing and outgoing null geodesics. In such a coordinate system the metric takes the form

\be\label{clmetric12}
\dd s^2=-\left(1-\frac{2M}{r}\right)\dd u\dd v+r^2\dd\Omega_2^2\;,
\ee

\noindent
where $r$ is determined by $(v-u)/2=r+2M\log\left(\frac{r-2M}{2M}\right)$. Kruskal coordinates in the exterior region $r>2M$ are then defined by

\be
T=\frac{1}{2}(V+U)\qquad,\qquad X=\frac{1}{2}(V-U)
\ee

\noindent
with $T \in \left(-\infty, \infty\right)$, $X > 0$ and $T^2-X^2 < 0$, and $r(X,T)$ defined by the implicit equation $T^2-X^2=UV=-\left(\frac{r-2M}{2M}\right)\exp(r/2M)$ with

\be
U=-\exp\left(-\frac{u}{4M}\right)\qquad,\qquad V=\exp\left(\frac{v}{4M}\right)
\ee

\noindent
$U < 0$ and $V > 0$ for all values of $r$. The metric \eqref{clmetric12} then takes the form

\be\label{clmetric3}
\dd s^2=\frac{32M^3}{r}\exp\left(-\frac{r}{2M}\right)(-\dd T^2+\dd X^2)+r^2\dd\Omega_2^2\;.
\ee

\noindent
The metric \eqref{clmetric3} is well-defined and non-singular for the whole range $T \in \mathbb{R}$ and $X \in \mathbb{R}$. In particular, the metric is non-singular at the horizon ($r=2M$) which in these coordinates is located at $T=\pm X$. The curvature singularity ($r=0$) is located at $T^2-X^2 =UV=1$. The maximally extended Schwarzschild geometry can be thus divided into four regions separated by event horizons: \texttt{I}) the black hole exterior region $-X<T<+X$ which is isometric to the exterior Schwarzschild solution ($r>2M$), \texttt{II}) the black hole interior region $|X|<T<\sqrt{1+X^2}$ which corresponds to $0<r<2M$ in Schwarzschild coordinates, \texttt{III}) the white hole exterior region $+X<T<-X$ which is again isometric to the exterior Schwarzschild solution and can be regarded as another asymptotically flat universe on the other side of the Schwarzschild throat, \texttt{IV}) the white hole interior region $-\sqrt{1+X^2}<T<-|X|$ corresponding to the region $0<r<2M$ on the other side. Light-like geodesics moving in a radial direction look like straight lines at a 45-degree angle in the $(X,T)$-plane. Therefore, any event inside the black hole interior region will have a future light cone that remains in this region, while any event inside the white hole interior region will have a past light cone that remains in this region. This means that there are no time-like or null curves which go from region \texttt{I} to region \texttt{III}. Curves of constant $r$ look like hyperbolas bounded by a pair of event horizons at 45 degrees, while lines of constant $t$-coordinate look like straight lines at various angles passing through the center $T=X=0$.

\begin{figure}[t!]
\centering\includegraphics[scale=0.35]{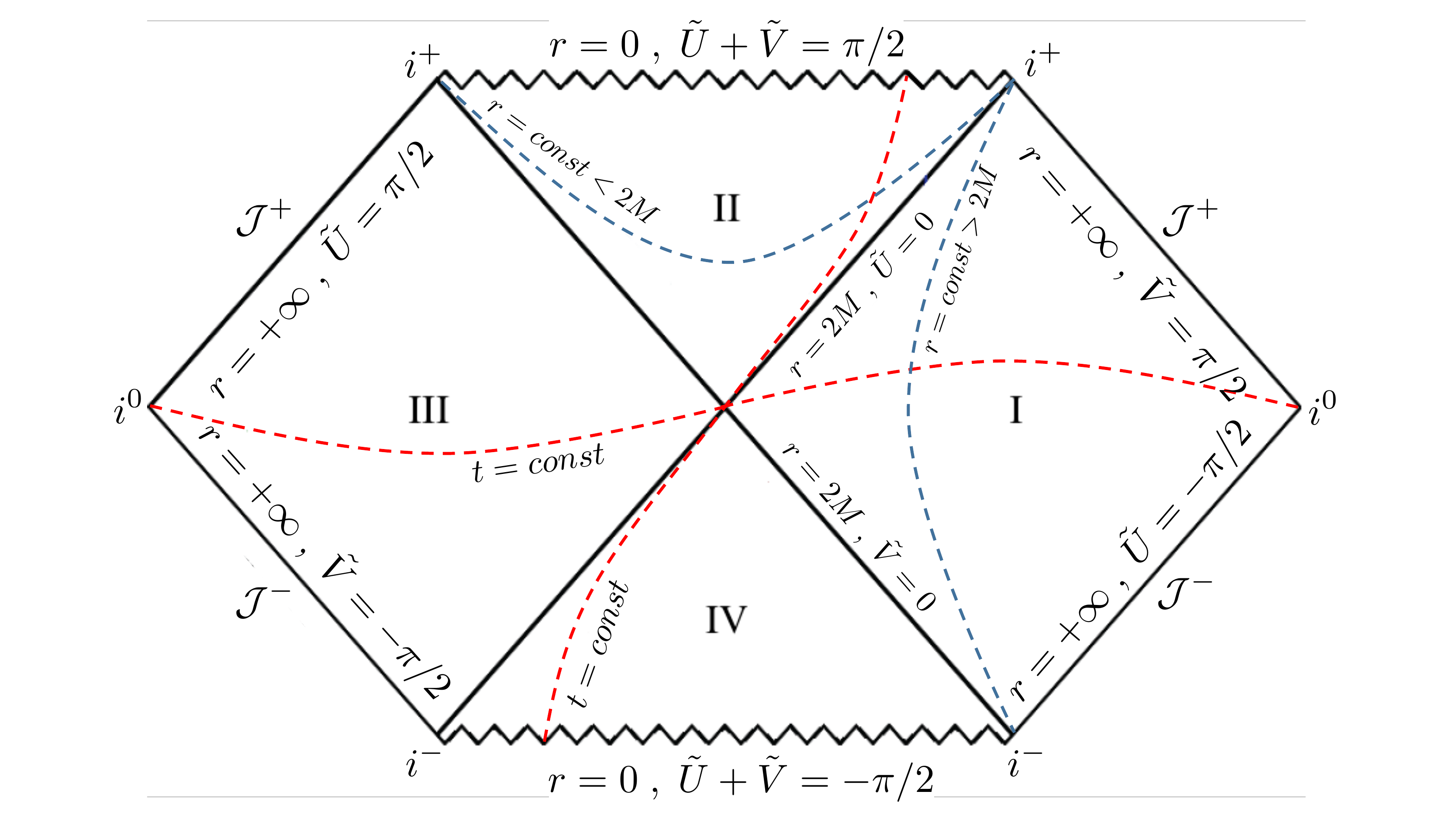}
\caption{Penrose diagram for the Kruskal extension of Schwarzschild spacetime. The angular coordinates $\theta,\phi$ are suppressed so that each point of the diagram can be thought of as representing a 2-sphere of radius $r$.}
\label{Penrosediag1}
\end{figure}

The causal structure of the Kruskal extension of the Schwarzschild geometry can be easily visualised by means of a Penrose diagram (Fig. \ref{Penrosediag1}). This is constructed by introducing a new set of null coordinates 

\be
\tilde U=\arctan U\;,\quad\tilde V=\arctan V\qquad(-\pi/2<\tilde U,\tilde V<\pi/2)
\ee

\noindent
and performing a conformal transformation of the metric such that the resulting line element is given by

\be
\dd\tilde s^2=4\cos^2\tilde U\cos^2\tilde V\dd s^2=-\frac{128M^3}{r}\exp\left(-\frac{r}{2M}\right)\dd\tilde U\dd\tilde V+4 r^2\cos^2\tilde U\cos^2\tilde V\dd\Omega_2^2\;,
\ee

\noindent
and the curvature singularity $UV=1$ corresponds to $\tilde U+\tilde V=\pm\frac{\pi}{2}$.

\subsection{Hamiltonian framework}\label{HamiltonainFramework}

To construct a Hamiltonian description of Schwarzschild spacetime we start from a generic static spherically symmetric line element of the form \cite{CavagliaHamiltonianFormalismfor,VakiliClassicalpolymerizationof}

\be\label{clmetric2}
\dd s^2=-\bar a(r)\dd t^2+N(r)\dd r^2+2\bar B(r)\dd t\dd r+\bar b^2(r)\dd\Omega_2^2\;,
\ee

\noindent
where $\bar a(r)$, $N(r)$, $\bar B(r)$ and $\bar b(r)$ are some functions of $r$. The function $N(r)$ plays the role of the lapse w.r.t. the foliation in $r$-slices \cite{CavagliaHamiltonianFormalismfor}. Substituting the metric \eqref{clmetric2} into the Einstein-Hilbert action ($G=c=1$)
\be
S_{EH}=\frac{1}{16\pi}\int\dd^4x\sqrt{-g}\,R\;,
\ee

\noindent
a straightforward calculation leads, up to boundary terms, to the action

\be\label{eq:action}
S=\frac{1}{4}\int\dd r L(\bar a,\bar b,\bar n)\;,
\ee

\noindent
with the effective Lagrangian given by

\be\label{lagrangian1}
L(\bar a,\bar b,\bar n)=2L_o\sqrt{\bar n}\left(\frac{\bar a'\bar b'\bar b}{\bar n}+\frac{\bar a\bar b'^2}{\bar n}+1\right)\;,
\ee

\noindent
where primes denote derivatives w.r.t. $r$ and $\bar n$ is a Lagrange multiplier given by

\be\label{lagrangemultiplier}
\bar n(r)=\bar a(r)N(r)+\bar B^2(r)\;,
\ee

\noindent
which reflects a gauge freedom in the definition of the coordinates $r$ and $t$. We introduced a fiducial cell $\mathcal C$ in the constant $r$ slices of topology $\mathbb R\times\mathbb S^2$ with an infrared cut-off $L_o$ in the non-compact $t$-direction (i.e., $t\in[0,L_o]$). The length $L_o$ of the fiducial cell in the expression \eqref{lagrangian1} of the Lagrangian can be absorbed by the following redefinition of the variables

\be\label{variableredef}
\sqrt{n}= \int_{0}^{L_o} \dd t \sqrt{\bar n} =L_o\sqrt{\bar n}\quad,\quad \sqrt{a} = \int_{0}^{L_o} \dd t \sqrt{\bar a} =L_o\sqrt{\bar a}\quad,\quad b=\bar b\quad,\quad B= \int_{0}^{L_o} \dd t \bar{B} =L_o\,\bar B\;,
\ee

\noindent
thus yielding the Lagrangian

\be\label{lagrangian2}
L(a,b,n)=2\sqrt{n}\left(\frac{a'b'b}{n}+\frac{ab'^2}{n}+1\right)\;.
\ee

\noindent
Note that $\sqrt{a}$ gives the physical length of the fiducial cell and as such it is independent of coordinate transformations, as we discuss later. 
Moreover, let us remark that $L_o$ is the coordinate length of the fiducial cell in $t$-direction and as such it depends on the choice of the chart. However, we can define the physical size of the fiducial cell in $t$-direction at a certain reference point $r^*$ by
	
\begin{equation}\label{L_o}
\mathscr{L}_o := \left.\sqrt{a}\right|_{r=r_{\text{ref}}} = L_o \left.\sqrt{\bar{a}}\right|_{r=r_{\text{ref}}}\;,
\end{equation}
\noindent	
which has the same behaviour of $L_o$ under fiducial cell rescaling, i.e. $\mathscr{L}_o \mapsto \alpha \mathscr{L}_o$ as $L_o \mapsto \alpha L_o$, but in contrast to $L_o$ does not transform under any coordinate transformation (preserving the form \eqref{clmetric2} of the metric), i.e. it is a spacetime scalar. The definition \eqref{L_o} depends explicitly on a reference point $r^*$. Nevertheless, $\mathscr{L}_o$ as well as $L_o$ are fiducial structures and hence what is physically relevant is that in the end all physical quantities would not depend on them.

The only independent variables that can be determined by the Einstein field equations are the functions $a(r)$ and $b(r)$ whose conjugate momenta are given by

\be\label{papb}
p_a=\frac{\del L}{\del a'}=\frac{2bb'}{\sqrt{n}}\qquad,\qquad p_b=\frac{\del L}{\del b'}=\frac{4ab'+2a'b}{\sqrt{n}}\;.
\ee 

\noindent
The momentum $p_n$ conjugate to $n$ vanishes, thus giving the primary constraint $p_n=\frac{\del L}{\del n'}\approx0$. The Hamiltonian associated to the Lagrangian \eqref{lagrangian2} is given by

\be\label{hamiltonian1}
H_{cl}=\sqrt{n} \mathcal{H}_{cl} +\Lambda p_n \quad, \quad \mathcal{H}_{cl} = \frac{p_ap_b}{2b}-\frac{ap_a^2}{2b^2}-2\;,
\ee

\noindent
where the primary constraint is implemented via the Lagrange multiplier $\Lambda(r)$. The stability algorithm of $p_n \approx 0$ gives furthermore the Hamiltonian constraint $ \mathcal{H}_{cl} \approx 0$. The equation of motion for $n$ yields $n'=\{n,H\}=\Lambda$ from which it follows that gauge-fixing $n$ to be constant is equivalent to setting $\Lambda=0$. With this gauge choice, the equations of motion for the other variables then read

\be\label{eom1}
\begin{cases}
a'=\sqrt{n}\left(\frac{p_b}{2b}-\frac{ap_a}{b^2}\right)\\
p_a'=\sqrt{n}\frac{p_a^2}{2b^2}\\
b'=\sqrt{n}\frac{p_a}{2b}\\
p_b'=\sqrt{n}\left(\frac{p_ap_b}{2b^2}-\frac{ap_a^2}{b^3}\right)\\
\mathcal{H}_{cl} = \frac{p_ap_b}{2b}-\frac{ap_a^2}{2b^2}-2 \approx 0
\end{cases}
\ee

\noindent
Note that although fiducial structures would explicitly enter the intermediate steps of the phase space analysis as all the otherwise-divergent integrals have to be restricted to $\mathcal C$, the physical quantities and the equations of motion have to be independent of the choice of the fiducial cell. In particular, under a rescaling of the fiducial length $L_o\mapsto\alpha\,L_o$ by a constant $\alpha$, the Lagrange multiplier transforms as

\be
\sqrt{n}\mapsto\alpha\,\sqrt{n}\;,
\ee

\noindent
while the variables \eqref{variableredef} and their conjugate momenta \eqref{papb} rescale as

\be\label{scaling}
a\mapsto\alpha^2\,a \quad,\quad b\mapsto b \quad,\quad p_a\mapsto\alpha^{-1}\,p_a \quad,\quad p_b\mapsto\alpha\,p_b\;,
\ee

\noindent
so that physical quantities can depend only on $b, a/L_o^2, L_o\,p_a, p_b/L_o$ or $b, a/\mathscr L_o^2, \mathscr L_o\,p_a, p_b/\mathscr L_o$ \footnote{Let us remark that $b, a/L_o^2, L_o\,p_a, p_b/L_o$ are fiducial cell independent but no spacetime scalars as $L_o$ depends directly on the choice of the $t$-coordinate. In contrast $b, a/\mathscr L_o^2, \mathscr L_o\,p_a, p_b/\mathscr L_o$ are fiducial cell independent as well as spacetime scalars. However, one has to be careful since a dependence on the reference point $r_{\text{ref}}$ may be introduced by the presence of $\mathscr L_o$ but in the end physical quantities will not depend on it.} and the equations of motion \eqref{eom1} are invariant under rescaling of the fiducial cell. 
  
As discussed e.g. in \cite{VakiliClassicalpolymerizationof}, solving these equations for $a(r)$, $b(r)$, $p_a(r)$, $p_b(r)$ and using the Hamiltonian constraint $\mathcal{H}_{cl}=0$, the Schwarzschild metric \eqref{clmetric} is recovered by choosing the lapse to be $N(r)=\left(1-\frac{2M}{r}\right)^{-1}$. However, in order to polymerise the model, we introduce a new set of canonical conjugate variables

\be\label{newvar}
\begin{aligned}
v_1=\frac{2}{3}b^3\qquad&,\qquad P_1=\frac{a'}{\sqrt{n}\,b}=\left(\frac{p_b}{2b^2}-\frac{ap_a}{b^3}\right)\;,\\
v_2=2ab^2\qquad&,\qquad P_2=\frac{b'}{\sqrt{n}\,b}=\frac{p_a}{2b^2}\;,
\end{aligned}
\ee

\noindent
satisfying the following Poisson brackets

\be\label{pvPB}
\begin{aligned}
&\{v_1,P_1\}=1\quad,\quad\{v_2,P_2\}=1\;,\\
&\{v_1,v_2\}=\{P_1,P_2\}=\{v_1,P_2\}=\{v_2,P_1\}=0\;,
\end{aligned}
\ee

\noindent
as can be checked by direct calculation. As it will be clear in the following, such variables turn out to be more simple to reasonably polymerise the model\footnote{A polymerisation scheme based on the variables $a(r)$, $b(r)$, $p_a(r)$, $p_b(r)$ has been discussed in \cite{VakiliClassicalpolymerizationof}. However, it does not resolve the classical singularity.}. Under a rescaling of the fiducial length $L_o\mapsto\alpha\,L_o$, the canonical variables \eqref{newvar} rescale as
\be\label{rescaling}
v_1\mapsto v_1\quad,\quad v_2\mapsto\alpha^2\,v_2\quad,\quad P_1\mapsto\alpha\,P_1\quad,\quad P_2\mapsto \alpha^{-1}\,P_2\,.
\ee

\noindent 
Therefore, physical quantities can depend only on the combinations $v_2/L_o^2,P_1/L_o,L_o P_2$ or $v_2/\mathscr L_o^2$, $P_1/\mathscr L_o, \mathscr L_o P_2$ and $v_1$. As we will discuss later in this section, the introduction of $\mathscr L_o$ will play a crucial role in the geometric interpretation of these variables. In particular, the scalar $P_1/\mathscr L_o$ can be related to the Kretschmann scalar. 

In the new variables, the Hamiltonian constraint acquires the remarkably simple expression

\be\label{hamiltonian2}
H_{cl}=\sqrt{n}\mathcal{H}_{cl} \quad,\quad \mathcal{H}_{cl} = 3v_1P_1P_2+v_2P_2^2-2 \approx 0\;.
\ee

\noindent
The corresponding equations of motion are thus given by
\be\label{eom2}
\begin{cases}
v_1'=3\sqrt{n}v_1P_2\\
v_2'=3\sqrt{n}v_1P_1+2\sqrt{n}v_2P_2\\
P_1'=-3\sqrt{n}P_1P_2\\
P_2'=-\sqrt{n}P_2^2\\
\mathcal{H}_{cl} = 3v_1P_1P_2+v_2P_2^2-2 \approx 0
\end{cases}
\ee

\noindent
which, as expected, are invariant under rescaling of the fiducial cell. Integrating now the fourth equation of \eqref{eom2}, choosing the gauge $\sqrt{n} = const. = \mathscr{L}_o$, we get

\be\label{p2solution}
P_2(r)=\frac{1}{\sqrt{n}(r+A)}\;,
\ee

\noindent
where we denote the integration constant by $A$. Substituting the expression \eqref{p2solution} of $P_2$ into the first and third equations of \eqref{eom2}, we get

\be
\frac{P_1'}{P_1}=-\frac{3}{(r+A)}\qquad,\qquad \frac{v_1'}{v_1}=\frac{3}{(r+A)}
\ee

\noindent
from which after integration it follows that

\be\label{p1v1solutions}
P_1(r)=\frac{C}{(r+A)^3}\qquad,\qquad v_1(r)=D\,(r+A)^3
\ee

\noindent
where $C$ and $D$ are integration constants. Finally, using the Hamiltonian constraint \eqref{hamiltonian2} together with the solutions \eqref{p2solution}, \eqref{p1v1solutions}, we find

\be\label{v2solution}
v_2(r)=\frac{2}{P_2^2(r)}-3v_1(r)\frac{P_1(r)}{P_2(r)}=  \sqrt{n} (r+A)^2\left(2\sqrt{n}-\frac{3CD}{r+A}\right)\;.
\ee

\noindent
Note that we have three integration constants $A,C,D$ for four equations of motion as we used the Hamiltonian constraint to solve one of them. Moreover, the integration constant $A$ just reflects the gauge freedom in shifting the $r$ coordinate and hence we can set $A=0$ without loss of generality.\footnote{This can be also seen on the one hand by rephrasing the equations of motion in a gauge independent way (i.e. solving them in terms of $b$) which reduces the number of equations to three and on the other hand by using the Hamiltonian constraint so that only two independent integration constants are left.} The remaining integration constants can be fixed in a gauge invariant way by means of Dirac observables. However, besides of the Hamiltonian itself, which we already used, there exists only one further independent Dirac observable given by

\begin{equation}\label{diracobs}
F = \left(\frac{3}{2} v_1\right)^\frac{4}{3} P_1 P_2 = \frac{b^2 a' b'}{n}\;, 
\end{equation}

\noindent
which, according to Eqs. \eqref{rescaling}, is also invariant under fiducial cell rescaling, as it should\footnote{As Dirac observables also determine the dynamics of the system, of course there exists one further independent phase space function which commutes with the Hamiltonian. However, as it turns out to be not invariant under fiducial cell rescaling, it cannot be physical.}. Inserting the solutions \eqref{p2solution}, \eqref{p1v1solutions} and \eqref{v2solution} into \eqref{diracobs}, we get

\begin{equation}
F = \left(\frac{3}{2} D\right)^\frac{4}{3} \frac{C}{\sqrt{n}} \;,
\end{equation}

\noindent
which then gives only one condition for a combination of both $C$ and $D$. Therefore, since there are no further fiducial cell independent Dirac observables, it is not possible to  find a second gauge invariant condition which allows to determine $C$ and $D$ individually. As discussed in the following, all other $C$ and $D$ dependences can be removed by choosing a gauge.

Now, in order to deal with coordinate independent quantities we need to express $a$ in terms of $b$, the latter being a scalar under $t$- and $r$-transformations\footnote{As in usual general relativity jargon, in what follows we will refer to $b$ as the physical radius to distinguish it from the radial coordinate $r$. However, properly speaking, the physical distance from the center to the surface of a $t = const$, $b=const$ sphere is given by $\int \sqrt{g_{rr}} dr$, while $b$ appears in the surface area $A = 4\pi b^2$.}.
Reversing the definitions \eqref{newvar} to express $a$ and $b$ in terms of $v_1$ and $v_2$ and expressing $a$ in terms of $b$, we get

\begin{equation}\label{aofbclass}
a(b) = \frac{\mathscr L_o^2}{ \left(\frac{3 D}{2}\right)^{\frac{2}{3}}} \left(1 - \left(\frac{3}{2} D\right)^{\frac{4}{3}} \frac{C}{\sqrt{n}} \frac{1}{b}\right) = \frac{\mathscr L_o^2}{ \left(\frac{3 D}{2}\right)^{\frac{2}{3}}} \left(1 - \frac{F}{b}\right) \;,
\end{equation}

\noindent
where $\sqrt{n} = \mathscr L_o$ is used. Note that, and this will play a crucial role in the following, $a(b)$ is gauge independent under $t$ and $r$ redefinitions. Indeed, recalling the definition of $\sqrt{a}$ as

$$
\sqrt{a} = \int_{0}^{L_o} dt \sqrt{\bar{a}} \; ,
$$
\noindent
we see that, using $\bar{a} = g_{tt}$, it is not sensitive to $r$-coordinate transformations. Nonetheless, since under a $t$-redefinition $t \mapsto \tau$ $\bar{a}$ transforms as $\bar{a} \mapsto \tilde{\bar{a}} = \bar{a} (dt/d\tau)^2$, $a$ remains unchanged, i.e.

\begin{equation}
\sqrt{a} = \int_{0}^{L_o} dt \sqrt{\bar{a}} = \int_{\tau(0)}^{\tau(L_o)} d\tau \sqrt{\tilde{\bar{a}}} \;.
\end{equation}
\noindent
In particular, for a constant rescaling $t \mapsto \tau = const. \cdot t$, we have

\begin{equation}
\sqrt{a} = L_o \sqrt{\bar{a}} = \int_{0}^{L_o} dt \sqrt{\bar{a}} = \int_{\tau(0)}^{\tau(L_o)} d\tau \sqrt{\tilde{\bar{a}}}  = \tau(L_o) \sqrt{\tilde{\bar{a}}}\; ,
\end{equation}
\noindent
with $\tau(L_o) = const. \cdot L_o$. Thus, for $a$ unaffected by this transformation, $\bar{a}$ as well as $L_o$ have to transform accordingly. The metric component $g_{\tau \tau}$ is then given as $a$ divided by the coordinate distance in $\tau$ which corresponds to the coordinate distance $L_o$ in the $t$-chart, i.e. $g_{\tau \tau} = a/\tau(L_o)^2$. Recalling Eq. \eqref{aofbclass}, this means that there exists a chart $\tau$ in which $\mathscr L_o/(3D/2)^{(1/3)} = \tau(L_o)$, such that $g_{\tau \tau} = 1-F/b$. Therefore, we get rid of the $D$ dependence in $g_{\tau \tau}$ by this gauge transformation. As already discussed, this reflects that $D$ cannot be physical as it was indicated before by the existence of only one $\mathscr L_o$-independent Dirac observable.

Analogous considerations hold also for $\sqrt{n}$ and $\sqrt{\bar{n}}$, hence in $\tau$-chart, we have $\bar{n} = n/\tau(L_o)^2$. In contrast to $a$, $n$ is sensitive to $r$-redefinitions as it depends on the metric components $N$ and $B$. Therefore, we can set $\bar{n} = 1$ in $\tau$-chart by transforming $r$ accordingly to cancel the additional factor.
 
We can now write down the line element \eqref{clmetric2} only in terms of $F$ and remove the remaining $C$-,  $D$-dependent prefactors by redefining coordinates as $t \mapsto \tau = \mathscr L_o(3D/2)^{-\frac{1}{3}} t/L_o$, $r \mapsto b = (3D/2)^\frac{1}{3} r$, thus yielding

\be\label{clmetric4}
\dd s^2=-\bar a(b)\dd \tau^2+N(b)\dd b^2+2\bar B(b)\dd \tau\dd b+ b^2\dd\Omega_2^2\;,
\ee

\noindent
with

\be\label{compclmetric}
\begin{aligned}
&\bar{a}(b)=1-\frac{F}{b}\;,\\
&\bar B(b)\stackrel{\eqref{lagrangemultiplier}}{=}\left[1-\left(1-\frac{F}{b}\right)N(b)\right]^{1/2}\;.
\end{aligned}
\ee

\noindent
The classical Schwarzschild solution \eqref{clmetric} is thus recovered by choosing $F=2M$ and $N(b)=\left(1-\frac{2M}{b}\right)^{-1}$ and all dependencies on fiducial structures $\mathscr L_o$, $L_o$ and the reference $r_{\text{ref}}$ are removed. 
This also allows us to relate the Dirac observable \eqref{diracobs} to the black hole mass $M$\footnote{Of course we could now just rename $b \mapsto r$, to come back to the usual notation. We could also fix $D = 2/3$, i.e. $\tau = \mathscr L_o t/L_o$ and $b = r$, which is then the gauge choice usually chosen in the classical Schwarzschild setting. Here it was important to show that the standard Schwarzschild result can be obtained even without fixing both integration constants. Later on, we will use the identification $b = r$ for the classical solution.}. 
Moreover, this provides us with an \textit{on-shell} geometric interpretation for the canonical momenta. Indeed, substituting the expressions \eqref{compclmetric} for the metric components into the definitions \eqref{newvar} of $P_1$ and $P_2$, we find
\be\label{interpretation}
\frac{P_1(b)}{\mathscr L_o}=\frac{2M}{b^3} \left(\frac{2}{3D}\right)^\frac{1}{3} \qquad,\qquad P_2(b) \mathscr L_o=\frac{1}{b} \left(\frac{3 D}{2}\right)^\frac{1}{3} 
\ee
from which it follows that, for mass independent $D$, $P_1(b)/\mathscr L_o$ is related to the square root of the Kretschmann scalar \eqref{clkretschmann}, while $P_2(b) \mathscr L_o$ is related to the angular components of the extrinsic curvature (w.r.t. $r$) by the relation $P_2(b) \mathscr L_o \left(\frac{3 D}{2}\right)^{-\frac{1}{3}}  = 1/b =\sqrt{N(b)}K^{\phi}_{\phi}=\sqrt{N(b)}K^{\theta}_{\theta}$. Let us remark that, consistently with the statement below Eq. \eqref{rescaling}, now the fiducial cell rescaling independent quantities are $P_1(b)/\mathscr L_o$, $P_2(b)\mathscr L_o$, which also are spacetime scalars. As we will discuss in the following sections, this on-shell interpretation guarantees that in the polymerised effective theory quantum effects become relevant at high curvatures and small radii for an appropriate fixing of the integration constants. Let us also remark that polymerizing $P_1$ with constant scale here plays the role of polymerising the spatial mean curvature (Hubble rate) in LQC with a constant scale as in \cite{AshtekarRobustnessOfKey}. There, the spatial mean curvature is proportional to the square root of the spacetime Ricci scalar.

\section{Effective quantum theory}\label{sec:effectivequantumtheory}

In the previous section we constructed a Hamiltonian function for a general spherically symmetric spacetime and showed that the resulting Hamiltonian equations yield the standard form of the Schwarzschild metric. Now, we will discuss the polymerisation of this minisuperspace model and solve the resulting effective Hamiltonian equations to get a polymer quantum corrected effective Schwarzschild metric. As also done in earlier approaches \cite{AshtekarQuantumGeometryand,ModestoLoopquantumblack,CorichiLoopquantizationof,ModestoSemiclassicalLoopQuantum,BoehmerLoopquantumdynamics,ChiouPhenomenologicalloopquantum,AshtekarQuantumTransfigurarationof,AshtekarQuantumExtensionof}, $r$ being a time-like coordinate in the interior of a black hole ($a < 0$, $N<0$), the interior region can be foliated by homogeneous space-like three-dimensional hypersurfaces and it is then isometric to the vacuum Kantowski-Sachs cosmological model. This allows to import techniques from homogeneous and anisotropic loop quantum cosmology and construct a Hamiltonian framework for the effective quantum theory. Moreover, in analogy to the classical case, the fact that we are considering canonical variables directly related to the metric coefficients will allow us to solve the effective quantum equations also in the exterior, where the interpretation of a time evolution fails. Once we have at our disposal explicit expressions for the effective solutions -- well-defined both in the interior and exterior regions -- we can check their reliability in the exterior.

\subsection{Polymerisation of the model}\label{Polymerisation}

In the spirit of LQC, the semi-classical features of the model should be captured by an effective Hamiltonian obtained by polymerising the canonical momenta. This amounts to replace them by functions of their point holonomies. The reasoning behind this procedure is to consider quantum theories of minisuperspaces that are constructed with similar techniques as full loop quantum gravity. It turns out that the dynamics of these theories is well approximated in the large quantum number limit by the above effective classical theory \cite{DienerNumericalSimulationsOf,TaverasCorrectionstothe,RovelliWhyAreThe} that includes corrections arranged in a power series with an expansion parameter $\lambda^2$ related to $\hbar$. Strictly speaking, such a theory should be considered as phenomenological model only unless shown to arise as the symmetry reduced sector of a fundamental quantum gravity theory\footnote{Much work has been invested into constructing examples for this, see e.g. \cite{BojowaldSphericallySymmetricQuantum, AlesciLoopQuantumCosmology, GielenHomogeneousCosmologiesAs, BeetleDiffeomorphismInvariantCosmological, BZI, BodendorferAnEmbeddingOf}. The main conceptual point is to bridge from a description involving many small quantum numbers to a description involving few large quantum numbers via the analogue of block-spin transformations, see e.g. \cite{BodendorferCoarseGrainingAs}. Also, going beyond the homogeneous sector may result in unexpected surprises such as signature change, see e.g. \cite{BojowaldSignatureChangeInTwo, BenAchourPolymerSchwarzschildBlack}.}. The current state of the field can therefore be described as an exploration of the physical consequences of certain choices which should eventually constrain the parameter space of physically viable models.

A commonly adopted and particularly simple choice in LQC-literature (see e.g. \cite{AshtekarMathematicalStructureOf, AshtekarLoopQuantumCosmology}) is the $\sin$ function, i.e.

\begin{align}
P_1 &\longmapsto \frac{\sin\left(\lambda_1 P_1\right)}{\lambda_1} \;,\label{polyP1}
\\
P_2 &\longmapsto \frac{\sin\left(\lambda_2 P_2\right)}{\lambda_2} \;, \label{polyP2}
\end{align}

\noindent
where $\lambda_1$ and $\lambda_2$ denote the polymerisation scales, but many other examples are possible. In fact, finding the ``correct'' polymerisation function is the main obstacle to deriving a unique model and most likely requires input from the anomaly-freedom of the constraint algebra, see e.g. \cite{LaddhaTheDiffeomorphismConstraint, LaddhaHamiltonianConstraintIn}, protection of classical algebraic structures, see e.g. \cite{BenAchourThiemannComplexifierIn}, or renormalization, see e.g. \cite{BodendorferCoarseGrainingAs}. In this paper, we use the above choice for simplicity while being well aware that other choices may lead to different phenomenology \cite{HellingHigherCurvatureCounter, DaporCosmologicalEffectiveHamiltonian}.

Both scales are related to the Planck length $\ell_p$. This polymerisation describes well the classical behaviour in the $\lambda_1 P_1 \ll 1$ and $\lambda_2 P_2 \ll 1$ regime, since

\begin{align*}
\frac{\sin\left(\lambda_1 P_1\right)}{\lambda_1} &\simeq P_1 + \mathcal{O}\left(\lambda_1^2\right), \quad \lambda_1 P_1 \ll 1 \\
\frac{\sin\left(\lambda_2 P_2\right)}{\lambda_2} &\simeq P_2 + \mathcal{O}\left(\lambda_2^2\right) , \quad \lambda_2 P_2 \ll 1 
\end{align*}

\noindent
As discussed in Sec. \ref{HamiltonainFramework}, $P_1$ is in the classical regime directly related to the square root of the Kretschmann scalar. Hence, the replacement \eqref{polyP1} leads to corrections in the Planck curvature regime, as we should expect from the quantum theory\footnote{We note that arguments involving the area gap such as \cite{AshtekarQuantumNatureOf} are also only heuristic because they refer to full LQG in the low spin regime, i.e. where the quanta of area are close to the area gap. It is unclear whether the effective LQC dynamics, which is successful for large volumes, is accurate here. On top, it just transfers the problem of choosing a polymerisation scheme to the full theory, but does not solve it. Without additional insights, one can better understand such schemes as demanding boundedness of curvature invariants as they cut off the integrated gravitational connection at order $1$ over distances of order $1$ in natural units. They are thus in similar spirit as ours.}.

In turn, $P_2$ is a measure of the angular components of the extrinsic curvature. Quantum effects play a role, i.e. \eqref{polyP2} will give corrections, in the regime where these components of the extrinsic curvature become large. This is the case for small areas of the 2-spheres ($b^2 \approx \ell_p^2$), which allows the interpretation of the polymerisation of $P_2$ as small distance corrections. We will analyse this features later on in Sec. \ref{sec:curvinvariants} in more detail.

Notice that, since $P_1$ and $P_2$ scale according to \eqref{rescaling} under rescaling of the fiducial cell $\mathscr L_o$, also $\lambda_1$ and $\lambda_2$ have to scale according to

\begin{align}\label{lambdascaling}
\lambda_1 \longmapsto \frac{1}{\alpha} \lambda_1 \quad ,\quad \lambda_2 \longmapsto \alpha \lambda_2 \;,
\end{align}

\noindent
from which it follows that the scale invariant, i.e. physical, quantities are $\mathscr L_o \lambda_1$ and $\lambda_2/\mathscr L_o$. As $\mathscr L_o$ is a spacetime scalar while $L_o$ is not, only the combinations of $\lambda_1$ and $\lambda_2$ with $\mathscr L_o$ are physically meaningful.
Further, we can study the dimension of the parameters $\lambda_1$ and $\lambda_2$. To this aim, let us recall that since $\bar{a}$ is dimensionless, $\left[a\right] = L^2$, while $\left[b\right] = L$, $[n] = L^2$, where $L$ denotes the dimension of length. Therefore, recalling the definitions \eqref{newvar} of $P_1$ and $P_2$, we find

\begin{align*}
\left[P_1\right] &= \left[\frac{a'}{\sqrt{n}b}\right] = \frac{1}{L} \; ,\\
\left[P_2\right] &= \left[\frac{b'}{\sqrt{n} b}\right] = \frac{1}{L^2} \;.
\end{align*}

\noindent
Being the products $\lambda_1 P_1$ and $\lambda_2 P_2$ dimensionless, the dimensions of $\lambda_1$ and $\lambda_2$ are then given by

\begin{align*}
\left[ \lambda_1 \right] &= \left[\frac{1}{P_1}\right] = L\quad , \quad \left[\mathscr L_o \lambda_1 \right] = L^2 \;,
\\
\left[\lambda_2 \right] &= \left[\frac{1}{P_2}\right] = L^2\quad , \quad \left[\frac{\lambda_2}{\mathscr L_o}\right] = L \;.
\end{align*}

\noindent
The physical scale $\mathscr L_o \lambda_1$ has dimension $L^2$ and hence can be interpreted as inverse curvature, i.e. should be related to the inverse Planck curvature. Instead $\lambda_2/\mathscr L_o$ has dimension length, i.e. $\lambda_2/\mathscr L_o$ should be related to the Planck length.

\subsection{Solution of the effective equations}\label{sec:solution}

The polymerised effective Hamiltonian\footnote{It is worth to remark that this Hamiltonian has a relatively simple structure thinking about a possible quantum theory. As will be sketched in Sec. \ref{quantumtheory}, assigning operators to the canonical variables is (up to the usual ordering ambiguities) straight forward, since all variables occur only with positive and at most quadratic powers.} reads now:

\begin{equation}\label{Heff1}
H_{\text{eff}} = \sqrt{n} \mathcal{H}_{\text{eff}}\quad , \quad \mathcal{H}_{\text{eff}} = 3v_1 \frac{\sin\left(\lambda_1 P_1\right)}{\lambda_1} \frac{\sin\left(\lambda_2 P_2\right)}{\lambda_2} + v_2 \frac{\sin\left(\lambda_2 P_2\right)^2}{\lambda_2^2} - 2 \approx 0
\end{equation}

\noindent
As in the classical case, we will use the gauge in which $\sqrt{n} = const. = \mathscr L_o$ and $B = 0$. The equations of motion for the Hamiltonian \eqref{Heff1} are given by

\begin{numcases}{}
v_1' = 3 \sqrt{n} v_1 \cos\left(\lambda_1 P_1\right) \frac{\sin\left(\lambda_2 P_2\right)}{\lambda_2} \label{v1quantEoM} \;,
\\
v_2' = 3 \sqrt{n} v_1 \frac{\sin\left(\lambda_1 P_1\right)}{\lambda_1} \cos\left(\lambda_2 P_2\right) + 2 \sqrt{n} v_2 \frac{\sin\left(\lambda_2 P_2\right)}{\lambda_2} \cos\left(\lambda_2 P_2\right) \label{v2quantEoM} \;,
\\
P_1' = -3 \sqrt{n} \frac{\sin\left(\lambda_1 P_1\right)}{\lambda_1} \frac{\sin\left(\lambda_2 P_2\right)}{\lambda_2} \label{P1quantEoM} \;,
\\
P_2' = - \sqrt{n} \frac{\sin\left(\lambda_2 P_2\right)^2}{\lambda_2^2} \label{P2quantEoM} \;,
\\
\mathcal{H}_{\text{eff}} = 3v_1 \frac{\sin\left(\lambda_1 P_1\right)}{\lambda_1} \frac{\sin\left(\lambda_2 P_2\right)}{\lambda_2} + v_2 \frac{\sin\left(\lambda_2 P_2\right)^2}{\lambda_2^2} - 2 \approx 0 \label{hamiltonianquantEoM} \;.
\end{numcases}

\noindent
Note that the structure of the equations with respect to their coupling with each other is similar to the classical case. Hence, we can apply the same solution strategy: first solving \eqref{P2quantEoM} for $P_2$, then inserting the result into \eqref{P1quantEoM}, which then only depends on $P_1$ and can be solved. Both, $P_1$ and $P_2$ can be inserted into \eqref{v1quantEoM} which then is also decoupled. Finally, all the previous results can be inserted into the Hamiltonian constraint \eqref{hamiltonianquantEoM} to find $v_2$.

Starting with $P_2$, the integration of \eqref{P2quantEoM} leads to

\begin{equation}\label{P2int}
\cot\left(\lambda_2 P_2\right) = \frac{\sqrt{n}}{\lambda_2} \left(r+A\right) \; ,
\end{equation}
\noindent
where $A$ is an integration constant. In analogy to the classical case the integration constant $A$ reflects the gauge freedom in shifting $r$. Without loss of generality we can then set $A = 0$. The more general case of non-zero $A$ can be recovered by shifting the coordinate $r \mapsto r +A$.
Solving this equation for $P_2$ requires some caution, since the function $\cot(x)$ has different branches, which are not taken care by the inverse function $\text{arccot}(x) = \cot^{-1}(x)$. To be continuous at $r = 0$\footnote{Note that $r$ is just a coordinate and does not necessarily coincide with the radius of the 2-sphere, which is strictly speaking $b(r)$. Hence, in principle also negative values are allowed, i.e. $r \in \left(-\infty,\infty\right)$. Whether this extension of the domain is meaningful or necessary has to be checked afterwards.}, the proper inversion is:

\begin{equation}\label{P2sol}
P_2(r) = \frac{1}{\lambda_2} \cot^{-1}\left(\frac{\sqrt{n}\; r}{\lambda_2}\right) + \frac{\pi}{\lambda_2} \theta\left(-\frac{\sqrt{n}\; r}{\lambda_2}\right) \;,
\end{equation}
\noindent
where $\theta(x)$ is the Heavyside-step-function. Taking the limit into the classical regime, i.e. $\sqrt{n} r/\lambda_2 \gg 1$, we find

\begin{equation}
P_2(r) = \frac{1}{\lambda_2} \cot^{-1}\left(\frac{\sqrt{n}\; r}{\lambda_2}\right) + \frac{\pi}{\lambda_2} \theta\left(-\frac{\sqrt{n}\; r}{\lambda_2}\right) \xrightarrow{ \frac{ \sqrt{n} r}{\lambda_2} \rightarrow \infty} \frac{1}{\sqrt{n}\; r} \;, 
\end{equation}
\noindent
which corresponds to the classical result \eqref{p2solution}. Note that the physically relevant quantities are 

\begin{align}\label{sinP2}
\frac{\sin\left(\lambda_2 P_2\right)}{\lambda_2} &= \frac{1}{\lambda_2} \frac{1}{\sqrt{1+\left(\frac{\sqrt{n}\;r}{\lambda_2}\right)^2}} \;,
\\
\label{cosP2}
\cos\left(\lambda_2 P_2\right) &= \frac{1}{\lambda_2} \frac{\sqrt{n}\;r}{\sqrt{1+\left(\frac{\sqrt{n}\;r}{\lambda_2}\right)^2}} \;.
\end{align}
\noindent
Inserting the expression \eqref{sinP2} into \eqref{P1quantEoM} yields

\begin{equation}\label{P1'}
P_1' = -3\sqrt{n}\; \frac{\sin\left(\lambda_1 P_1\right)}{\lambda_1} \frac{1}{\lambda_2} \frac{1}{\sqrt{1+\left(\frac{\sqrt{n}\;r}{\lambda_2}\right)^2}} \;,
\end{equation}
\noindent
which can be integrated to

\begin{equation}\label{eq:P1sol1}
P_1(r) = \frac{2}{\lambda_1} \cot^{-1}\left(\frac{\lambda_2^3}{4 C \lambda_1 \sqrt{n}^3}\left( \frac{\sqrt{n}\;r}{\lambda_2} + \sqrt{1+\frac{n\;r^2}{\lambda_2^2}} \right)^3\right) \; ,
\end{equation}
\noindent
where $C$ is an integration constant. Note that since $\sqrt{n}/\lambda_2$ is fiducial cell independent and $\lambda_1$ scales according to \eqref{lambdascaling}, $C$ has to scale as $C \mapsto \alpha C$. Moreover, the argument of $\cot^{-1}(x)$ is always positive, i.e. there are no continuity issues here. Taking the limit into the classical regime $\sqrt{n} r/\lambda_2 \gg 1$ and $2 r^3/C \lambda_1 \gg 1$, we get

\begin{equation}\label{eq:P1limit}
P_1(r) \xrightarrow{\frac{\sqrt{n} r}{\lambda_2} \rightarrow \infty} \frac{2}{\lambda_1} \cot^{-1}\left(\frac{2 r^3}{C \lambda_1}\right) \xrightarrow{\frac{2 r^3}{C \lambda_1} \rightarrow \infty} \frac{C}{r^3}\;,
\end{equation} 
\noindent
which is in agreement with the classical solution \eqref{p1v1solutions}. Again, the physically relevant quantities are

\begin{align}\label{sinP1}
\frac{\sin\left(\lambda_1 P_1\right)}{\lambda_1} &=  \frac{\lambda_2^3}{2 C \lambda_1^2 \sqrt{n}^3}  \frac{\left( \frac{\sqrt{n} r}{\lambda_2} + \sqrt{1+\frac{n r^2}{\lambda_2^2}} \right)^3}{\frac{\lambda_2^6}{16 C^2 \lambda_1^2 n^3} \left( \frac{\sqrt{n} r}{\lambda_2} + \sqrt{1+\frac{n r^2}{\lambda_2^2}} \right)^6 +1}\;,
\\
\label{cosP1}
\cos\left(\lambda_1 P_1\right) &= \frac{\frac{\lambda_2^6}{16 C^2 \lambda_1^2 n^3} \left( \frac{\sqrt{n} r}{\lambda_2} + \sqrt{1+\frac{n r^2}{\lambda_2^2}} \right)^6 - 1}{\frac{\lambda_2^6}{16 C^2 \lambda_1^2 n^3} \left( \frac{\sqrt{n} r}{\lambda_2} + \sqrt{1+\frac{n r^2}{\lambda_2^2}} \right)^6 + 1}\;.
\end{align}
\noindent
Inserting \eqref{sinP2} and \eqref{cosP1} into \eqref{v1quantEoM}, we get

\begin{equation}
\frac{v_1'}{v_1} = 3 \frac{\sqrt{n}}{\lambda_2} \frac{1}{\sqrt{1+\left(\frac{n r}{\lambda_2}\right)^2}} \frac{\frac{\lambda_2^6}{16 C^2 \lambda_1^2 n^3} \left( \frac{\sqrt{n} r}{\lambda_2} + \sqrt{1+\frac{n r^2}{\lambda_2^2}} \right)^6 - 1}{\frac{\lambda_2^6}{16 C^2 \lambda_1^2 n^3} \left( \frac{\sqrt{n} r}{\lambda_2} + \sqrt{1+\frac{n r^2}{\lambda_2^2}} \right)^6 + 1} \;,
\end{equation}
\noindent
which can be integrated to

\begin{equation}
v_1(r) = \frac{2 C^2 \lambda_1^2 \sqrt{n}^3}{\lambda_2^3} D \frac{\frac{\lambda_2^6}{16 C^2 \lambda_1^2 n^3} \left( \frac{\sqrt{n} r}{\lambda_2} + \sqrt{1+\frac{n r^2}{\lambda_2^2}} \right)^6 +1 }{\left( \frac{\sqrt{n} r}{\lambda_2} + \sqrt{1+\frac{n r^2}{\lambda_2^2}} \right)^3} \; ,
\end{equation}
\noindent
where $D$ is an integration constant. Again note that according to Eq. \eqref{rescaling}, $D$ is fiducial cell independent. The limit into the classical regime yields

\begin{equation}
v_1(r) \xrightarrow{\frac{\sqrt{n} r}{\lambda_2} \rightarrow \infty} D \frac{C^2 \lambda_1^2}{4 r^3} \left( \frac{4 r^6}{C^2 \lambda_1^2} +1 \right) \xrightarrow{\frac{2 r^3}{C \lambda_1} \rightarrow \infty} D r^3 \; ,
\end{equation}
\noindent
in agreement with the classical behaviour \eqref{p1v1solutions}.
By using now the Hamiltonian constraint \eqref{hamiltonianquantEoM} together with Eq. \eqref{sinP2} and noticing that

\begin{equation}
v_1 \frac{\sin\left(\lambda_1 P_1\right)}{\lambda_1} = const. = C D \;,
\end{equation}
\noindent
we can calculate $v_2(r)$ as

\begin{align}
v_2(r) &= \frac{\lambda_2^2}{\sin\left(\lambda_2 P_2\right)^2} \left( 2- 3v_1 \frac{\sin\left(\lambda_1 P_1\right)}{\lambda_1} \frac{\sin\left(\lambda_2 P_2\right)}{\lambda_2} \right)
\notag
\\
&=2 n \left(\frac{\lambda_2}{\sqrt{n}}\right)^2 \left(1+\frac{n r^2}{\lambda_2^2}\right) \left( 1 - \frac{3 C D}{ 2 \lambda_2} \frac{1}{\sqrt{1+\frac{n r^2}{\lambda_2^2}}} \right) \; ,
\end{align}
\noindent
which has the desired scaling behaviour (cfr. \eqref{rescaling}) as all the combinations $\lambda_2/\sqrt{n}$ and $C/\lambda_2$ are fiducial cell independent and only the overall $n$-factor scales with $\alpha^2$. Again the classical behaviour \eqref{v2solution} is recovered in the limit $\sqrt{n} r/\lambda_2 \gg 1$, namely

\begin{equation}
v_2(r) \xrightarrow{\frac{\sqrt{n} r}{\lambda_2}\rightarrow \infty} 2 n r^2 \left( 1 - \frac{3 C D}{2 \sqrt{n} r} \right) \; . 
\end{equation}

To sum up, the solutions of the effective equations are given by

\begin{align}
v_1(r) &= \frac{2 C^2 \lambda_1^2 \sqrt{n}^3}{\lambda_2^3} D \frac{\frac{\lambda_2^6}{16 C^2 \lambda_1^2 n^3} \left( \frac{\sqrt{n} r}{\lambda_2} + \sqrt{1+\frac{n r^2}{\lambda_2^2}} \right)^6 +1 }{\left( \frac{\sqrt{n} r}{\lambda_2} + \sqrt{1+\frac{n r^2}{\lambda_2^2}} \right)^3} \; ,\label{solutionv1quant}
\\
v_2(r) &= 2 n \left(\frac{\lambda_2}{\sqrt{n}}\right)^2 \left(1+\frac{n r^2}{\lambda_2^2}\right) \left( 1 - \frac{3 C D}{2 \lambda_2} \frac{1}{\sqrt{1+\frac{n r^2}{\lambda_2^2}}} \right) \; , \label{solutionv2quant}
\end{align}
\begin{align}
P_1(r) &= \frac{2}{\lambda_1} \cot^{-1}\left( \frac{\lambda_2^3}{4 C \lambda_1 \sqrt{n}^3} \left( \frac{\sqrt{n} r}{\lambda_2} + \sqrt{1+\frac{n r^2}{\lambda_2^2}} \right)^3\right) \; , \label{solutionP1quant}
\\
P_2(r) &= \frac{1}{\lambda_2} \cot^{-1}\left(\frac{\sqrt{n} r}{\lambda_2}\right) + \frac{\pi}{\lambda_2} \theta\left(-\frac{\sqrt{n}\; r}{\lambda_2}\right) \;, \label{solutionP2quant}
\end{align}

\noindent
which, according to the scaling behaviours \eqref{lambdascaling} and 

\begin{equation}
C \longmapsto \alpha C \quad , \quad D \longmapsto D \;,
\end{equation}

\noindent
all have the desired behaviour under fiducial cell rescaling, and agree with the classical solutions in the classical regime $\sqrt{n} r/\lambda_2\gg 1$, $2 r^3/C \lambda_1 \gg 1$\footnote{As can be easily checked the two limits commute.}.

Given the solutions \eqref{solutionv1quant}-\eqref{solutionP2quant} of the effective equations it is easy to reconstruct the metric components $a$ and $b$ due to the relations \eqref{newvar}. Specifically, we find

\begin{align}
b &= \left(\frac{3 v_1}{2}\right)^{\frac{1}{3}} = \frac{\sqrt{n}}{\lambda_2}\left( 3 D C^2 \lambda_1^2\right)^\frac{1}{3} \frac{\left(\frac{\lambda_2^6}{16 C^2 \lambda_1^2 n^3} \left( \frac{\sqrt{n} r}{\lambda_2} + \sqrt{1+\frac{n r^2}{\lambda_2^2}} \right)^6 +1 \right)^\frac{1}{3}}{\left( \frac{\sqrt{n} r}{\lambda_2} + \sqrt{1+\frac{n r^2}{\lambda_2^2}} \right)}\; , \label{bquant}\\
a &=  \frac{v_2}{2 b^2}  = \frac{v_2}{2} \left(\frac{2}{3 v_1}\right)^{\frac{2}{3}} \notag \\ 
&= n \left(\frac{\lambda_2}{\sqrt{n}}\right)^4 \left(1+\frac{n r^2}{\lambda_2^2}\right) \left( 1 - \frac{3 C D}{2 \lambda_2} \frac{1}{\sqrt{1+\frac{n r^2}{\lambda_2^2}}} \right) \frac{ \left(\frac{1}{3 D C^2 \lambda_1^2}\right)^{\frac{2}{3}} \left( \frac{\sqrt{n }r}{\lambda_2} + \sqrt{1+\frac{n r^2}{\lambda_2^2}} \right)^2}{\left(\frac{\lambda_2^6}{16 C^2 \lambda_1^2 n^3} \left( \frac{\sqrt{n} r}{\lambda_2} + \sqrt{1+\frac{n r^2}{\lambda_2^2}} \right)^6 +1 \right)^{\frac{2}{3}}} \; , \label{aquant} 
\end{align}

\noindent
and the line element then reads

\begin{equation}\label{effmetric}
\dd s^2 = -\frac{a(r)}{L_o^2} \dd t^2 + \frac{\mathscr L_o^2}{a(r)} \dd r^2 + b(r)^2 \left(\dd\theta^2 + \sin\left(\theta\right)^2 \dd\phi^2\right) \;,
\end{equation}

\noindent
where we used the expression of the metric coefficient $\bar{a} = a/L_o^2$ (cfr. \eqref{variableredef}) and the fact that $\sqrt{n} = \mathscr L_o$ as stated in the beginning of the section. Note that all the solutions \eqref{solutionv1quant}-\eqref{solutionP2quant} as well as \eqref{bquant} and \eqref{aquant} are smoothly well-defined in the whole $r$-domain $r \in \left(-\infty,\infty\right)$. As will be discussed in the next section, this observation will play a crucial role in determining the integration constants $C$ and $D$ by means of Dirac observables.

\subsection{Fixing the integration constants}\label{sec:intconst}

In the previous section we explicitly solved the effective equations of motion and rewrote the effective spacetime metric in terms of the corresponding solutions \eqref{solutionv1quant}-\eqref{solutionP2quant} in which two still undetermined integration constants, $C$ and $D$, occur\footnote{In analogy with the classical case, we might expect four integration constants. However, as discussed above, one of them is set to zero as it encodes the freedom in shifting the radial coordinate, while we get rid of another integration constant by using the Hamiltonian constraint.}. To fix them in a gauge independent way, we use the following two Dirac observables

\begin{align}
F_Q =&\; 3 v_1 \frac{\sin\left(\lambda_1 P_1\right)}{\lambda_1} \frac{\left( \frac{3}{2} v_1 \cos^2\left(\frac{\lambda_1 P_1}{2}\right) \right)^\frac{1}{3}}{\lambda_2 \cot\left(\frac{\lambda_2 P_2}{2}\right)} \label{F} \;,
\\
\bar{F}_Q =&\; 3 v_1 \frac{\sin\left(\lambda_1 P_1\right)}{\lambda_1} \left( \frac{3}{2} v_1 \sin^2\left(\frac{\lambda_1 P_1}{2}\right) \right)^\frac{1}{3} \frac{\cot\left(\frac{\lambda_2 P_2}{2}\right)}{\lambda_2} \label{Fbar} \;,
\end{align}

\noindent
which can be easily constructed by looking at the solutions of the effective equations and, as can be checked by direct computation, commute with the Hamiltonian and are also fiducial cell independent. Note that $F_Q$ reduces to $F$ (cfr. Eq. \eqref{diracobs}) in the limit $\lambda_1, \lambda_2 \rightarrow 0$, while $\bar{F}_Q$ is not well-defined in this limit coherently with it not being present at the classical level. Indeed, it is possible to multiply $\bar{F}_Q$ by a suitable power of $\lambda_1$ and $\lambda_2$ such that the limit exists and yields a classical Dirac observable. Nevertheless, this introduces a fiducial cell dependence, as $\lambda_1$ and $\lambda_2$ scale with $\mathscr L_o$ (cfr. \eqref{lambdascaling}). Let us also remark that in the classical case there is only one independent non-zero curvature invariant, e.g. the Kretschmann scalar. Consistently, as discussed in Sec. \ref{HamiltonainFramework}, there is only one Dirac observable related to the black hole mass which can be used to fix the initial value of the Kretschmann scalar and hence completely determining the system. In the effective quantum theory, instead, there are two independent non-zero curvature invariants, say the Kretschmann scalar and the Ricci scalar, which in turn means that two Dirac observables ($F_Q$ and $\bar{F}_Q$) have to be specified to uniquely determine the system.

Being Dirac observables, $F_Q$ and $\bar{F}_Q$ are constant along the solutions of the effective dynamics and their on-shell evaluation reads

\be
F_Q = \left(\frac{3}{2} D\right)^{\frac{4}{3}} \frac{C}{\sqrt{n}} \quad , \quad
\bar{F}_Q = \frac{3 C D \sqrt{n}}{\lambda_2^2} \left(3 D C^2 \lambda_1^2\right)^{\frac{1}{3}}  \; . \label{Fonshell}
\ee

\noindent
As both Dirac observables $F_Q$ and $\bar{F}_Q$ are gauge independent and do not scale under fiducial cell rescaling, it is possible to give them a physical interpretation. To this aim, we adopt the following strategy. As already stressed before, $r$ is just a coordinate and has no physical meaning, hence in order to get gauge (coordinate) independent expressions we should rephrase all the quantities in terms of the physical radius $b$. Therefore, we first calculate $a(b)$, then take the limit $b \rightarrow \infty$ corresponding to the classical regime, and use the resulting expression to recast the metric in a coordinate-free Schwarzschild-like form, thus providing an interpretation for $F_Q$ and $\bar{F}_Q$. Specifically, inverting Eq. \eqref{bquant} leads to
\begin{equation}\label{rofb}
r^{(\pm)}(b) = \frac{\lambda_2}{2\sqrt{n}} \frac{z_{\pm}^2(b)-1}{z_{\pm}(b)}\quad ,\quad z_{\pm}(b) = \left(\frac{8}{3 D } \left(\frac{\sqrt{n} b}{\lambda_2}\right)^3 \pm \frac{4 C \lambda_1 \sqrt{n}^3}{\lambda_2^3} \sqrt{\frac{4 b^6}{9 \lambda_1^2 D^2 C^2}-1}\right)^{\frac{1}{3}}
\end{equation}
\noindent
which has two distinct branches in the positive and negative $r$ range, respectively. As will be discussed in Sec. \ref{CSPD}, this indicates two distinct asymptotic regions of the effective spacetime. Indeed, the $b \rightarrow \infty$ limit of Eq. \eqref{rofb} yields

\begin{equation}
z_\pm(b) \xrightarrow{b \rightarrow \infty} \begin{cases}
z_+ \simeq \left(\frac{16}{3 D }\right)^{\frac{1}{3}} \frac{\sqrt{n} b}{\lambda_2}  \\
z_- \simeq \left(3 D C^2 \lambda_1^2\right)^{\frac{1}{3}} \frac{\sqrt{n}}{\lambda_2 b}
\end{cases} \;, 
\end{equation}  

\noindent
respectively corresponding to $r^{(+)} \to + \infty$ and $r^{(-)} \to - \infty$. Plugging now Eq. \eqref{rofb} into the expression \eqref{aquant} of $a(r)$ allows us to express $a$ as a function of $b$. As can be checked by direct calculation, the resulting expression for $a(b)$ also exhibits two branches $a_\pm(b)\equiv a(r^{(\pm)}(b))$, which for $b\to\infty$ are given by

\begin{equation}
a_\pm(b) \xrightarrow{b\to\infty} \begin{cases}
a_+ \simeq \frac{n}{4} \left(\frac{16 }{3 D }\right)^{\frac{2}{3}} \left(1- \frac{F_Q}{b}\right) \\
a_- \simeq \frac{n}{4} \left(\frac{\lambda_2}{\sqrt{n}}\right)^4 \left(\frac{1}{3 D C^2 \lambda_1^2}\right)^{\frac{2}{3}} \left(1- \frac{\bar{F}_Q}{b}\right)
\end{cases} \; ,
\end{equation}

\noindent
where we used the on-shell expressions \eqref{Fonshell} of $F_Q$ and $\bar{F}_Q$. Note that the point $b_{\mathcal{T}}$ where $z_{+}(b_{\mathcal{T}})=z_{-}(b_{\mathcal{T}})$ corresponds to the minimal value of $b$ and, as can be easily seen from Eq. \eqref{rofb}, we have $b_{\mathcal{T}} = (3\lambda_1 CD/2)^{\frac{1}{3}}$. In what follows, the 3-dimensional surface $b = b_{\mathcal{T}}$ will be called transition surface, while its meaning as well as its physical interpretation will be clear once the structure of the effective spacetime is studied (Sec. \ref{effectivestruc}). 

Finally, we are in the same situation as in the classical case and by rescaling $t \mapsto \tau = \mathscr L_o(3D/2)^{-\frac{1}{3}} t/L_o$  and accordingly $r \mapsto b = (3D/2)^{\frac{1}{3}} r$ for the positive branch as well as $t \mapsto \tau = \mathscr L_o (24 D C^2 \lambda_1^2 \mathscr L_o^6/\lambda_2^6)^{-\frac{1}{3}} t/L_o$, $r \mapsto b = (24 D C^2 \lambda_1^2 \mathscr L_o^6/\lambda_2^6)^{\frac{1}{3}} (-r)$ for the negative branch, the line element \eqref{effmetric} takes the form

\begin{align}
\dd s_+^2 \simeq& -\left(1-\frac{F_Q}{b}\right) \dd \tau^2 + \frac{1}{1-\frac{F_Q}{b}} \dd b^2 + b^2 \dd\Omega_2^2 \;, \\
\dd s_-^2 \simeq& -\left(1-\frac{\bar{F}_Q}{b}\right) \dd\tau^2 + \frac{1}{1-\frac{\bar{F}_Q}{b}} \dd b^2 + b^2 \dd\Omega_2^2 \;.
\end{align}
\noindent
This shows that the two asymptotic regions are described by Schwarzschild spacetimes with the different asymptotic masses $F_Q/2$ and $\bar{F}_Q/2$. We will refer to the positive branch as black hole exterior and the negative branch as white hole exterior and hence $M_{BH} = F_Q/2$ as black hole and $M_{WH} = \bar{F}_Q/2$ as white hole masses, respectively. Therefore, the integration constants $C$ and $D$ can be completely fixed by giving the independent boundary data $M_{BH}$ and $M_{WH}$\footnote{Note that already in \cite{BoehmerLoopquantumdynamics} a dependence of the white hole mass (more precise: white hole horizon) on the initial conditions was observed. However, there the white hole mass was fiducial cell dependent and no phase space expression as \eqref{Fbar} for the white hole mass observable was exhibited.}, namely

\begin{equation}
M_{BH}= \left(\frac{3}{2} D\right)^{\frac{4}{3}} \frac{C}{2 \sqrt{n}} \quad , \quad
M_{WH} = \frac{3 C D \sqrt{n}}{2 \lambda_2^2} \left(3 D C^2 \lambda_1^2\right)^{\frac{1}{3}}  \;,
\end{equation}
\noindent
and the inverse relations

\begin{equation}\label{CandD}
 C = \frac{\lambda_2^3}{4 \lambda_1 \sqrt{n}^3}\left(\frac{M_{WH}}{M_{BH}}\right)^{\frac{3}{2}} \quad , \quad D = \left(\frac{2 \sqrt{n}}{\lambda_2}\right)^3\left( \frac{2}{3}\left(\frac{\lambda_1 \lambda_2}{3}\right)^3 M_{BH}^3 \left(\frac{M_{BH}}{M_{WH}}\right)^{\frac{9}{2}} \right)^{\frac{1}{4}}\;.
\end{equation}

This is a clear difference compared to the classical case where there was only one $\mathscr L_o$-independent Dirac observable for two integration constants. However, as we will discuss in Sec. \ref{sec:curvinvariants}, studying the behaviour of the Kretschmann scalar constrains the boundary data to be related to each other if certain physical viability criteria are met.

Fixing the constants and studying the asymptotic behaviour allowed us already to get some insight into the spacetime structure. Leaving a detailed discussion for Sec. \ref{effectivestruc}, here we just summarise its main aspects as it will be useful for the discussion of the next section as well as to fix some terminology. First of all, we found that the range of the coordinate $r$ can be extended to the full range $r \in \mathbb{R}$, which tells us again that $r$ is just a coordinate and not the physical radius, which is $b$. Furthermore, as opposed to the classical case, $b(r)$ does not reach zero but it has a minimal value $b_{\mathcal{T}}= (3\lambda_1CD/2)^{\frac{1}{3}}$ (see Sec. \ref{sec:transitionsurf} for more details). The transition surface, where $b$ has a non-zero minimal value separating the two branches $b_\pm(r)$, replaces the classical singularity, as shown for instance by plotting the Kretschmann scalar (cfr. Fig. \ref{fig:kretschmann} below). In the two branches, for large radii (i.e. into the classical regime), the spacetime is asymptotically a Schwarzschild spacetime with masses $M_{BH}$ and $M_{WH}$, respectively. Without further discussion, which will be provided in Sec. \ref{sec:curvinvariants}, the two masses can be chosen arbitrarily. We refer to the mass of the positive branch $M_{BH}$ as black hole mass and the mass of the negative branch $M_{WH}$ as white hole mass. This will be more clear once the Penrose diagram is constructed (cfr. Sec. \ref{CSPD}). Note that the names `black' and `white' hole have no definite meaning and can be completely exchanged as an observer in the `white hole' asymptotic region would experience this region as the exterior Schwarzschild spacetime of a `black hole'. As discussed in detail in Sec. \ref{CSPD}, this can be made more precise. We keep however this terminology as it helps to distinguish the two branches and it will acquire more meaning by studying the causal structure of the effective spacetime. Finally, as it will be discussed in detail in Sec. \ref{sec:horizon}, coherently with the asymptotic regions being black hole spacetimes, they admit a black hole horizon for each side, which will be now modified by quantum corrections. 

\begin{figure}[t!]
	\centering
	\subfigure[]
	{\includegraphics[width=7.75cm,height=5.5cm]{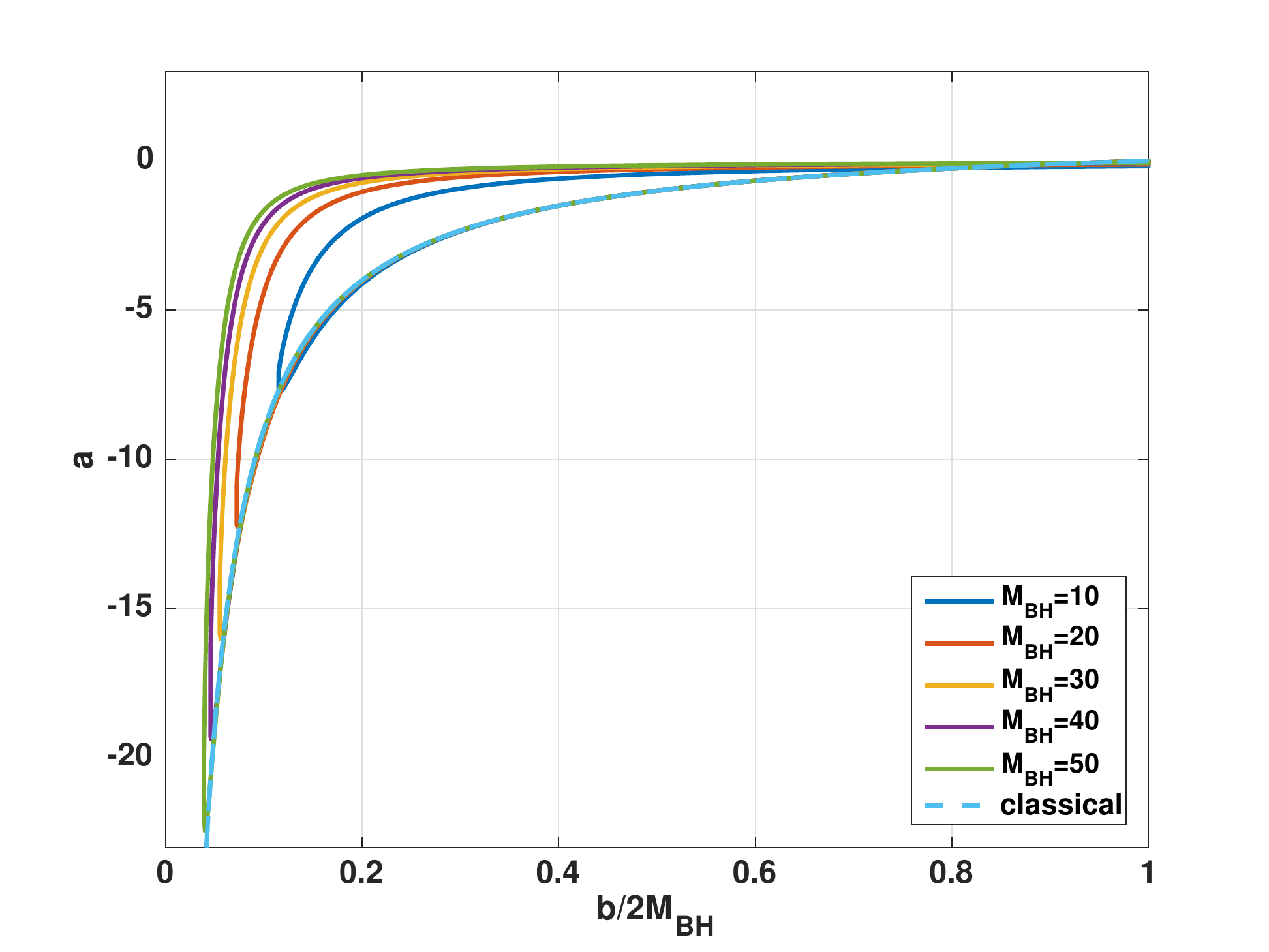}}
	\hspace{2mm}
	\subfigure[]
	{\includegraphics[width=7.75cm,height=5.5cm]{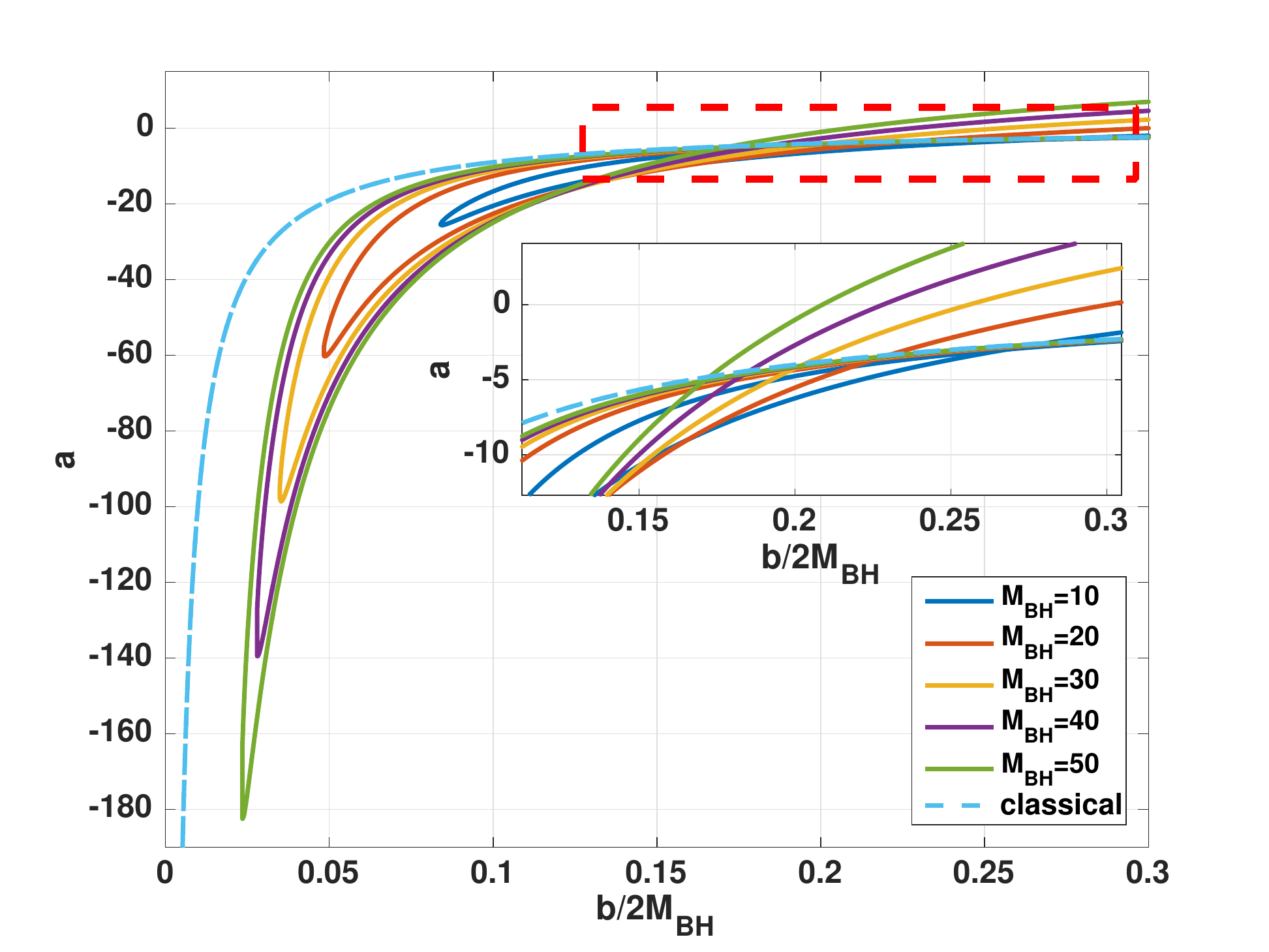}}
	\caption{Plot of $a$ as a function of $b$ for different black hole masses for the parameters $\mathscr L_o\lambda_1= \lambda_2/\mathscr L_o =1$. We choose $M_{WH} = M_{BH} \left(\frac{M_{BH}}{m}\right)^{\beta-1}$ for $m = 1$ and $\beta = \frac{5}{3}$ in (a) as well as $\beta = \frac{3}{5}$ in (b). In the plots, $a$ takes only negative values indicating that the interior region of the black hole is depicted. The plot shows only the interior of the black hole and shows already there a good agreement with the classical solution for larger $b$. Furthermore, a minimal value for $b$, i.e. $b_{\mathcal{T}}$ is visible.}
	\label{fig:aofb2}
\end{figure}

For visualising our effective metric, Fig. \ref{fig:aofb2} shows the plot of $a(b)$. As discussed, in principle we are free to choose $M_{BH}$ and $M_{WH}$ arbitrarily. For the plots we fixed the white hole mass being a function of the black hole mass due to the relation $M_{WH} = M_{BH} \left(\frac{M_{BH}}{m}\right)^{\beta-1}$, where $m$ is a constant of dimension mass. We choose $m = 1$ and $\beta = \frac{5}{3}$ as well as $\beta = \frac{3}{5}$. This choice is discussed in detail in Sec. \ref{sec:curvinvariants}. The plots nicely show that the effective spacetime approaches the classical result already inside the black hole, and coincides with the classical result for larger $b$. The bouncing behaviour in $b$ at the transition surface is also visible.

\section{Curvature invariants and onset of quantum effects}\label{sec:curvinvariants}

A still remaining but important question is at which scale quantum effects become relevant. For phenomenologically viable models, we expect quantum effects to be small and negligible in the classical (i.e. the low curvature) regime. In turn, we expect quantum effects to be relevant at high curvatures. It is possible to specify more precisely the meaning of $\lambda_1$ and $\lambda_2$ and consequently when quantum effects really become relevant by asking when the approximation $\sin(\lambda_1 P_1) \simeq \lambda_1 P_1$ holds, i.e. which limits correspond to the classical regime. Recalling e.g. Eqs. \eqref{eq:P1sol1} and \eqref{eq:P1limit}, we find that for positive and large $r$ the classical regime is given by

\begin{equation}\label{limits+}
\frac{\mathscr L_o r}{\lambda_2} \gg 1 \quad , \quad \frac{2 r^3}{C \lambda_1} \gg 1 \;,
\end{equation}

\noindent
while for negative $r$, we reach the asymptotic classical Schwarzschild spacetime for 

\begin{equation}\label{limits-}
\frac{\mathscr L_o |r|}{\lambda_2} \gg 1 \quad , \quad \frac{32 C \lambda_1 \mathscr L_o^6 |r|^3}{\lambda_2^6} \gg 1 \;.
\end{equation}

\noindent
These expressions depend on the choice of the $r$-coordinate, and hence to get coordinate-free conditions it is convenient to re-express $r$ in terms of $b$. This can be done for both the positive and negative $r$-branches. Let then consider them separately. For the positive branch we find

$$
b(r \rightarrow \infty) \simeq \left(\frac{3 D}{2}\right)^{\frac{1}{3}} r =: b_+ \;,
$$

\noindent
for which the conditions \eqref{limits+} read

\begin{equation}\label{limitsb+}
\frac{\mathscr L_o}{\lambda_2} \left(\frac{2}{3 D}\right)^{\frac{1}{3}} b_+ \gg 1 \quad , \quad \frac{1}{\mathscr L_o^2 \lambda_1^2} \gg \frac{9 C^2 D^2}{16 \mathscr L_o^2 b_+^6}  \quad , \;
\end{equation}

\noindent
where we squared the second condition. As discussed in Sec. \ref{sec:Classical}, the classical quantity $P_1/\mathscr L_o$ can be related to the Kretschmann scalar only if the integration constant $D$ is mass independent. Hence, for quantum effects to become relevant at a unique curvature scale, we expect to have to relate the initial data $M_{BH}$ and $M_{WH}$ with each other. The r.h.s. of the second equation of \eqref{limitsb+} can be related to the classical Kretschmann scalar of the black hole side by demanding

\begin{equation}\label{MBH+}
\frac{9 C^2 D^2}{16 \mathscr L_o^2} \propto M_{BH}^2 \;.
\end{equation}

\noindent
Assuming then a simple relation of the kind 

\begin{equation}
M_{WH} = M_{BH} \left(\frac{M_{BH}}{\bar{m}_{(\beta)}}\right)^{\beta-1} \sim M_{BH}^\beta \;,
\label{eq:BHWH}
\end{equation}
\noindent
where $\bar{m}_{(\beta)}$ is an arbitrary constant of dimension mass, condition \eqref{MBH+} is satisfied for $\beta = \frac{5}{3}$ for which we have

\begin{equation}\label{amplification}
M_{WH} = M_{BH} \left(\frac{M_{BH}}{\bar{m}_{\left(\frac{5}{3}\right)}}\right)^{\frac{2}{3}} \;,
\end{equation}

\noindent
which describes a mass dependent amplification of the white hole side.
For such a value of $\beta$, Eq. \eqref{CandD} yields

\begin{equation}\label{CDbeta5/3}
\frac{C}{\mathscr L_o} = \frac{\lambda_2^3}{4 \lambda_1 \mathscr L_o^4} \frac{M_{BH}}{\bar{m}_{\left(\frac{5}{3}\right)}} = 2 \frac{M_{BH}}{m_{\left(\frac{5}{3}\right)}} \quad , \quad D = \frac{1}{3} \left(\frac{2 \mathscr L_o}{\lambda_2}\right)^3 \left[2 (\bar{m}_{\left(\frac{5}{3}\right)} \lambda_1 \lambda_2)^3\right]^{\frac{1}{4}} = \frac{2}{3} \left(m_{\left(\frac{5}{3}\right)}\right)^\frac{3}{4} \;,
\end{equation}

\noindent
where we defined the dimensionless constant $m_{\left(\beta\right)} = 8 \lambda_1 \mathscr L_o^4 \bar m_{\left(\beta\right)}/\lambda_2^3$. Note that as $C$ and $D$ remain finite in the limit $\lambda_1, \lambda_2 \rightarrow 0$ (cfr. Sec. \ref{sec:solution}), $m_{\left(\beta\right)}$ remains finite as well. The conditions \eqref{limitsb+} then become

\begin{equation}
b_+ \gg \left(m_{\left(\frac{5}{3}\right)}\right)^{\frac{1}{4}} \frac{\lambda_2}{\mathscr L_o} \quad , \quad \frac{1}{\mathscr L_o^2 \lambda_1^2} \gg \frac{M_{BH}^2}{\left(m_{\left(\frac{5}{3}\right)}\right)^{\frac{1}{2}} b_+^6} = \frac{1}{48 \left(m_{\left(\frac{5}{3}\right)}\right)^{\frac{1}{2}}} \mathcal{K}_{BH}^{class} \;,
\end{equation}

\noindent
from which it follows that the critical length and the curvature scale where quantum effects get relevant are given by

\begin{equation}
\ell^{\left(\frac{5}{3}\right)}_{crit} = \left(m_{\left(\frac{5}{3}\right)}\right)^{\frac{1}{4}} \frac{\lambda_2}{\mathscr L_o} \quad , \quad \mathcal{K}^{\left(\frac{5}{3}\right)}_{crit} = \frac{48 \left(m_{\left(\frac{5}{3}\right)}\right)^{\frac{1}{2}}}{\mathscr L_o^2 \lambda_1^2 }\;.
\end{equation}

Note that the relation \eqref{amplification} is consistent with $\sin(\lambda_1 P_1)/\mathscr L_o \lambda_1$ being related to the classical Kretschmann scalar on the black hole exterior as 

$$
\frac{\sin\left(\lambda_1 P_1\right) }{\mathscr L_o \lambda_1}\stackrel{r \gg 1}{\simeq}  \frac{C}{\mathscr L_o r^3}  =  \frac{3 C D}{2 \mathscr L_o b_{+}^3} = 2 \left(m_{\left(\frac{5}{3}\right)}\right)^{-\frac{1}{4}} \frac{M_{BH}}{b_+^3} \propto \sqrt{\mathcal{K}_{BH}^{class}}\;.
$$

Moreover, given the relation \eqref{amplification}, we can ask when quantum effects become relevant on the white hole side. To this aim, re-expressing \eqref{limits-} in terms of $b_-:=(24 D C^2 \lambda_1^2 \mathscr L_o^6/\lambda_2^6)^{\frac{1}{3}} |r|$ $ \simeq b(r \rightarrow - \infty)$ yields 

\begin{equation}\label{limitsb-}
b_- \gg \frac{\mathscr L_o}{\lambda_2} \left(24 D C^2 \lambda_1^2\right)^{\frac{1}{3}} \quad , \quad \frac{1}{\mathscr L_o^2 \lambda_1^2} \gg \frac{9 D^2 C^2}{16 \mathscr L_o b_-^6} \;,
\end{equation}

\noindent
which together with Eq. \eqref{CDbeta5/3} leads to

\begin{equation}\label{critscalesWH+}
b_- \gg \frac{M_{WH}}{M_{BH}} \ell^{\left(\frac{5}{3}\right)}_{crit} \quad , \quad \frac{M_{WH}^2}{M_{BH}^2}\mathcal{K}^{\left(\frac{5}{3}\right)}_{crit} \gg \mathcal{K}_{WH}^{class} \;.
\end{equation}

\noindent
As $M_{WH} > M_{BH}$ for $\beta = \frac{5}{3}$, both scales are larger than the critical scales on the black hole side derived above. Therefore, on the white hole side curvature effects become relevant only at higher curvatures while small area effects become relevant already at larger areas.

Let us now consider the $b$-branch in negative $r$ domain. The conditions \eqref{limits-} corresponding to the classical regime can be now rewritten in terms of $b_- := (24 D C^2 \lambda_1^2 \mathscr L_o^6/\lambda_2^6)^{\frac{1}{3}} |r|$ thus yielding 

\begin{equation}\label{limitsb-2}
b_- \gg \frac{\mathscr L_o}{\lambda_2} \left(24 D C^2 \lambda_1^2\right)^{\frac{1}{3}} \quad , \quad \frac{1}{\mathscr L_o^2 \lambda_1^2} \gg \frac{9 D^2 C^2}{16 \mathscr L_o b_-^6} \;.
\end{equation}

\noindent
Following the same logic as before, we can now relate the r.h.s. of the second equation of Eq. \eqref{limitsb-2} with the classical Kretschmann scalar on the white hole side by setting

\begin{equation}\label{MWH-}
\frac{9 C^2 D^2}{16 \mathscr L_o^2} \propto M_{WH}^2 \;.
\end{equation}

\noindent
The ansatz \eqref{eq:BHWH} then satisfies the condition \eqref{MWH-} for $\beta = \frac{3}{5}$ thus yielding

\begin{equation}\label{deamplification}
M_{WH} = M_{BH} \left(\frac{M_{BH}}{\bar{m}_{\left(\frac{3}{5}\right)}}\right)^{-\frac{2}{5}} \;,
\end{equation}

\noindent
i.e. a de-amplified white hole mass. For such a value of $\beta$, Eq. \eqref{CandD} yields

\begin{equation}\label{CDbeta3/5}
\frac{C}{\mathscr L_o} = \frac{\lambda_2^3}{4 \lambda_1 \mathscr L_o^4} \frac{\bar{m}_{\left(\frac{3}{5}\right)}}{M_{WH}}  \quad , \quad D= \frac{2}{3} \left[ \left(\bar{m}_{\left(\frac{3}{5}\right)}\right)^{-5} \left(\frac{8 \mathscr L_o^4 \lambda_1}{\lambda_2}\right)^3\right]^{\frac{1}{4}} M_{WH}^2 \;.
\end{equation}

\noindent
Defining again $m_{\left(\beta\right)} = 8 \lambda_1 \mathscr L_o^4 \bar m_{\left(\beta\right)}/\lambda_2^3$ leads to

\begin{equation}
b_- \gg \left(m_{\left(\frac{3}{5}\right)}\right)^{\frac{1}{4}} \frac{\lambda_2}{\mathscr L_o}  \quad, \quad \frac{48 \left(m_{\left(\frac{3}{5}\right)}\right)^{\frac{1}{2}}}{\mathscr L_o^2 \lambda_1^2} \gg \frac{48 M_{WH}^2}{b_-^6} = \mathcal{K}_{WH}^{class} \;, 
\end{equation}

\noindent
from which the critical scales where quantum effects become relevant are then given by

\begin{equation}
\ell^{\left(\frac{3}{5}\right)}_{crit} = \left(m_{\left(\frac{3}{5}\right)}\right)^{\frac{1}{4}} \frac{\lambda_2}{\mathscr L_o} \quad , \quad \mathcal{K}^{\left(\frac{3}{5}\right)}_{crit} = \frac{48 \left(m_{\left(\frac{3}{5}\right)}\right)^{\frac{1}{2}}}{\mathscr L_o^2 \lambda_1^2 }\;.
\end{equation}

Inserting \eqref{CDbeta3/5} into \eqref{limitsb+}, we find that the classical regime of the black hole side in the $\beta = \frac{3}{5}$ case corresponds to 

\begin{equation}\label{critscalesBH-}
b_+ \gg \frac{M_{BH}}{M_{WH}} \ell^{\left(\frac{3}{5}\right)}_{crit} \quad , \quad \frac{M_{BH}^2}{M_{WH}^2}\mathcal{K}^{\left(\frac{3}{5}\right)}_{crit} \gg \mathcal{K}_{WH}^{class} \;,
\end{equation}

\noindent
which is perfectly consistent with \eqref{critscalesWH+}. Thus, being now $M_{BH} > M_{WH}$, both the scales \eqref{critscalesBH-} are shifted to higher values on the black hole side, leading to curvature effects relevant at higher curvatures and finite volume effects relevant at larger volumes.

Let us remark that Eq. \eqref{CDbeta3/5} leads us now to relate $\sin(\lambda_1 P_1)/\mathscr L_o \lambda_1$ with the Kretschmann scalar on the white hole side as

\begin{equation}
\frac{\sin(\lambda_1 P_1)}{\mathscr L_o \lambda_1} \stackrel{r \ll -1}{\simeq} \frac{\lambda_2^6}{16 C \lambda_1^2 \mathscr L_o^7} \frac{1}{|r|^3} = \frac{3 C D}{2 \mathscr{L}_o b_-^3} = 2 \left(m_{\left(\frac{3}{5}\right)}\right)^{-\frac{1}{4}} \frac{M_{WH}}{b_-^3} \propto \sqrt{\mathcal{K}_{WH}^{class}}\;.
\end{equation} 

We can furthermore study whether the amplification we found in \eqref{amplification} with $\beta = \frac{5}{3}$ is consistent with the de-amplification we found in \eqref{deamplification} for $\beta = \frac{3}{5}$. Inverting \eqref{deamplification} yields 

\begin{equation}\label{inverseamplification}
M_{BH} = M_{WH} \left(\frac{M_{WH}}{\bar{m}_{\left(\frac{3}{5}\right)}}\right)^{\frac{2}{3}} \;,
\end{equation}

\noindent
which for $\bar{m}_{\left(\frac{3}{5}\right)} = \bar{m}_{\left(\frac{5}{3}\right)} =: \bar{m}$, i.e. $m_{\left(\frac{3}{5}\right)} = m_{\left(\frac{5}{3}\right)} =: m$ , is exactly \eqref{amplification} with $M_{BH}$ and $M_{WH}$ exchanged. Therefore, Eq. \eqref{inverseamplification} describes exactly the inverse amplification of \eqref{amplification} and hence both values of $\beta$ are consistent with each other. Finally, the identification $m_{\left(\frac{3}{5}\right)} = m_{\left(\frac{5}{3}\right)} =: m$ leads to the following $\beta$-independent scales

\begin{equation}\label{scales}
\ell_{crit} = m^{\frac{1}{4}} \frac{\lambda_2}{\mathscr L_o} \quad , \quad \mathcal{K}_{crit} = \frac{48 m^{\frac{1}{2}}}{\mathscr L_o^2 \lambda_1^2 }\;.
\end{equation}

Let us summarise the above analysis. To achieve quantum effects at a unique, mass-independent Kretschmann-curvature scale $\mathcal{K}_{crit}$ we need to fix a relation between $M_{BH}$ and $M_{WH}$ according to
\begin{equation}\label{massfinal}
M_{WH} = M_{BH} \left(\frac{M_{BH}}{\bar{m}}\right)^{\beta-1} \quad ,\quad \beta = \frac{5}{3} \;,\; \frac{3}{5} \;,
\end{equation}
\noindent
where one value of $\beta$ describes exactly the inverse relation as the other and this leads to relate in the classical regime $\sin(\lambda_1 P_1)/\mathscr L_o \lambda_1$ to the square-root of the Kretschmann scalar with the smaller mass (respectively $M_{BH}$, $M_{WH}$ for $\beta = \frac{5}{3}$, $\frac{3}{5}$). On the side with lower mass (denoted by subscript $1$), quantum effects become relevant when 
\begin{equation}
b_1 \sim \ell_{crit} = m^{\frac{1}{4}} \frac{\lambda_2}{\mathscr L_o} \quad , \quad \mathcal{K}_{1} \sim \mathcal{K}_{crit} = \frac{48 m^{\frac{1}{2}}}{\mathscr L_o^2 \lambda_1^2 } \;,
\end{equation}  
\noindent
where $m$ is a dimensionless number, which is related to $\bar{m}$ (and for $\beta = 5/3$ to $D$) due to
\begin{equation}\label{mmbar}
m = \frac{8 \lambda_1 \mathscr L_o^4}{\lambda_2^3} \bar{m} \stackrel{\beta = \frac{5}{3}}{=} \left(\frac{3}{2} D \right)^{\frac{4}{3}} \;.
\end{equation}
This means that for an onset of quantum effects around the Planck curvature and Planck area, we need to choose $D$ at the order of $1$. Very large or small $D$ would lead to an onset of quantum effects that is too early in one of the sectors. Indeed, as alredy stressed in Sec. \ref{HamiltonainFramework}, for $D=2/3$ we recover again the classical gauge for which $b \simeq r$ for $r \rightarrow \infty$.

\noindent
On the other hand, on the amplified side (denoted by subscript $2$) we have

\begin{equation}\label{scaleamplification}
b_2 \sim \frac{M_2}{M_1}\;\ell_{crit} \quad , \quad \mathcal{K}_{2} \sim \frac{M_2^2}{M_1^2}\; \mathcal{K}_{crit}\;,
\end{equation}  

\noindent
where $M_2 > M_1$. As discussed in Sec. \ref{Polymerisation}, $\lambda_2/\mathscr L_o$ set (up to a number) the critical length $\ell_{crit}$ and gives corrections when the volume becomes small. Furthermore $\mathscr L_o \lambda_1$ is directly related to an inverse curvature and sets the critical curvature scale $\mathcal{K}_{crit}$, i.e. it controls quantum corrections in the high curvature regime.

\begin{figure}[t!]
	\centering
	\subfigure[]
	{\includegraphics[width=7.75cm,height=5.5cm]{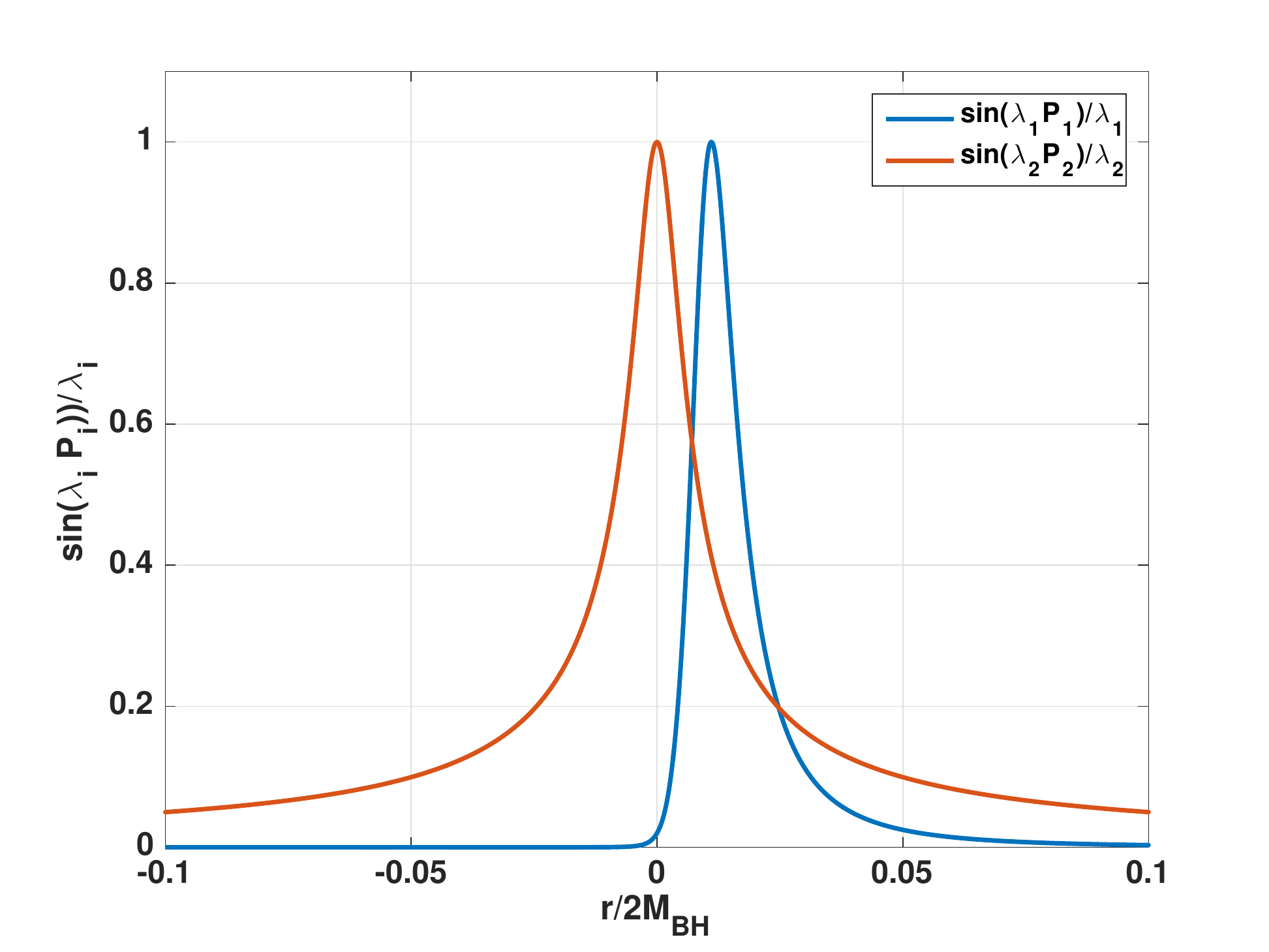}}
	\hspace{2mm}
	\subfigure[]
	{\includegraphics[width=7.75cm,height=5.5cm]{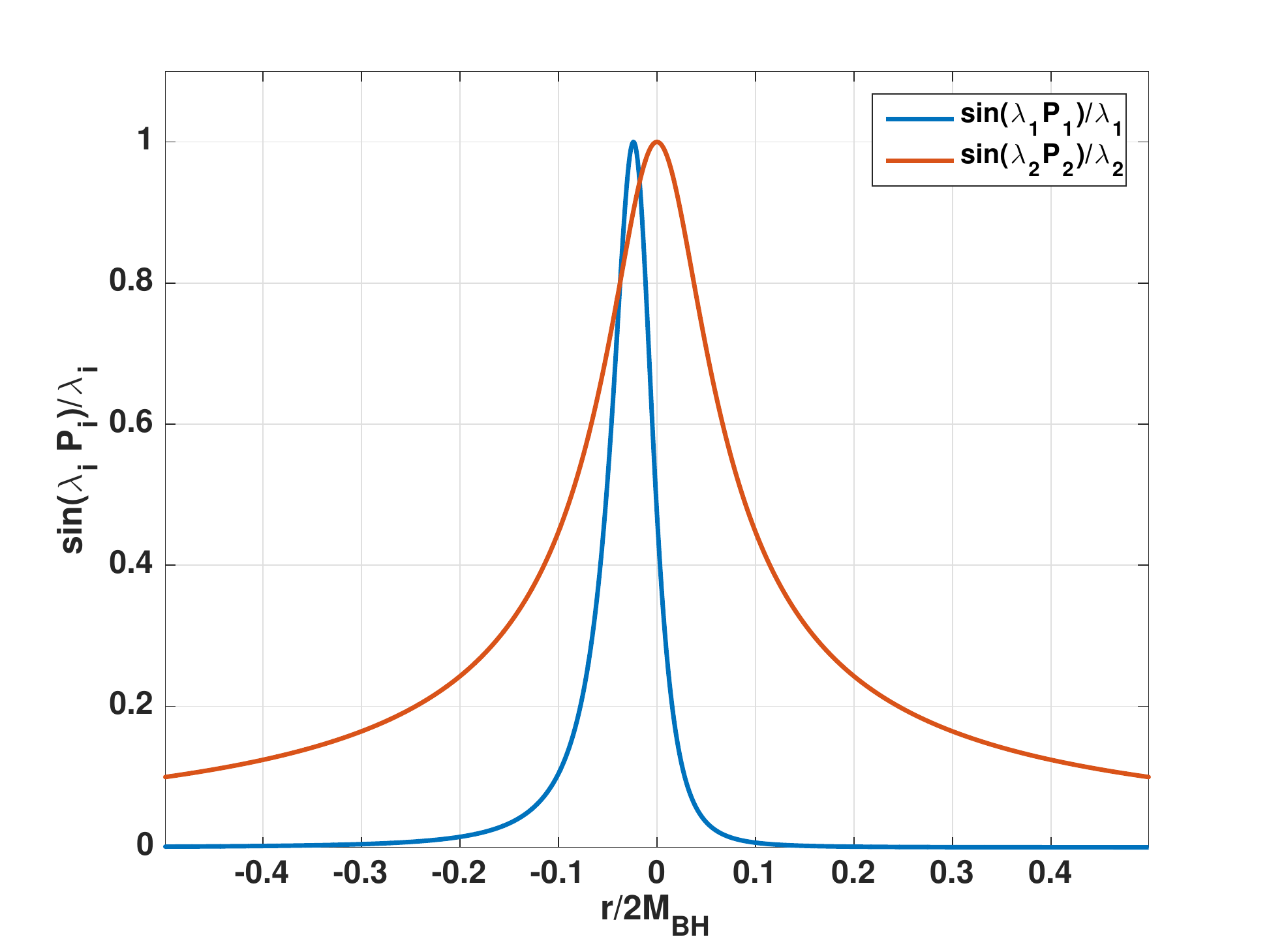}}
	\caption{Plot of $\sin(\lambda_1 P_1)/\mathscr L_o \lambda_1$ and $\sin(\lambda_2 P_2) \mathscr L_o/\lambda_2$ for $\beta = 5/3$ in (a) and $\beta = 3/5$ in (b) for the parameters $\mathscr L_o = \lambda_1 = \lambda_2 = \bar{m} = 1$ and $M_{BH} = 100$. The plot shows that the order of high curvature corrections and finite volume corrections is exchanged coming from the other side or changing $\beta$.}
	\label{fig:scales}
\end{figure}

Finally we want to discuss the change of the scales on the amplified side given in Eq. \eqref{scaleamplification}. The curvature scale is shifted to higher curvatures, such that curvature corrections become relevant later, i.e. closer to the transition surface. On the other hand, the length scale is shifted to larger lengths, such that finite volume effects become relevant earlier. Nevertheless, they will never be relevant at the horizon as $b(r_s) \sim M_{2}$ (cfr. Sec. \ref{sec:horizon}) for large masses, while $M_2/M_1 \sim M_2^\frac{3}{5}$, hence $b(r_s)$ grows faster with the mass than $M_2/M_1 \ell_{crit}$. Moreover, the change of scales on the amplified side leads to an exchange of when curvature effects or volume effects become relevant. While coming from the lower mass side an in-falling observer would first observe high curvature corrections and then finite volume corrections, an observer falling in from the other side would first see finite volume corrections and afterwards high curvature corrections. In Fig. \ref{fig:scales}, we plot $\sin(\lambda_1 P_1)/\mathscr L_o \lambda_1$ and $\sin(\lambda_2 P_2) \mathscr L_o/\lambda_2$ for both values of $\beta$. Exchanging the two $\beta$-values corresponds to exchanging whether $P_1$ or $P_2$-corrections become relevant first.

\begin{figure}[t!]
	\centering
	\includegraphics[width=7.75cm,height=5.5cm]{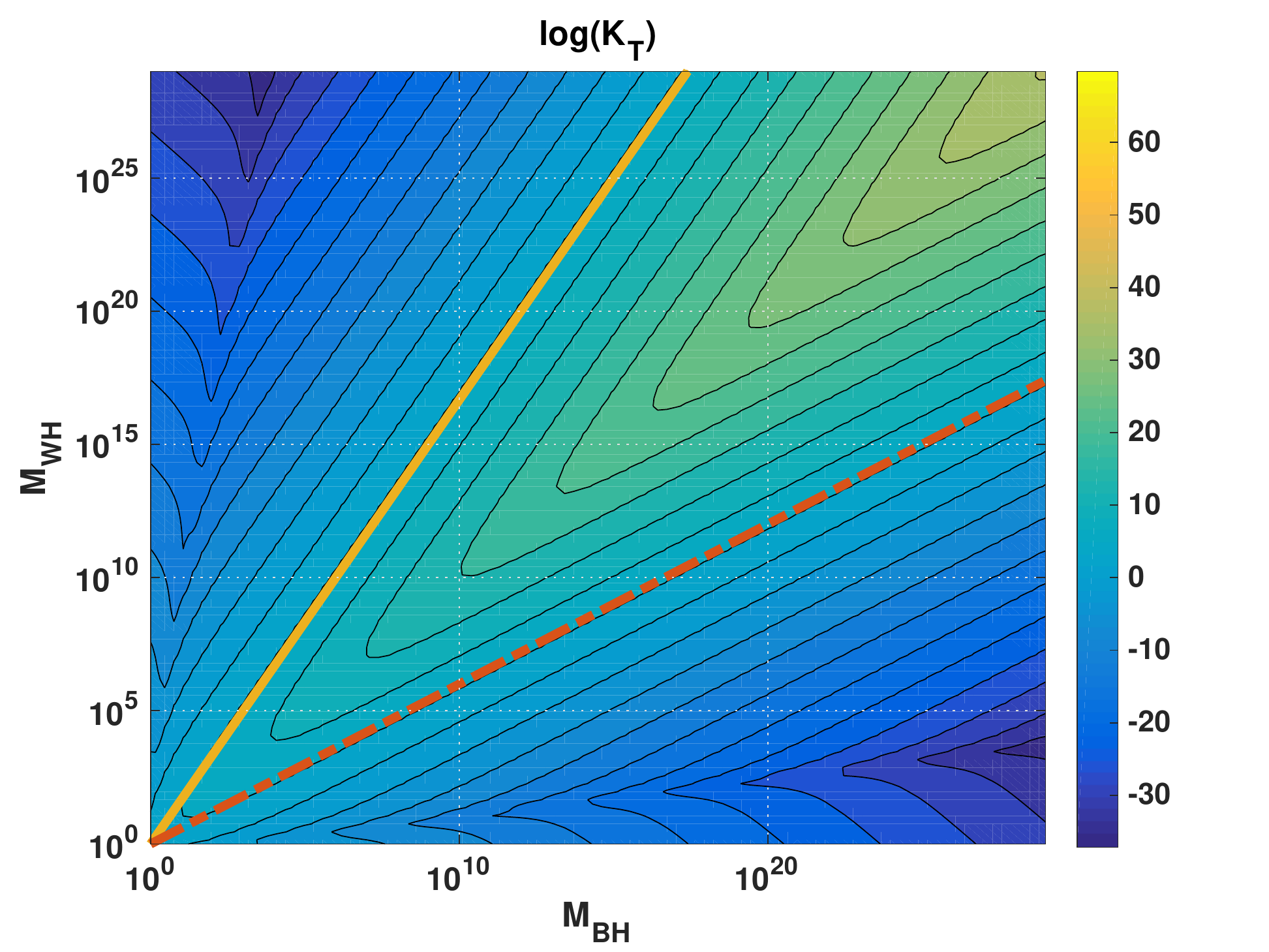}
	\caption{The color scale encodes the value of the logarithm of the Kretschmann scalar at the transition surface as a function of the black hole $M_{BH}$ and white hole mass $M_{WH}$ for $\mathscr L_o\lambda_1= \lambda_2/\mathscr L_o =1$. Both axis are logarithmically. Finite non-zero curvatures for large masses can only be achieved by following a level line asymptotically given by Eq. \eqref{eq:BHWH} for $\beta = \frac{5}{3}$ and $\beta = \frac{3}{5}$. Different values of $\bar{m}$ correspond to different choices of the level line. The yellow line corresponds to $\beta = \frac{5}{3}$ and the red dashed line to $\beta = \frac{3}{5}$.}
	\label{fig:Colorkretschmann}
\end{figure}

A further important question is whether the curvature invariants have a unique upper bound. For this purpose, we study the Kretschmann scalar at the transition surface, where quantum effects are large and the Kretschmann scalar reaches almost its maximal value. The explicit expression of the Kretschmann scalar can be calculated easily with computer algebra software, but is quite involved and not insightful, so we will not report it here. Instead, we focussed at the transition surface, where quantum effects are large. In Fig. \ref{fig:Colorkretschmann}, we show as a colorplot the logarithm of the Kretschmann scalar at the transition surface as a function of the two masses $M_{BH}$ and $M_{WH}$. Immediately from there, we can read off that the value of the curvature at the transition surface is different for different relations between the black hole and the white hole mass. Physically plausible are relations where for all masses, especially in the large mass limit, the Kretschmann scalar remains non-zero and finite. Studying the plot leads to the conclusion that this can only be achieved if the relation between black hole and white hole mass follows, at least in the large mass limit, a level line. As the plot and also computations show for large masses, this holds exactly for a relation of the kind $M_{WH} = M_{BH} \left(\frac{M_{BH}}{\bar{m}}\right)^{\beta-1} \sim M_{BH}^\beta$ for $\beta = \frac{5}{3}$ and $\beta = \frac{3}{5}$. Different values of $\bar{m}$ simply pick different level lines. Hence, demanding an upper bound of the Kretschmann scalar is consistent with the previous discussion of a unique curvature scale where quantum effects become relevant.

\begin{figure}[t!]
	\centering
	\subfigure[]
	{\includegraphics[width=7.75cm,height=5.5cm]{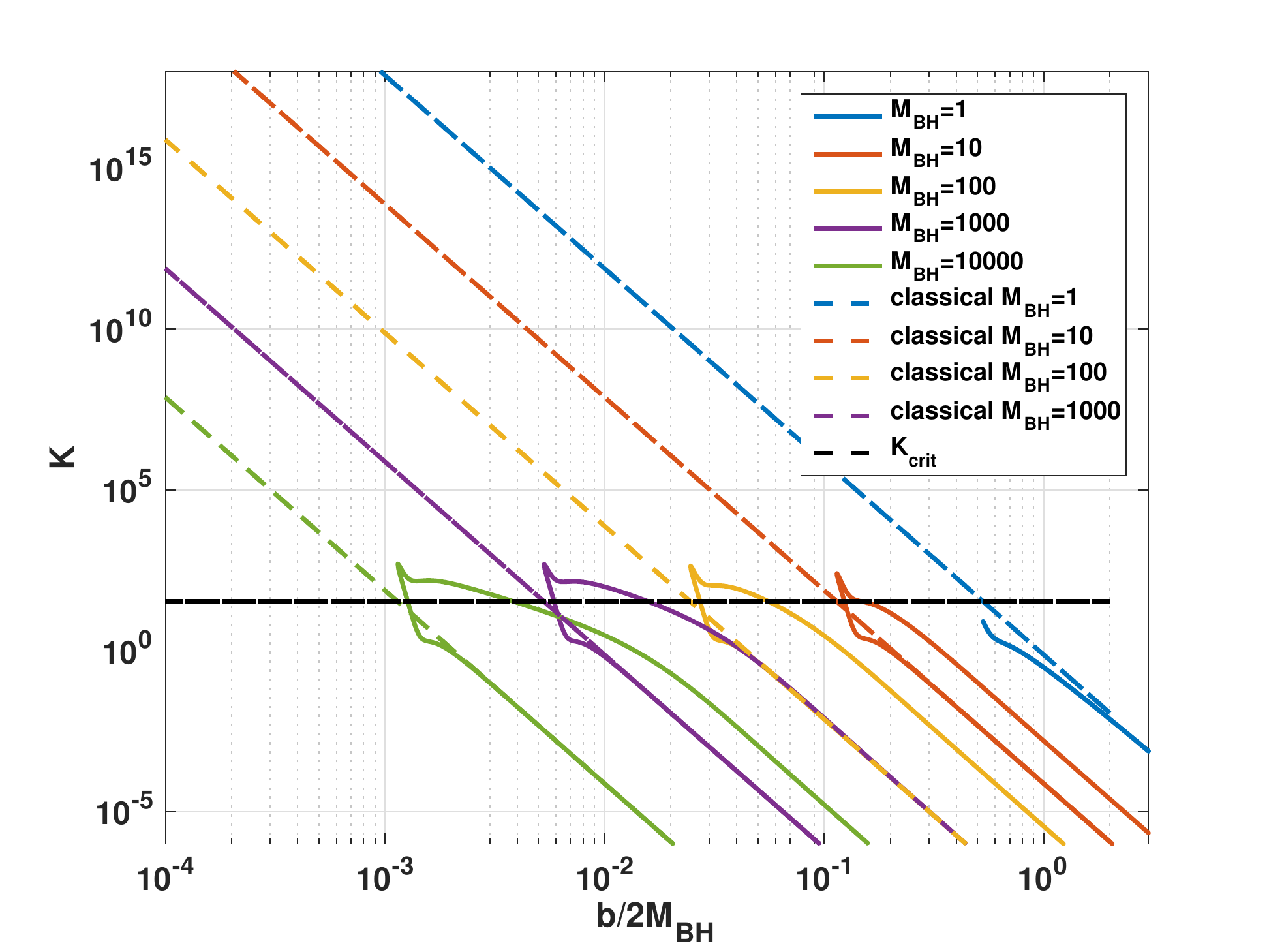}}
	\hspace{2mm}
	\subfigure[]
	{\includegraphics[width=7.75cm,height=5.5cm]{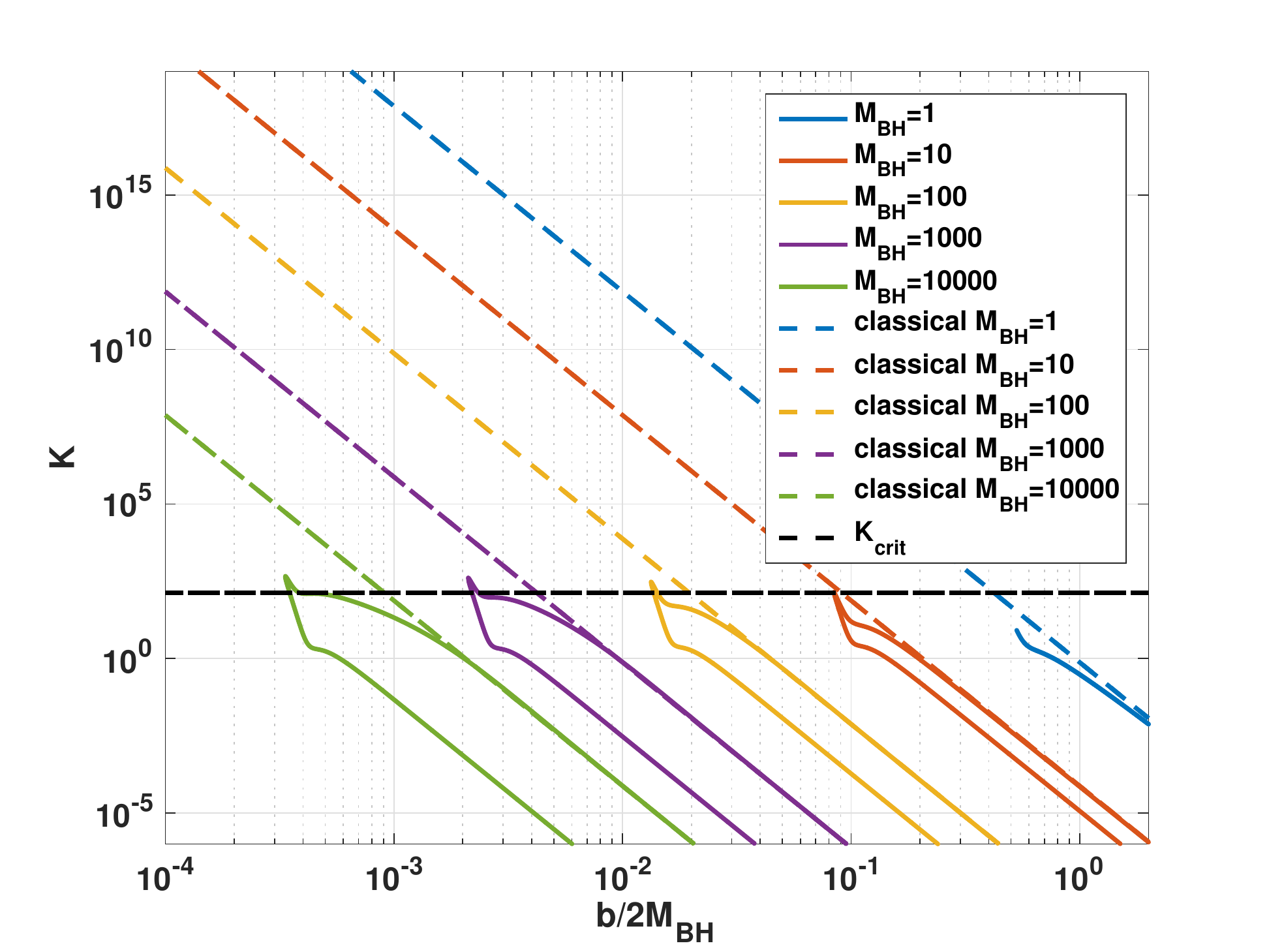}}
	\caption{Kretschmann scalar $\mathcal{K}$ against $b$ in a log-log scale for different masses. The dashed lines correspond to the classical result. We choose the parameters $\mathscr L_o\lambda_1= \lambda_2/\mathscr L_o =1$, $\bar{m} = 1$ and $\beta = \frac{5}{3}$ in (a) as well as $\beta = \frac{3}{5}$ in (b). Quantum effects become relevant always at the same scale. The horizontal dashed line corresponds to $\mathcal{K}_{crit}$ given in Eq. \eqref{scales}. Differences occur only for Planck sized black holes ($M_{BH}=1$), for which quantum effects due to the polymerisation of $P_2$ become relevant first.}
	\label{fig:kretschmann}
\end{figure}

Fig. \ref{fig:kretschmann} shows the plot of the full Kretschmann scalar $\mathcal{K}$ as function of $b$ for the two values we found for $\beta$ and for different masses. Indeed, as required, quantum effects become relevant always at the same scale for both $\beta$ values\footnote{As can be directly checked by computer algebra software, also the other curvature invariants exhibit a unique mass independent upper bound.}. The only exception is the case of small masses $M_{BH} = 1$ which corresponds to a Planck sized black hole and hence quantum effects caused by the polymerisation of $P_2$ become relevant earlier. The critical curvature $\mathcal{K}_{crit}$ is in the plots indicated by a vertical line and shows that it is close to the maximal curvature. Note that for both cases before and after the bounce, the Kretschmann scalar approaches the classical behaviour, but with a different mass. As mentioned above for $\beta = \frac{5}{3}$ we see a mass dependent amplification, while for $\beta = \frac{3}{5}$ a de-amplification is visible. As Eq. \eqref{amplification} and \eqref{deamplification} show, only for $M_{BH}/m = 1$ we see a symmetric bounce, but we do not consider our equations to be reliable in this ``Planck mass black hole'' regime.

\section{Effective spacetime structure}\label{effectivestruc}

\subsection{Horizon structure}\label{sec:horizon}

Let us now analyse the structure of the spacetime geometry described by the quantum corrected effective metric \eqref{effmetric}. First of all, the black hole horizon is characterised by the vanishing of $a(r)$, which in the classical case occurs at $r=r_s=2M$. Similarly, in the polymer effective model, the quantum corrected metric is again spherically symmetric and hence the resulting spacetime will still be foliated by homogeneous space-like Cauchy surfaces. The horizon will now be characterised by the vanishing of $a(r)$ given in Eq. \eqref{aquant} (i.e., as in the classical case, by the divergence of $N(r)=\mathscr L_o^2/a(r)$) which in turn corresponds to the vanishing of $v_2(r)$ in the phase space description. Therefore, using the expression \eqref{solutionv2quant} for $v_2$, we get

\be
v_2(r)=0\qquad\Leftrightarrow\qquad 1-\frac{3CD}{2 \lambda_2}\frac{1}{\sqrt{1+\frac{n r^2}{\lambda_2^2}}}=0\;,
\ee

\noindent
from which it follows that

\be\label{rhorizon}
r_s^{(\pm)}=\pm \frac{3CD}{2\sqrt{n}} \sqrt{1- \left(\frac{2 \lambda_2}{3 C D}\right)^2} \; \text{.}
\ee

\noindent
The corresponding values of the physical radius are given by

\be\label{bofrspm}
b(r_s^{(\pm)})=\left(\frac{3}{2}\,v_1(r_s^{(\pm)})\right)^{1/3}\;,
\ee  

\noindent
where, according to the expression \eqref{solutionv1quant}, we have

\be
\begin{aligned}
	v_1(r_s^{(\pm)})&=\frac{2 C^2 \lambda_1^2 \sqrt{n}^3}{\lambda_2^3}D\,\frac{\frac{\lambda_2^6}{16 C^2 \lambda_1^2 n^3}\left(\pm\frac{3CD}{2 \lambda_2}\sqrt{1-\left(\frac{2\lambda_2}{3CD}\right)^2}+\frac{3CD}{2\lambda_2}\right)^6+1}{\left(\pm\frac{3CD}{2\lambda_2}\sqrt{1-\left(\frac{2\lambda_2}{3CD}\right)^2}+\frac{3CD}{2\lambda_2}\right)^3}\nonumber\\
	&=\frac{\lambda_2^3}{8 \sqrt{n}^3}D f^{(\pm)}(x)+\frac{2 \lambda_1^2 \sqrt{n}^3 D C^2}{\lambda_2^3 f^{(\pm)}(x)}\;,\label{v1rs}
\end{aligned}
\ee

\noindent
with

\be
f^{(\pm)}(x)=\frac{1}{x^3}\left(1\pm\sqrt{1-x^2}\right)^3\qquad,\qquad x=\frac{2\lambda_2}{3CD}\;.
\ee

In order to study the relation between the two horizons as well as their dependence from the black hole and white hole masses, we consider the two cases discussed in Sec. \ref{sec:curvinvariants}. Let us start with the $\beta=5/3$ case for which $M_{WH}=\bar{m}^{-2/3}M_{BH}^{5/3}$. By using the expressions \eqref{CandD} for $C$ and $D$, Eq. \eqref{v1rs} can be written in terms of the black hole and white hole masses

\be
v_1(r_s^{(\pm)})=\left[\frac{2}{3}\left(\frac{\bar{m}\,\lambda_1\lambda_2}{3}\right)^{3}\right]^{\frac{1}{4}}\left(f^{(\pm)}(x)+\frac{M_{BH}^2}{\bar{m}^2f^{(\pm)}(x)}\right)\;,
\ee

\noindent
with

\be
x=\left(\frac{\bar{m}\,\lambda_1\lambda_2}{2}\right)^{1/4}\frac{1}{M_{BH}}\;.
\ee

\noindent
Therefore, expanding around $x=0$, which corresponds to a large mass expansion, we have

\be\label{v1r+BH}
v_1(r_s^{(+)})\simeq\frac{16M_{BH}^3}{3}-2\sqrt{2\bar{m}\lambda_1\lambda_2}\,M_{BH}+\mathcal O\left(\frac{(\lambda_1\lambda_2)^{3/2}}{\bar{m}^{1/2} M_{BH}}\right)\;,
\ee

\noindent
which corresponds to the classical result plus quantum corrections suppressed
in the limit $\lambda_1,\lambda_2\to0$ as well as $M_{BH}\to +\infty$, and

\begin{align}\label{v1r-BH}
v_1(r_s^{(-)})&\simeq\frac{16M_{BH}^5}{3\bar{m}^2}-2\,\sqrt{\frac{2\lambda_1\lambda_2}{\bar{m}^3}}\,M_{BH}^3+\mathcal{O}\left(\frac{(\lambda_1\lambda_2)^{3/2}}{\bar{m}^{1/2} M_{BH}}\right)\nonumber\\
&=\frac{16M_{WH}^3}{3}-2\sqrt{\frac{2\lambda_1\lambda_2}{\bar{m}^{3/5}}}\,M_{WH}^{9/5}+\mathcal O\left(\frac{(\lambda_1\lambda_2)^{3/2}}{\bar{m}^{9/10} M_{WH}^{3/5}}\right)\;,
\end{align}

\noindent
which at leading order shows a perfect symmetry between the black hole and white hole sides consistently with having two asymptotically classical Schwarzschild geometries. Expanding in a similar way the physical radius $b$, we have

\begin{align}
&b(r_s^{(+)})\simeq 2M_{BH}-\sqrt{\frac{\bar{m}\lambda_1\lambda_2}{8}}\frac{1}{M_{BH}}+\mathcal O\left(\frac{\bar{m} \lambda_1\lambda_2}{M_{BH}^3}\right)\;,\\
&b(r_s^{(-)})\simeq\frac{2M_{BH}^{5/3}}{\bar{m}^{2/3}}-\sqrt{\frac{\lambda_1\lambda_2}{8\bar{m}^{1/3}}}\frac{1}{M_{BH}^{1/3}}+\mathcal{O}\left(\frac{\bar{m}^{1/3} \lambda_1\lambda_2}{M_{BH}^{7/3}}\right)\nonumber\\
&\quad\quad\quad=2M_{WH}-\sqrt{\frac{\lambda_1\lambda_2}{8\bar{m}^{3/5}}}\frac{1}{M_{WH}^{1/5}}+\mathcal{O}\left(\frac{\lambda_1\lambda_2}{\bar{m}^{3/5} M_{WH}^{7/5}}\right)\;,
\end{align}

\noindent
while the ratio $b(r_s^{(-)})/b(r_s^{(+)})$ yields

\be
\frac{\mathcal{R}_{WH}}{\mathcal{R}_{BH}} :=\frac{b(r_s^{(-)})}{b(r_s^{(+)})}\simeq\left(\frac{M_{BH}}{\bar{m}}\right)^{2/3}-\mathcal O\left(\frac{\bar{m}^{1/3}\lambda_1\lambda_2}{M_{BH}^{10/3}}\right)\;.
\ee

\noindent
Thus, the classical Schwarzschild radius gets modified by quantum corrections and we now have two solutions respectively in the positive and negative $r$ regions. As it will be clear later on in this section by studying the Penrose diagram for the quantum extended effective geometry, these represent the past and future boundaries of the black hole ($b(r_s^{(+)})$) and white hole ($b(r_s^{(-)})$) interior regions connected by a transition surface at which $b(r)$ reaches its minimal value. The classical singularity is replaced by an asymmetric bounce interpolating between the black hole and white hole interior regions and the radius of the white hole horizon grows with the mass of the black hole. This may be interpreted as a quantum gravity induced mass amplification similarly to what happens in the generalised $\mu_o$-scheme of \cite{CorichiLoopquantizationof}. We note that due to the absence of a time-like Killing vector, we may not exclude such a phenomenon due to energy conservation.

However, as will be clear from the structure of the Kruskal extension of the effective quantum spacetime, there is no indefinite mass amplification. Indeed, if we consider now $M_{WH}=\bar{m}^{2/5}M_{BH}^{3/5}$ for which $\beta=3/5$, i.e. $M_{BH}=\bar{m}^{-2/3}M_{WH}^{5/3}$, we have

\be
v_1(r_s^{(\pm)})=\left[\frac{2}{3}\left(\frac{\bar{m}\,\lambda_1\lambda_2}{3}\right)^{3}\right]^{\frac{1}{4}}\left(\frac{M_{WH}^2}{\bar{m}^2}f^{(\pm)}(x)+\frac{1}{f^{(\pm)}(x)}\right)\,,\qquad x=\left(\frac{\bar{m}\,\lambda_1\lambda_2}{2}\right)^{1/4}\frac{1}{M_{WH}}
\ee

\noindent
whose corresponding large mass expansions ($x\to0$) yield 

\begin{align}
&v_1(r_s^{(+)})\simeq\frac{16M_{WH}^5}{3\bar{m}^2}-2\,\sqrt{\frac{2\lambda_1\lambda_2}{\bar{m}^3}}\,M_{WH}^3+\mathcal{O}\left(\frac{(\lambda_1\lambda_2)^{3/2}}{\bar{m}^{1/2} M_{WH}}\right)\nonumber\\
&\qquad\quad\;\;=\frac{16M_{BH}^3}{3}-2\sqrt{\frac{2\lambda_1\lambda_2}{\bar{m}^{3/5}}}\,M_{BH}^{9/5}+\mathcal O\left(\frac{(\lambda_1\lambda_2)^{3/2}}{\bar{m}^{9/10} M_{BH}^{3/5}}\right)\;,\\
&v_1(r_s^{(-)})\simeq\frac{16M_{WH}^3}{3}-2\sqrt{2\bar{m}\lambda_1\lambda_2}\,M_{WH}+\mathcal O\left(\frac{(\lambda_1\lambda_2)^{3/2}}{\bar{m}^{1/2} M_{WH}}\right)\;,
\end{align}

\noindent
which are exactly the same as Eq. \eqref{v1r+BH} and \eqref{v1r-BH} just with $M_{BH}$ and $M_{WH}$ exchanged. Moreover 

\be
\frac{\mathcal{R}_{BH}}{\mathcal{R}_{WH}}:=\frac{b(r_s^{(+)})}{b(r_s^{(-)})}\simeq\left(\frac{M_{WH}}{\bar{m}}\right)^{2/3}-\mathcal O\left(\frac{\bar{m}^{1/3} \lambda_1\lambda_2}{M_{WH}^{10/3}}\right)\;,
\ee

\noindent
which corresponds to a mass de-amplification of the same magnitude.

\subsection{Transition Surface}\label{sec:transitionsurf}

As already anticipated in Fig. \ref{fig:aofb2}, studying the components of the effective metric, we see a special property in $b$ already found in other loop quantisation schemes of black hole spacetimes: the physical radius $b$ does not reach zero. Hence, the black hole has a minimal size at which quantum geometry effects become relevant and the classical singularity is resolved. A quick calculation shows that the value of the radial coordinate at which $b'=0$ is given by

$$
r_\mathcal{T} = \frac{\lambda_2}{2\mathscr L_o} \left(\left(\frac{\lambda_2^3}{4 C \lambda_1 \mathscr L_o^3}\right)^{-\frac{1}{3}}-\left(\frac{\lambda_2^3}{4 C \lambda_1 \mathscr L_o^3}\right)^{\frac{1}{3}}\right)\;,
$$
\noindent
from which, evaluating \eqref{bquant} at $r=r_\mathcal T$, it follows that the corresponding physical (minimal) radius is given by

\begin{equation}\label{eq:btransition}
b_\mathcal{T} := b(r_\mathcal{T}) = \left(\frac{3\lambda_1CD}{2}\right)^{\frac{1}{3}}\;.
\end{equation}
\noindent

Using then the expressions \eqref{CandD} for $C$ and $D$ to rewrite the above minimal value of the physical radius in terms of the black hole and white hole masses, we get together with Eqs. \eqref{massfinal} and \eqref{mmbar}

\be\label{eq:btransition2}
b_\mathcal T=2^{1/12}(\lambda_1\lambda_2)^{1/4}(M_{BH}M_{WH})^{1/8}=\begin{cases}
	\left(\frac{2 \lambda_1 \mathscr L_o}{m^{\frac{1}{4}}}\right)^{\frac{1}{3}}M_{BH}^{1/3} &\;\text{for}\;\;\beta=\frac{5}{3}\\
	&\\
\left(\frac{2 \lambda_1 \mathscr L_o}{m^{\frac{1}{4}}}\right)^{\frac{1}{3}}M_{WH}^{1/3} &\;\text{for}\;\;\beta=\frac{3}{5}
\end{cases}
\ee

\noindent
which goes to zero as $\lambda_1\to0$ as expected in the classical regime. The expressions of the minimal radius are identical up to the exchange of the black and white hole masses so that, as we will discuss in Sec. \ref{CSPD}, the occurrence of the two $\beta$-values just reflects a certain choice of initial conditions in the black hole (resp. white hole) exterior region.
Thus, the point $b_\mathcal T$ denotes the minimal size of the interior region and, as it is also common in the loop quantum cosmology framework, it describes a bounce interpolating between two asymptotically classical Schwarzschild spacetimes. Furthermore, it is easy to see that $b_\mathcal T$ identifies a space-like 3-dimensional surface smoothly connecting a trapped and a anti-trapped region. This can be explicitly checked by computing the expansions of the future pointing null normals $u_{\pm}$ to the $t=const.$ and $r=const.$ metric 2-spheres. Specifically, in the region $r_s^{(-)}<r<r_s^{(+)}$, these are given by
\be
u_{\pm}=u_{\pm}^a\frac{\partial}{\partial x^a}=\frac{1}{\sqrt{-2N}}\frac{\partial}{\partial r}\pm\frac{1}{\sqrt{-2a}}\frac{\partial}{\partial t}\;,
\ee
satisfying the normalisation conditions $g(u_{\pm},u_{\pm})=0$ and $g(u_{\pm},u_{\mp})=-1$. The expansions $\theta_{\pm}$ of these null vectors then read
\be
\theta_{\pm}=S^{ab}\nabla_au^{\pm}_b=-\sqrt{-\frac{2}{N}}\frac{\dot b}{b}\;, \label{eq:ExpansionNullRays}
\ee
where $S^{ab}=g^{ab}+u_{+}^au_{-}^b+u_{-}^au_{+}^b$ is the projector on the metric 2-spheres. Therefore, since $b(r)$ is always positive and $N$ cannot vanish, we see that $\theta_{\pm}$ vanish if and only if $\dot b=0$, i.e. at $r=r_\mathcal T$. Moreover, both expansions are negative (resp. positive) for $r_\mathcal T<r<r_s^{(+)}$ (resp. $r_s^{(+)}<r<r_\mathcal T$) and hence the space-like 3-dimensional surface $b=b(r_\mathcal T)$ smoothly connects a trapped and a anti-trapped region respectively interpreted as black hole and white hole interior regions. This transition between black hole and white hole interior regions, occurring when spacetime curvature enters the Planck scale, replaces the classical singularity. This property will be immediately clear once the Penrose diagram is constructed (see section \ref{CSPD}).

\subsection{Causal structure and Penrose diagram}\label{CSPD}

We are now ready to construct the Penrose diagram for the quantum corrected effective spacetime and study its causal structure. To this aim, we have to construct the Kruskal extension for our polymer Schwarzschild geometry. As usual, the starting point is to define Kruskal-Szekeres coordinates $(X,T)$ by (cfr. \cite{KloeschClassicalandquantum})

\begin{equation}\label{KScoordinates}
T^2-X^2=\exp\left[\left(\left.\frac{\dd \bar{a}}{\dd r}\right|_{r=r_s^{(\pm)}}\right) r_*(b)\right]\quad , \quad \frac{T}{X} = \begin{cases}
\tanh\left(\frac{t}{2}\left(\left.\frac{\dd \bar{a}}{\dd r}\right|_{r=r_s^{(\pm)}}\right)\right) & -1<\frac{T}{X}<1 \\
\coth\left(\frac{t}{2}\left(\left.\frac{\dd \bar{a}}{\dd r}\right|_{r=r_s^{(\pm)}}\right)\right) & -1<\frac{X}{T}<1 \\
\end{cases} \;,
\end{equation}

\noindent
where the definition now refers to the physical radius $b$ instead of the radial coordinate $r$ (which unlike the classical case do not coincide in the effective quantum theory), $r_s^{(\pm)}$ is the radial coordinate of the horizon respectively in the positive and negative $r$-ranges given in Eq. \eqref{rhorizon}, and $r_*$ is the so-called tortoise coordinate defined by

\begin{equation}\label{rstar}
r_*(b)=\int_{b_0}^{b}\dd b\,\frac{\frac{\dd r^{(\pm)}}{\dd b}}{\bar{a}(b)}=\int_{r_\mathcal T}^{r^{(\pm)}(b)}\dd r\,\frac{L_o^2}{a(r)}\;.
\end{equation}

\noindent
where we set the reference value $b_0$ to be at the transition surface, i.e., $b_0=b_\mathcal T\equiv b(r_\mathcal T)$, where $b(r)$ takes its minimal value and the bounce occurs. At the transition surface $r_*(b_\mathcal T)=0$ by construction and hence $T^2-X^2=1$. Note that we fixed the reference point to be at the transition surface for both the interior and exterior regions. As we will see soon, by performing the integrals in the complex domain, the sign switch in the definition \eqref{KScoordinates} going from the interior to the exterior region is provided by the imaginary part of the integral \eqref{rstar} which corresponds to the residue at the pole occurring at the horizon. Moreover, the definition \eqref{KScoordinates} implies that we need two $(X,T)$-coordinate charts to cover the whole range $r\in(-\infty,+\infty)$. Indeed, as discussed in Sec. \ref{sec:intconst} (cfr. Fig. \ref{fig:aofb2}), $b(r)$ as a function of $r$ exhibits two branches for $r>r_\mathcal T$ and $r<r_\mathcal T$ where respectively it increases and decreases monotonously. This means that we have to split the construction of the Penrose diagram in these two regions where $b(r)$ is invertible and show that they can be smoothly glued afterwards. 

The explicit evaluation of the integral \eqref{rstar} is quite involved. However, to construct the maximal extension of the polymer Schwarzschild spacetime we need to understand the behaviour of $r_*$ for some specific values of $b(r)$, e.g., at the horizons and asymptotically far at infinity. Let us then start by considering the $r>r_\mathcal T$ region. At the horizon $b(r_s^{(+)})$, we have

\be\label{rstaratrs}
r_*(b(r_s^{(+)}))=\int_{b_{\mathcal T}}^{b(r_s^{(+)})}\dd b\,\frac{\frac{\dd r^{(+)}}{\dd b}}{\bar a(b)}=\int_{r_\mathcal T}^{r_s^{(+)}}\dd r\,\frac{1}{\bar a(r)}\;.
\ee

\noindent
In analogy to the classical case, we expect $r_*$ to be divergent for $r\to r_s^{(+)}$. In particular, being $\bar a(r)<0$ for $r_\mathcal T\leq r\leq r_s^{(+)}$, we expect the integral \eqref{rstaratrs} to yield $-\infty$. To see this, let us rewrite the integral as follows

\be\label{rstarsplit1}
r_*(b(r_s^{(+)}))=\int_{r_\mathcal T}^{r_s^{(+)}-\epsilon}\dd r\,\frac{1}{\bar a(r)}+\int_{r_s^{(+)}-\epsilon}^{r_s^{(+)}}\dd r\,\frac{1}{\bar a(r)}\;.
\ee

\noindent
for some $\epsilon>0$. The first integral in Eq. \eqref{rstarsplit1} is finite while, for $\epsilon$ small enough, the function in the second integral can be approximated with its series expansion around $r_s^{(+)}$ thus yielding

\begin{align}
r_*(b(r_s^{(+)}))&\simeq\text{finite terms}+\int_{r_s^{(+)}-\epsilon}^{r_s^{(+)}}\dd r\,\left(\frac{1}{\bar a'(r_s^{(+)})(r-r_s^{(+)})}+\mathcal O(r-r_s^{(+)})\right)\nonumber\\
&=\text{finite terms}+\frac{1}{\bar a'(r_s^{(+)})}\log{\left(|r-r_s^{(+)}|\right)}\Bigl|_{r=r_s^{(+)}-\epsilon}^{r=r_s^{(+)}}\;,
\end{align}

\noindent
from which we see that the (finite) pre-factor in front of the logarithm cancels the derivative in the exponential of Eq.\eqref{KScoordinates}, $r_*(r_s^{(+)})\to-\infty$ logarithmically, and hence $T^2-X^2=0$ (i.e., $T=\pm X$) at $b=b(r_s^{(+)})$. 

For the exterior region $b(r_s^{(+)})<b(r)<+\infty$ instead we have that

\be
r_*(b(r))=\int_{r_\mathcal T}^{r}\dd r\,\frac{1}{\bar a(r)}\qquad,\qquad r>r_s^{(+)}
\ee

\noindent
which can be split as

\be\label{rstarexterior}
r_*(b(r))=\int_{r_\mathcal T}^{r_s^{(+)}-\epsilon}\dd r\,\frac{1}{\bar a(r)}+\int_{r_s^{(+)}-\epsilon}^{r_s^{(+)}+\epsilon}\dd r\,\frac{1}{\bar a(r)}+\int_{r_s^{(+)}+\epsilon}^{r}\dd r\,\frac{1}{\bar a(r)}\;,
\ee

\noindent
with $\epsilon>0$. Let consider the first two integrals separately. The first one is finite. Concerning the second integral, for $\epsilon$ arbitrarily small (say $\epsilon\to0$), we can approximate it again by expanding the integrand function around $r_s^{(+)}$ thus yielding

\begin{align}
\int_{r_s^{(+)}-\epsilon}^{r_s^{(+)}+\epsilon}\dd r\,\frac{1}{\bar a(r)}&\underset{\epsilon\to0}{\simeq}\int_{r_s^{(+)}-\epsilon}^{r_s^{(+)}+\epsilon}\dd r\,\left(\frac{1}{\bar a'(r_s^{(+)})(r-r_s^{(+)})}+\mathcal O(r-r_s^{(+)})\right)\nonumber\\
&\;=\frac{1}{\bar a'(r_s^{(+)})}\left(\frac{1}{2}\oint_{\mathcal C}\frac{\dd r}{r-r_s^{(+)}}\right)+\text{finite terms}\nonumber\\
&\;=-\frac{i\pi}{\bar a'(r_s^{(+)})}+\text{finite terms}\;,\label{2ndintegral}
\end{align}

\noindent
where in the second line $\mathcal C$ denotes an infinitesimally small contour in the complex plane encircling $r=r_s^{(+)}$ where the integrand function has a first order pole, and in the third line we used Cauchy's residue theorem with the minus sign coming from the clockwise orientation of the integration contour.

Therefore, substituting the above result into \eqref{rstarexterior}, we get

\be\label{rstarexterior2}
r_*(b(r))=-\frac{i\pi}{\bar a'(r_s^{(+)})}+\int_{r_s^{(+)}+\epsilon}^{r}\dd r\,\frac{1}{\bar a(r)}+\text{finite terms}\;,
\ee

\noindent
from which, recalling the definition \eqref{KScoordinates}, it follows that

\be\label{extKScoordinates}
T^2-X^2=-\exp\left(\frac{\dd\bar a}{\dd r}\biggl|_{r=r_s^{(+)}}\int_{r_s^{(+)}+\epsilon}^{r}\dd r\,\frac{1}{\bar a(r)}+\text{finite terms}\right)\;,
\ee

\noindent
where, as already anticipated in the beginning of this section, the minus sign for the $r>r_s^{(+)}$ region comes from the imaginary term in Eq. \eqref{rstarexterior2} which, after cancellation of the $\dd\bar a/\dd r$ factors, gives $e^{-i\pi}=-1$. In particular, for $r\to+\infty$, the integral in \eqref{extKScoordinates} can be written as

\begin{align}
\int_{r_s^{(+)}+\epsilon}^{+\infty}\dd r\,\frac{1}{\bar a(r)}&=\int_{r_s^{(+)}+\epsilon}^{\tilde r}\dd r\,\frac{1}{\bar a(r)}+\int_{\tilde r}^{+\infty}\dd r\,\frac{1}{\bar a(r)}\nonumber\\
&=\text{finite terms}+\int_{\tilde r}^{+\infty}\dd r\,\frac{1}{\bar a(r)}\label{integral3}\;,
\end{align}

\noindent
for some finite $\tilde r$ large enough, the integrand in the second term of \eqref{integral3} is well approximated by its classical expression thus yielding\footnote{We omit constant pre-factors multiplying $r$ which can be absorbed by rescaling the integration variable accordingly as they do not affect the divergent logarithmic behaviour.}

\begin{align}
\int_{r_s^{(+)}+\epsilon}^{+\infty}\dd r\,\frac{1}{\bar a(r)}&\simeq\text{finite terms}+\int_{\tilde r}^{+\infty}\dd r\,\left(1-\frac{2M}{r}\right)^{-1}\nonumber\\
&=\text{finite terms}+\Bigl[2M\log(r-2M)+r\Bigr]\Bigl|_{r=\tilde r}^{r=+\infty}\longrightarrow+\infty\;.
\end{align}

\noindent
Therefore, $\frac{\dd\bar a}{\dd r}\bigl|_{r=r_s^{(+)}}$ being finite and positive, from Eq. \eqref{extKScoordinates} we see that $T^2-X^2\to-\infty$ as $r\to+\infty$. Analogous computations can be repeated for the region $r<r_\mathcal T$ where, taking into account that $b'(r)<0$ as $b(r)$ monotonously decreases for $r<r_\mathcal T$, we find that $r_*(b(r_s^{(-)}))\to-\infty$ and hence $T^2-X^2=0$ at $r=r_s^{(-)}$, while $r_*\to+\infty$ as $r\to-\infty$ and consequently $T^2-X^2\to-\infty$ asymptotically far in the negative $r$ range.

Finally, by introducing the null coordinates $(\tilde U,\tilde V)$ defined by

\be
\tilde U=\arctan\left(T-X\right)\quad,\quad \tilde V=\arctan\left(T+X\right)\qquad(-\pi/2<\tilde U,\tilde V<\pi/2)
\ee

\noindent
we have that\footnote{With a slight abuse of notation we use $(\tilde U,\tilde V)$-coordinates for both the $r>r_\mathcal T$ and $r<r_\mathcal T$ regions. However, as already noticed, it should be kept in mind that these two regions are covered by different $(T,X)$-coordinate charts and hence also the corresponding $(\tilde U,\tilde V)$ charts are different.} 

\begin{itemize}
	
	\item $b=b_\mathcal T$ corresponds to $T^2-X^2=1$ in $(T,X)$-coordinates which in turn corresponds to $\tilde U+\tilde V=\pm\frac{\pi}{2}$ in the $(\tilde U,\tilde V)$-coordinates;
	
	\item $b=b(r_s^{(\pm)})$ corresponds to $T^2-X^2=0$ and hence to $\tilde U \cdot\tilde V=0$;
	
	\item $b\to\pm\infty$ corresponds to $T^2-X^2\to-\infty$ and hence to $\tilde U=\mp\frac{\pi}{2}$, $\tilde V=\pm\frac{\pi}{2}$.
	
\end{itemize}

\begin{figure}[t!]
	\centering
	\subfigure[]
	{\includegraphics[width=8cm,height=5.5cm]{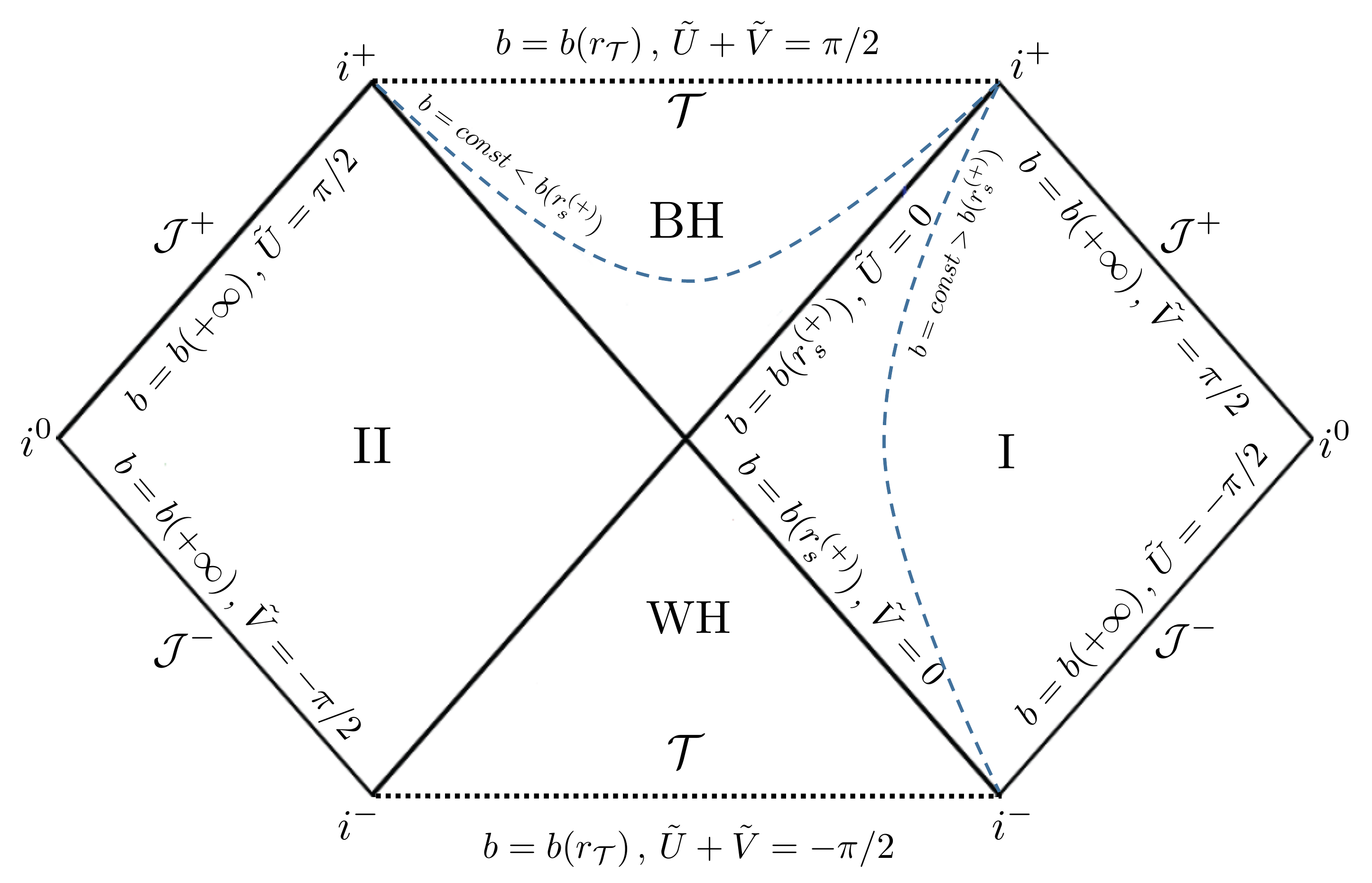}}
	\hspace{1mm}
	\subfigure[]
	{\includegraphics[width=8cm,height=5.5cm]{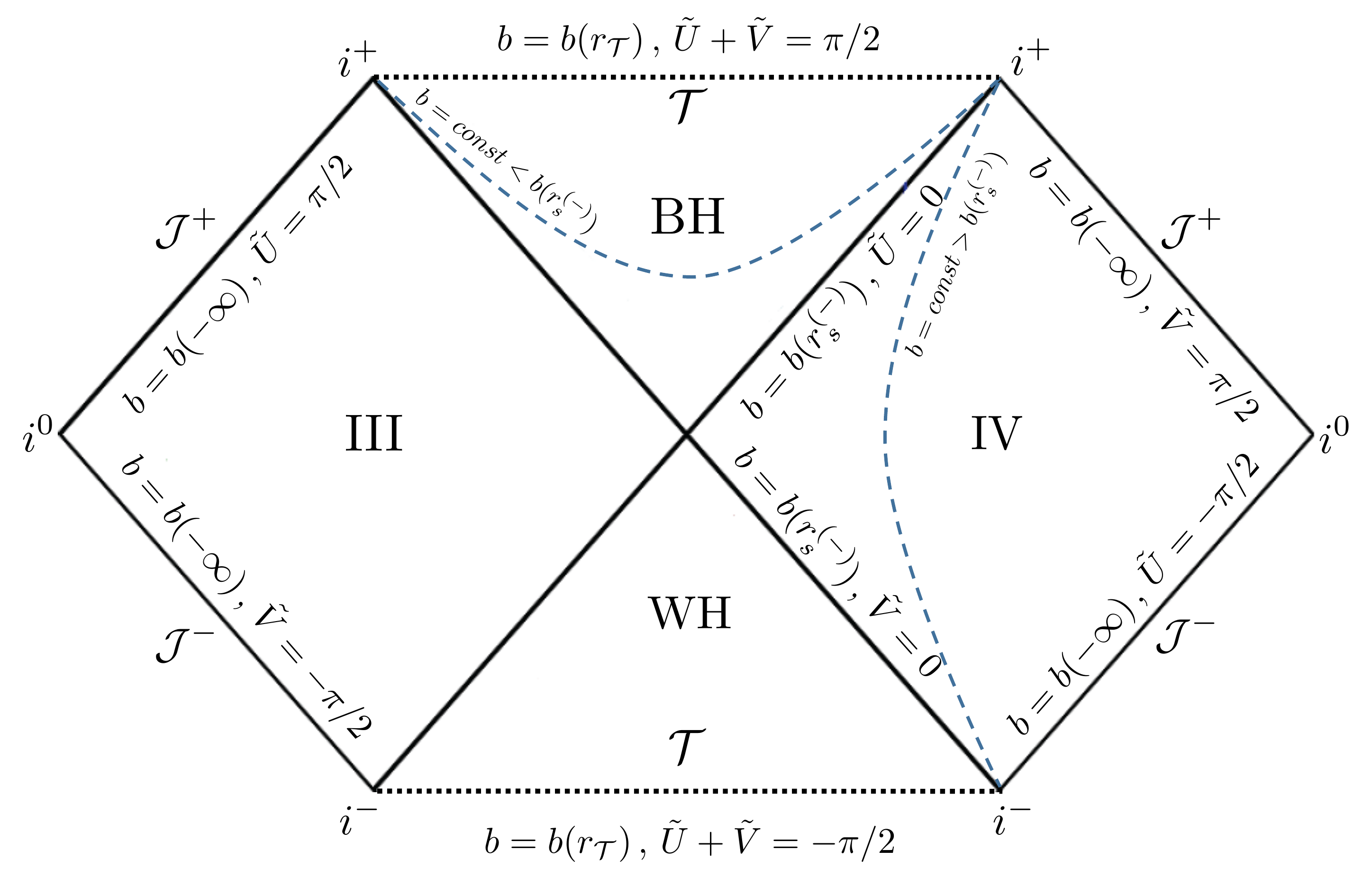}}
	\caption{Penrose diagrams for the $r>r_\mathcal T$ (a) and $r<r_\mathcal T$ (b) regions. We recall again that, although we use the same notation for both regions, they are covered by different $(\tilde U,\tilde V)$-coordinate charts. As usual, the angular coordinates are suppressed so that each point of the diagram can be thought of as representing a 2-sphere of radius $b$.}
	\label{PD}
\end{figure}

\begin{figure}[t!]
	\centering\includegraphics[scale=0.35]{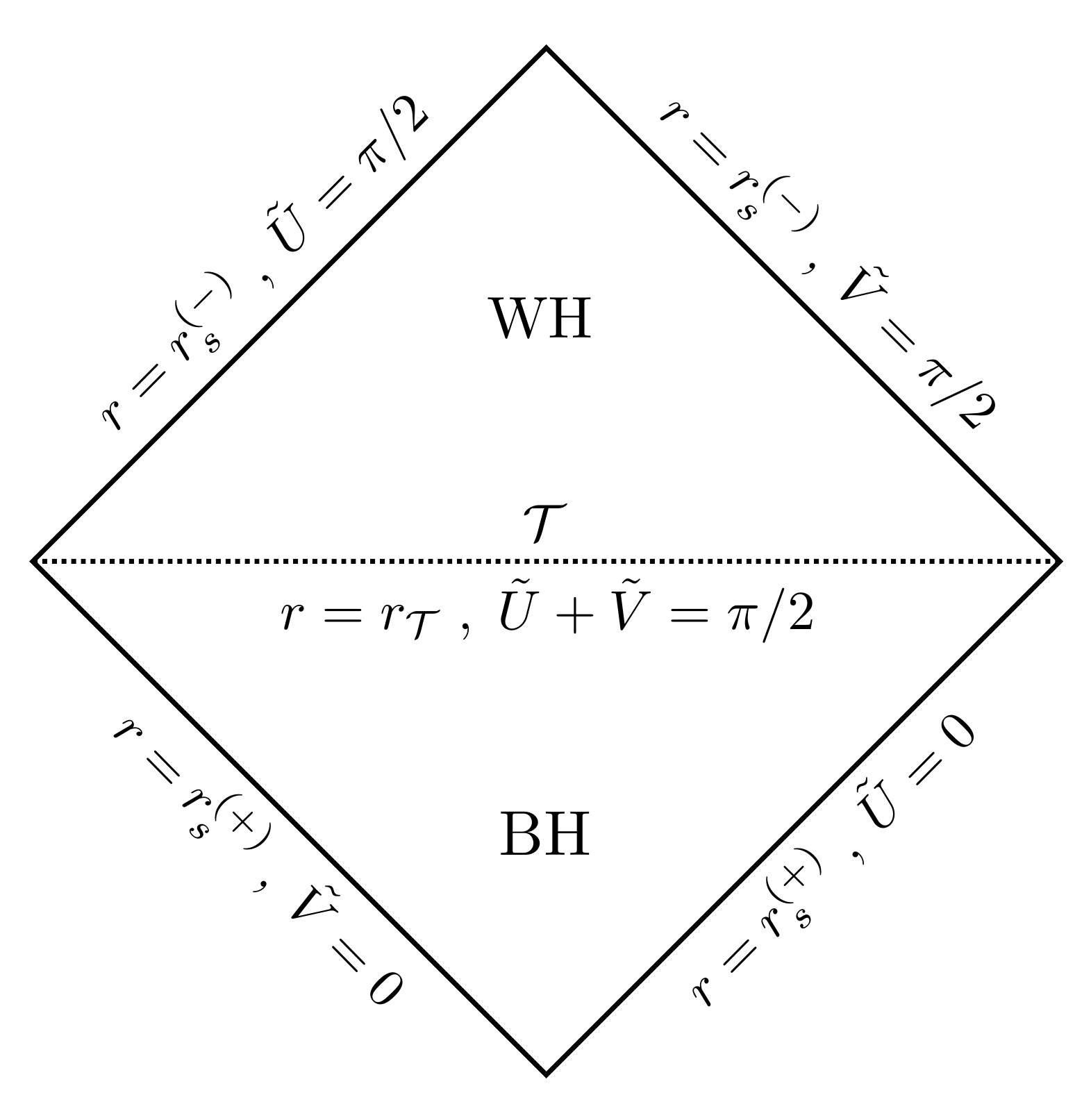}
	\caption{Penrose diagram for the interior region $BH\cup WH$ ($r_s^{(-)}<r<r_s^{(+)}$) given by the union of the trapped and anti-trapped regions BH and WH separated by a transition surface $\mathcal T$ (dotted line). The past boundary is a black hole type horizon while the future boundary is a white hole type horizon.}
	\label{glueint}
\end{figure}

Therefore, as summarized in the Penrose diagrams of Fig. \ref{PD} (a), the $r>r(r_\mathcal T)$ side of the effective quantum corrected Schwarzschild spacetime is divided in the following regions separated by event horizons located at $b=b(r_s^{(+)})$: the black hole exterior region (\texttt{I}) $-X<T<+X$ (i.e., $b>b(r_s^{(+)})$) which reduces to the classical asymptotically flat solution at infinity, the black hole interior region (BH) for which $|X|<T<\sqrt{1+X^2}$ (i.e., $b_\mathcal T<b<b(r_s^{(+)})$), the white hole exterior region (\texttt{II}) $+X<T<-X$ which is again asymptotically flat, and the white hole interior region (WH) $-\sqrt{1+X^2}<T<-|X|$. Similarly, for the $r<r_\mathcal T$ side we have two asymptotically flat regions \texttt{III} and \texttt{IV} where $b>b(r_s^{(-)})$ respectively corresponding to the white hole and black hole exterior regions, and the two interior regions BH and WH for which $b_\mathcal T<b<b(r_s^{(-)})$. Light-like geodesics moving in a radial direction correspond to straight lines at a 45-degree angle in the $(X,T)$-plane. Therefore, according to the direction of the future-pointing unit normal, any event inside the BH region will have a future light cone that remains in that region, while any event inside the WH region will have a past light cone that remains in that region until hitting $r=r_{\mathcal{T}}$. This means that there are no time-like or null curves which go from region \texttt{I} to region \texttt{II} or from region \texttt{III} to \texttt{IV}. Moreover, the BH and WH regions correspond to a trapped and anti-trapped region and we can interpret the event horizons $b=b(r_s^{(\pm)})$ as a black hole and a white hole type horizons, respectively. The BH interior region $b_\mathcal T<b<b(r_s^{(+)})$ and the WH interior region $b_\mathcal T<b<b(r_s^{(-)})$ are causally connected through the transition surface $\mathcal T$ which replaces the classical singularity. Indeed, it is possible to introduce a local $(T,X)$-chart defined by

\be\label{interiorchart}
T^2-X^2=\exp\left[\bar a'(r_s^{(+)})r_*(r)\right]\qquad,\qquad r_*(r)=\int_{r_\mathcal T}^{r}\dd r\,\frac{1}{\bar a(r)}
\ee

\noindent
which covers both interior regions\footnote{$b(r)$ being smooth in the two branches $r>r_\mathcal T$ and $r<r_\mathcal T$, the overlapping map between the chart \eqref{interiorchart} and the corresponding chart \eqref{KScoordinates} in one of the two interior regions (BH or WH) is smooth for any two intersecting open neighborhoods in that region. Furthermore, one can show that the chart \eqref{interiorchart} covers the entire region $BH\cup WH$.} as schematically showed in the portion of the Penrose diagram of Fig. \ref{glueint} where we report also the corresponding values of $\tilde U$ and $\tilde V$.

\begin{figure}[t!]
	\centering\includegraphics[scale=0.75]{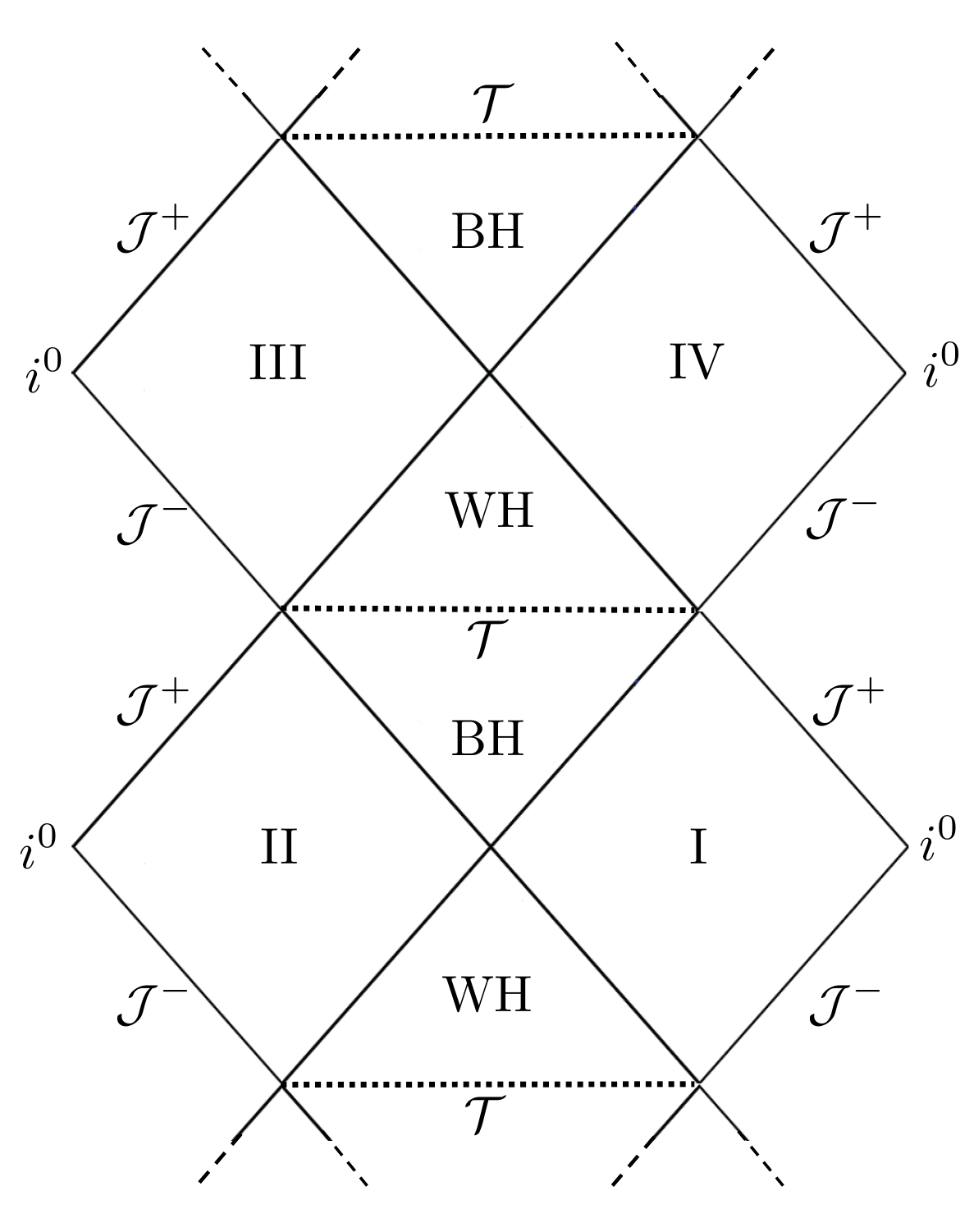}
	\caption{Penrose diagram for the Kruskal extension of the full quantum corrected polymer Schwarzschild spacetime.}
	\label{Penrosediag2}
\end{figure}

The two diagrams of Fig. \ref{PD} can be then glued together at the transition surface so that a light ray originating at the past boundary of region \texttt{I} will reach the future asymptotic boundary of region \texttt{III} passing trough the black hole and white hole interiors which are smoothly connected via the singularity resolution induced by quantum geometry effects. Similarly, region \texttt{II} is causally connected to region \texttt{IV}. Therefore, the Kruskal extension of the quantum corrected spacetime spans the whole range $r\in(-\infty,+\infty)$ corresponding to the entire region $\texttt{I}\cup BH\cup WH\cup\texttt{III}$ over which the metric coefficients are such that the effective 4-metric is smooth and well-defined. Furthermore, since the spacetime topology is $\mathbb R\times\mathbb R\times\mathbb S^2$, the above considerations can be repeated over the whole non compact $T$-direction. Hence, as indicated by the dashed lines in Fig. \ref{Penrosediag2}, the Penrose diagram for the Kruskal extension of the full quantum corrected effective Schwarzschild spacetime continues to infinitely many trapped and anti-trapped and asymptotic regions to the past and future. Choosing instead $\mathbb{S}^1$-topology in $T$-direction allows to glue an even number of Penrose diagrams Fig. \ref{PD} by identifying uppermost transition surface with the lowest one.

According to the considerations of Sec. \ref{sec:curvinvariants} and \ref{sec:horizon} each time we pass through an interior region the ADM mass changes according to Eq. \eqref{amplification} and \eqref{deamplification}, respectively. Hence assuming region \texttt{I} and \texttt{II} having ADM mass $M_{BH}$, region \texttt{III} and \texttt{IV} admit a (de)-amplified ADM mass $M_{WH}$. Crossing then the transition surface in the next future interior region we have a mass (amplification) de-amplification such that the regions \texttt{V} and \texttt{VI} (in the future of \texttt{III} and \texttt{IV} in Fig. \ref{PD}) are characterised by the ADM mass $M_{BH}$ again. In total, going through the Penrose diagram the mass oscillates between $M_{BH}$ and $M_{WH}$, i.e. an indefinite mass amplification or de-amplification is avoided. We can make this statement more precise by considering an observer 1 starting in region \texttt{I}. At a certain distance $b^{o}$, this observer will provide initial conditions $v_1^o$, $P_1^{o}$, $P_2^o$ and $v_2^{o}$. Being observer 1 in the black hole exterior, the initial data will be $P_1^{o}\sim 0$, $P_2^o \sim 0$, and especially this observer will fix $M_{BH}^{(1)}$ and also $M_{WH}^{(1)}$. If observer 1 falls into the black hole and exits into region \texttt{III}, the momenta will evolve from $\sim 0$ to $P_1 \sim \pi/\lambda_1$ and $P_2 \sim \pi/\lambda_2$ at the same distance $b^o$. Observer 1 will experience for instance a mass amplification $M_{WH}^{(1)}> M_{BH}^{(1)}$, i.e. $\beta = \frac{5}{3}$. Now, an observer 2 in region \texttt{III} will provide the initial conditions $\tilde{v}_1^o=v_1^o$, $\tilde{P}_1^o \sim 0$, $\tilde{P}_2^o \sim 0$ and some $\tilde{v}_2^o$ at the same value $b^o$. Therefore, the values of $v_1, P_1, v_2, P_2$ resulting from the evolution of observer 1 from region \texttt{I} to region \texttt{III} can be mapped into the corresponding values of observer 2 at the same $b_o$ by means of the transformation

\begin{equation}\label{map}
v_1 \longmapsto v_1 \quad , \quad P_1 \longmapsto \frac{\pi}{\lambda_1} - P_1 \quad , \quad P_2 \longmapsto \frac{\pi}{\lambda_2} - P_2\quad,\quad v_2 \longmapsto v_2\;.
\end{equation}

\noindent
Recalling then the expressions \eqref{F} and \eqref{Fbar} of the Dirac observables, the transformation \eqref{map} maps

\begin{equation}\label{mapMass}
F_Q \longmapsto \bar{F}_{Q} \quad , \quad \bar{F}_Q \longmapsto F_Q \qquad \text{i.e.} \qquad M_{BH} \longmapsto M_{WH} \quad , \quad M_{WH} \longmapsto M_{BH} \;.
\end{equation}

\noindent
Therefore observer 2 would fix $M_{BH}^{(2)} = M_{WH}^{(1)}$ and $M_{WH}^{(2)} = M_{BH}^{(1)}$, which shows that for observer 2 the white hole side of observer 1 actually looks like a black hole side, which is exactly what was already discussed in Sec. \ref{sec:intconst}. Moreover, for observer 2, $M_{WH}^{(2)}$ is now smaller than $M_{BH}^{(2)}$, i.e. observer 2 would experience exactly the other $\beta$ value, namely $\beta=\frac{3}{5}$. Due to the transformations \eqref{map} and \eqref{mapMass}, an indefinite mass amplification is thus avoided throughout the quantum-extended effective spacetime as the mass oscillates between $M_{BH}$ and $M_{WH}$.

\section{Discussion and previous work}\label{sec:comparison}

Our result can be translated in the language usually used in the loop quantum black hole context \cite{AshtekarQuantumGeometryand,ModestoLoopquantumblack,CorichiLoopquantizationof,ModestoSemiclassicalLoopQuantum,BoehmerLoopquantumdynamics,ChiouPhenomenologicalloopquantum,OlmedoFromblackholesto,AshtekarQuantumTransfigurarationof,AshtekarQuantumExtensionof}.
First of all, the variables we used $(v_1, P_1, v_2, P_2)$ can easily be related to connection variables (geometric one-forms) for the interior of the black hole. Comparing with the line element Eq. (2.8) in \cite{AshtekarQuantumExtensionof}\footnote{We focus here on \cite{AshtekarQuantumExtensionof}, but same can be found in the other references mentioned above.} we find

$$
-a = \frac{\left(p^{(conn)}_{b}\right)^2}{|p^{(conn)}_c|} \quad , \quad b^2 = |p^{(conn)}_c| \;,
$$
\noindent
where, to avoid confusion with the metric variables $b$ and $p_b$, the superscript ${(conn)}$ indicates connection variables. Note that by definition $a < 0$ in the interior of the black hole. This allows to relate $(v_1, P_1)$, $(v_2,P_2)$ with the connection variables $(c^{(conn)},p_c^{(conn)})$, $(b^{(conn)},p_b^{(conn)})$ via\footnote{Note that in the action \eqref{eq:action}, we did not include the factor $\frac{1}{4}$ into the Lagrangian, while in LQBH literature it is. Dynamically this is not a problem, but has to be taken into account if we want to compare the canonical structure. To this aim, we first need to perform a coordinate transformation, which rescales $v_i \mapsto 16 v_i$ and $P_i \mapsto P_i/16$. Including this rescaling leads to the relations above.}

\begin{align}
\left(p_b^{(conn)}\right)^2 = -8 v_2 \quad &, \quad |p^{(conn)}_c| = \left(24 v_1\right)^{\frac{2}{3}}\;, 
\label{connvaraibles1}
\\
b^{(conn)} = \text{sign}(p_b^{(conn)})\; \frac{\gamma}{4}\; \sqrt{-8 v_2} \;P_2 \quad &, \quad c^{(conn)} = -\text{sign}(p_c^{(conn)})\; \frac{\gamma}{8}\; \left(24 v_1\right)^{\frac{1}{3}}\; P_1  \;.
\label{connvaraibles2}
\end{align}
\noindent
A quick calculation shows that this mapping results in the proper symplectic structure of the connection variables and correspondingly to the Poisson brackets

$$
\left\{c^{(conn)},p^{(conn)}_c\right\}_{(v,P)} = 2 \gamma \quad, \quad \left\{b^{(conn)},p^{(conn)}_b\right\}_{(v,P)} = \gamma \;,
$$ 
\noindent
where $\left\{\cdot , \cdot \right\}_{(v,P)}$ denotes the Poisson bracket w.r.t. the canonical variables $\left\{(v_1, P_1), (v_2,P_2)\right\}$. The classical Hamiltonian \eqref{hamiltonian2} then reads 

\begin{equation}
H = N \left( c^{(conn)} p_c^{(conn)} + \left(b^{(conn)} + \frac{\gamma^2}{b^{(conn)}}\right) p_b^{(conn)} \right) \;.
\end{equation}
\noindent
Note that there is an ambiguity in choosing the sign of $p_b^{(conn)}$ and $p_c^{(conn)}$, which reflects the choice of an orientation of the triad basis.

Finally, this allows us to ask to which kind of polymerisation scheme our polymerisation of $P_1$, $P_2$ with constant $\lambda_1$ and $\lambda_2$ corresponds to. For this we set
\begin{align}
\lambda_1 P_1 = \delta_c c^{(conn)} \quad , \quad \lambda_2 P_2 = \delta_b b^{(conn)} \; ,
\end{align}
\noindent
from which, using Eq. \eqref{connvaraibles1},\eqref{connvaraibles2}, it follows that
\begin{align}
\delta_c &= \pm \frac{8}{\gamma} \frac{\lambda_1}{\sqrt{|p_c^{(conn)}|}}\;,\\
\delta_b &= \pm \frac{4 \lambda_2}{\gamma |p_b^{(conn)}|}\;,
\end{align}
\noindent
where the choice of the signs is due to the signs chosen in the square roots. The choice of the sign is a matter of convenience, since the polymerised Hamiltonians are invariant under sign changes of the polymerisation scales. In total, being in connection variables and using the polymerisation scales above, this will lead to the same effective spacetime as we found in $(v_i,P_i)$-variables with constant $\lambda_i$. This choice of $\delta_c$ and $\delta_b$ corresponds to a $\bar{\mu}$-like scheme.

Let us first compare our polymerisation with other $\bar{\mu}$ proposals as in \cite{BoehmerLoopquantumdynamics,ChiouPhenomenologicalloopquantum}. In these references, the quantum parameters $\delta_b$ and $\delta_c$ are chosen to be

$$
\delta_b \propto \frac{1}{\sqrt{p_c^{(conn)}}} \quad , \quad L_o \delta_c \propto \frac{L_o \sqrt{p_c^{(conn)}}}{p_b^{(conn)}}\;.
$$
\noindent
Within this scheme, the equations for $b^{(conn)}$ and $c^{(conn)}$ do not decouple and hence only numerical results are available \cite{BoehmerLoopquantumdynamics,ChiouPhenomenologicalloopquantum} (see also \cite{ChiouPhenomenologicaldynamicsof,JoeKantowski-Sachsspacetimein} for the cosmological Kantowski-Sachs setting). This scheme has the disadvantage that quantum effects become relevant at the horizon, which is for large masses the low curvature regime. This can be easily understood by studying the combinations
\begin{align}
&\delta_b b^{(conn)} \propto \delta_b p_b^{(conn)} P_2 \propto \frac{p_b^{(conn)}}{\sqrt{p_c^{(conn)}}} P_2 = \sqrt{|a|} \frac{b'}{\sqrt{n} b} \;,
\notag
\\
&\delta_c c^{(conn)} \propto \delta_c \sqrt{|p_c^{(conn)}|} P_1 \propto \frac{|p_c^{(conn)}|}{p_b^{(conn)}} P_1 = \frac{1}{\sqrt{|a|}} \frac{a'}{\sqrt{n}} \;,
\end{align}
\noindent
in metric variables. Quantum effects become relevant if these combinations become large. Since by definition $a \rightarrow 0$ at the horizon, while $n$, $b$, $b'$ and $a'$ remain finite, quantum effects become necessarily large at the horizon.
As discussed above, this is clearly distinct from our approach, where quantum effects at the horizon remain negligible.

Furthermore, we can compare to $\mu_o$-schemes \cite{AshtekarQuantumGeometryand,ModestoLoopquantumblack,CampigliaLoopquantizationof,ModestoSemiclassicalLoopQuantum} and generalised $\mu_o$-scheme \cite{CorichiLoopquantizationof,OlmedoFromblackholesto,AshtekarQuantumTransfigurarationof,AshtekarQuantumExtensionof}. For comparing we focus on the generalised $\mu_o$-scheme, where we look in detail on \cite{CorichiLoopquantizationof} from now on referred to as CS approach as the first of these generalised $\mu_o$-schemes and on \cite{AshtekarQuantumExtensionof} referred to as AOS in the following as the latest one. The important difference between them is that in CS the polymerisation scales are given by

$$
\delta_b = \frac{\sqrt{\Delta}}{r_o} \quad , \quad \delta_c = \frac{\sqrt{\Delta}}{L_o} \;,
$$
\noindent
where $r_o$ and $L_o$ are fiducial cell parameters and the mass dependence comes from identifying $r_o = 2 M_{BH}$ as a physical scale, while for the AOS generalised $\mu_o$-scheme the polymerisation scales are
\begin{figure}[t!]
	\centering
	\subfigure[]
	{\includegraphics[width=7.75cm,height=5.5cm]{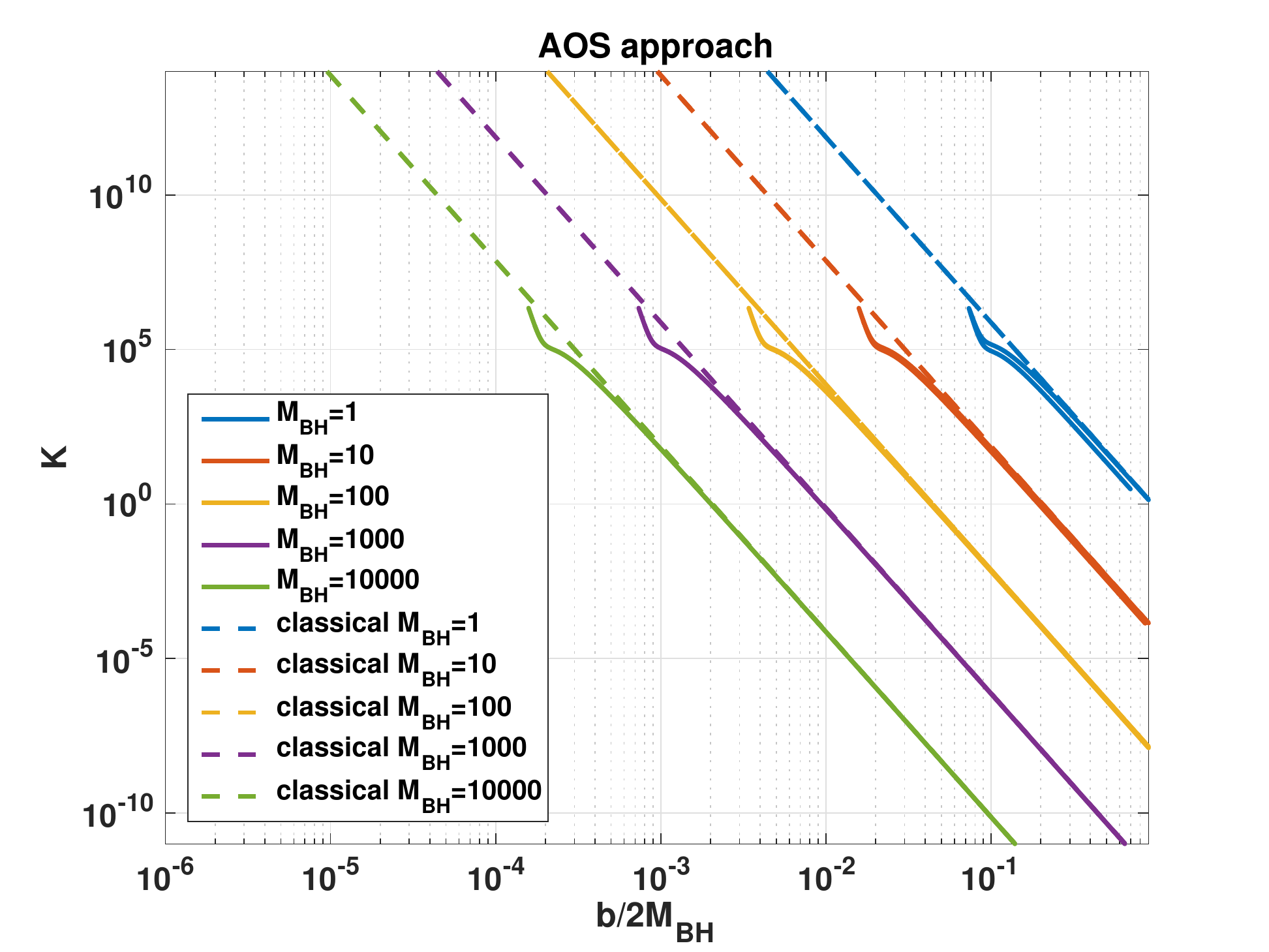}}
	\hspace{2mm}
	\subfigure[]
	{\includegraphics[width=7.75cm,height=5.5cm]{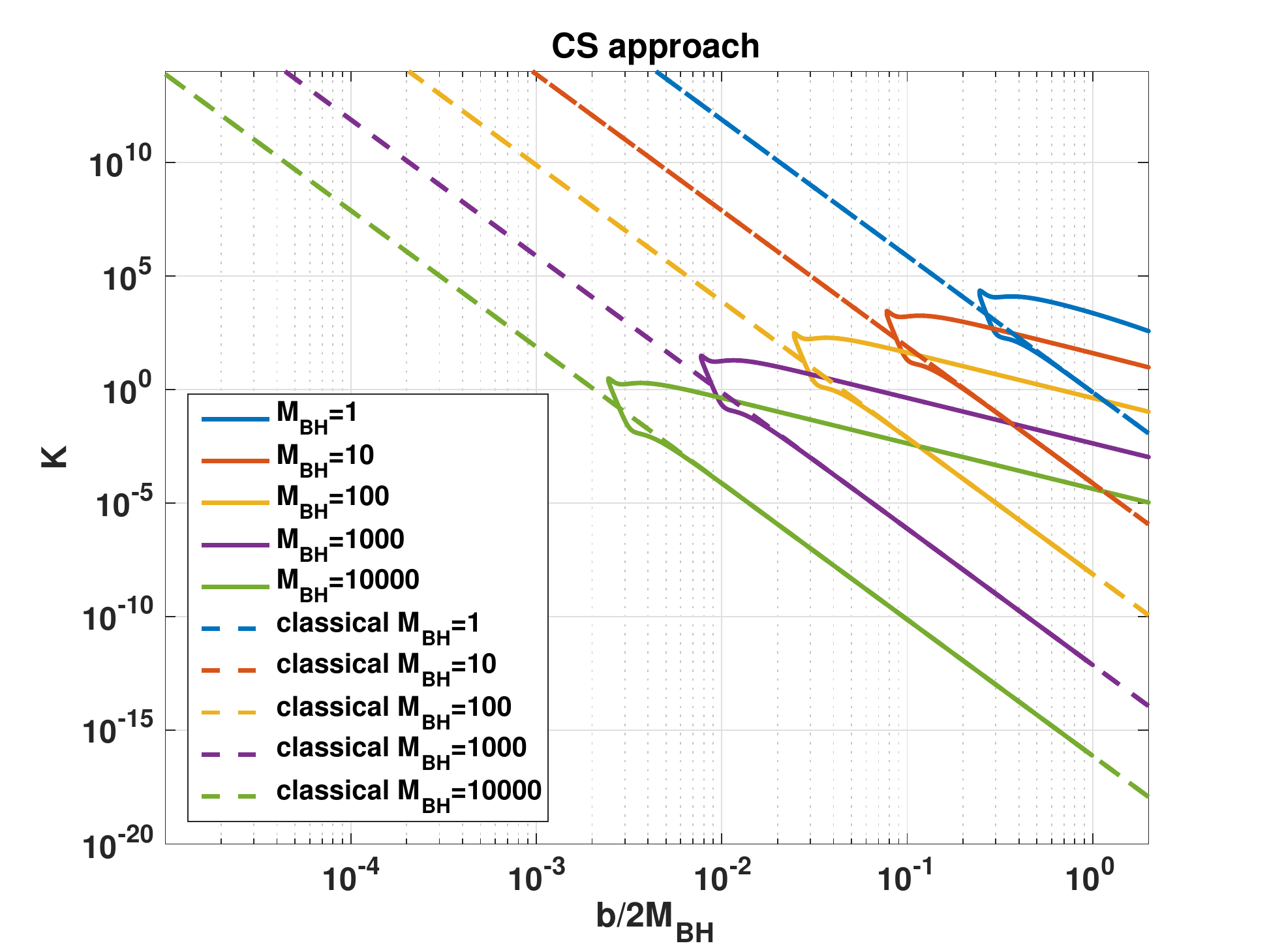}}
	\caption{Plot of the Kretschmann scalar $\mathcal{K}$ against $b$ in the AOS approach (a) and the CS approach (b) in log-log scale for different masses. While in the AOS approach the onset of quantum effects is always at the same scale, in the CS approch this scale decreases with increasing mass. The parameters are set to $\gamma = 0.2375$ (cfr. \cite{DomagalaBlack-holeentropy,MeissnerBlack-holeentropy}), $\Delta = 1$, $L_o =1$, $r_o = 2 M_{BH}$.}
	\label{fig:kretschmanncomp}
\end{figure} 
$$
\delta_b = \left(\frac{\sqrt{\Delta}}{\sqrt{2\pi} \gamma^2 M_{BH}}\right)^{\frac{1}{3}} \quad , \quad L_o \delta_c = \frac{1}{2} \left(\frac{\gamma \Delta^2}{4 \pi^2 M_{BH}}\right)^{\frac{1}{3}} \;,
$$

\noindent
where now both polymerisation scales depend non-trivially on the black hole mass due to a specific requirement based on rewriting the curvature in terms of the holonomies of the gravitational connection along suitably chosen plaquettes enclosing the minimal area at the transition surface (cfr. \cite{AshtekarQuantumTransfigurarationof,AshtekarQuantumExtensionof} for more details).
First of all, we can compare where quantum effects become relevant expressed in terms of the Kretschmann scalar. Fig. \ref{fig:kretschmanncomp} shows the Kretschmann scalar as a function of $b = \sqrt{|p_c^{(conn)}|}$ in the AOS and CS setting for different black hole masses.
While in the CS approach the scale at which quantum effects become relevant decreases with higher black hole masses, in the AOS setting, this scale remains constant and mass independent. Comparing this to Fig. \ref{fig:kretschmann} shows that our polymerisation achieves the onset of quantum effects at a mass independent Kretschmann scalar scale without choosing mass dependent polymerisation scales, but via restricting the initial conditions, i.e. via fixing the while hole mass as a function of the black hole mass.
\begin{figure}[t!]
	\centering
	\subfigure[]
	{\includegraphics[width=7.75cm,height=5.5cm]{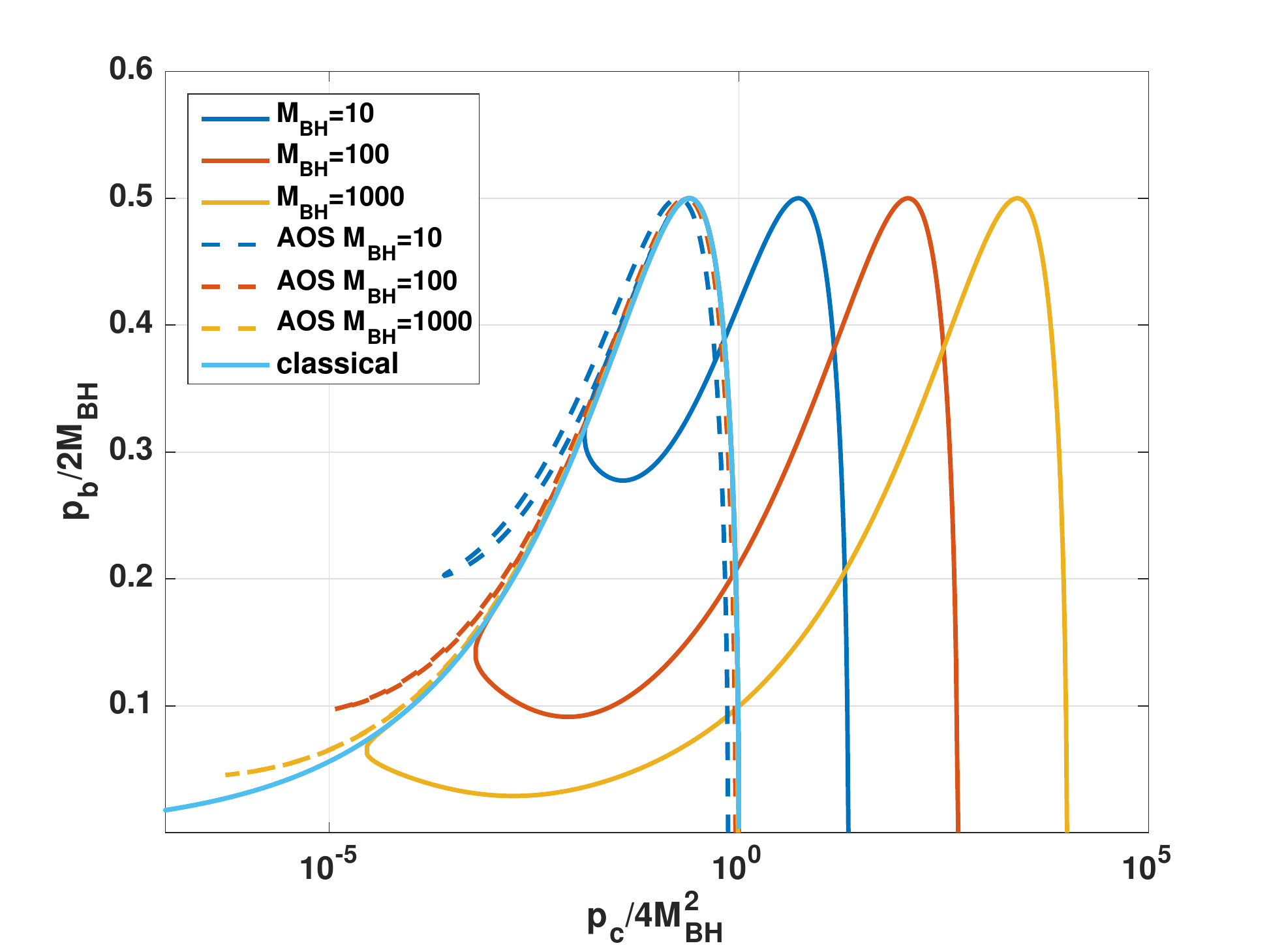}}
	\hspace{2mm}
	\subfigure[]
	{\includegraphics[width=7.75cm,height=5.5cm]{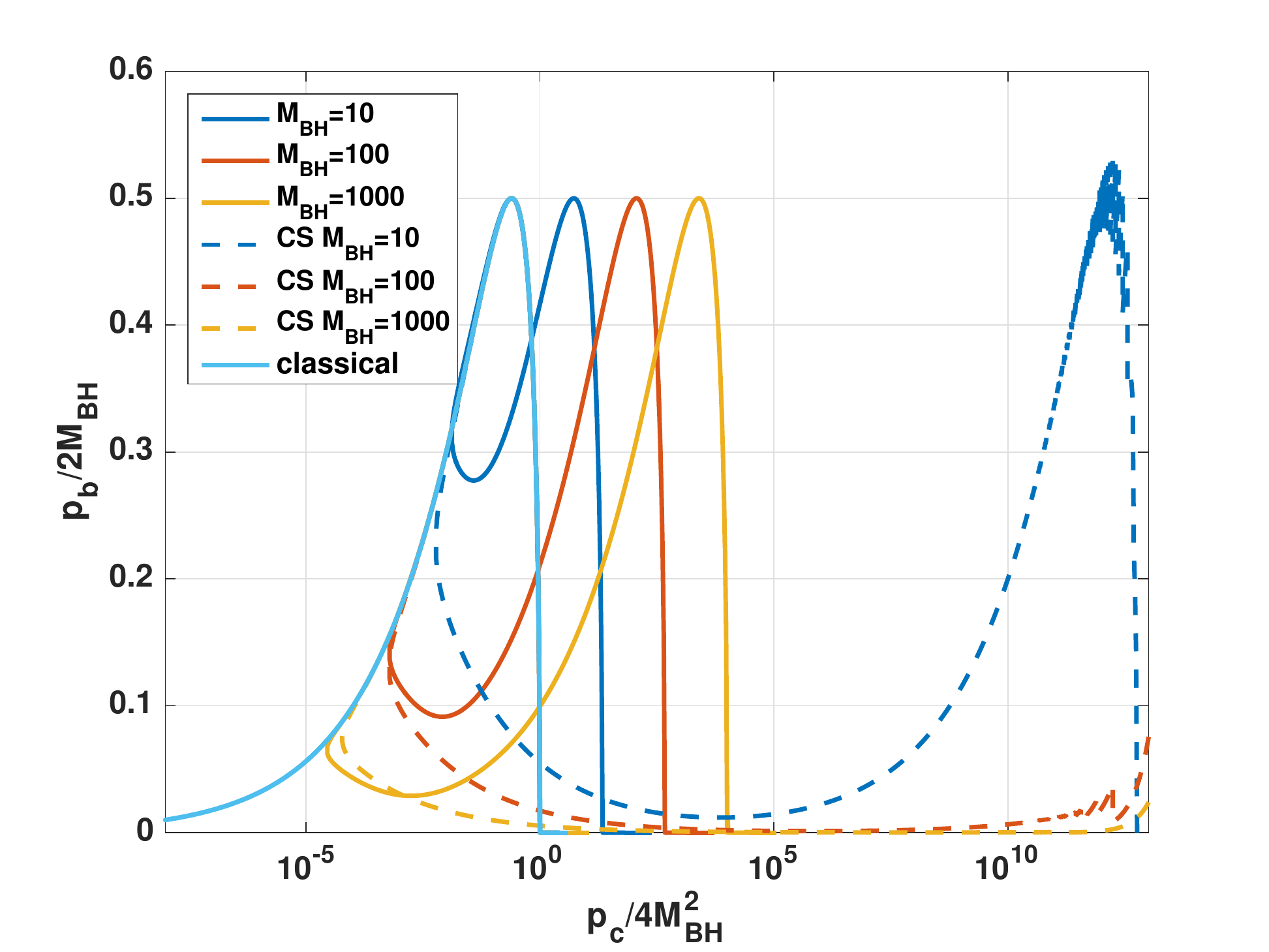}}
	\caption{Plot of $p_b^{(conn)}$ versus $p_c^{(conn)}$ with logarithmic x-axis for different masses. In (a) our polymerisation is compared to the AOS approach, while (b) compares it with the CS approach. Different behaviours regarding symmetry, ratio of black hole and white hole radius and location of transition surface are visible. The parameters are set to $\gamma = 0.2375$ (cfr. \cite{DomagalaBlack-holeentropy,MeissnerBlack-holeentropy}), $\Delta = 1$, $\mathscr L_o = L_o =1$, $r_o = 2 M_{BH}$, $\mathscr L_o\lambda_1= \lambda_2/\mathscr L_o =1$ and $\beta = \frac{5}{3}$ with $\bar{m} = 1$. Note that we can not plot the solutions of CS in (b) for larger $p_c$-values as numerical difficulties occur.}
	\label{fig:pbvspc}
\end{figure}

As a next step, we can compare the dynamical trajectories in a gauge independent plot of $p_b^{(conn)}$ against $\log(p_c^{(conn)})$, as reported in Fig. \ref{fig:pbvspc}. The plot shows for all three polymerisations a good accordance with the classical solution close to the horizon, while the quantum effects are quite different from case to case. All approaches share similar qualitative features as having two horizons and the existence of a transition surface. As the plot shows, all approaches have also in common that the mass normalised bouncing point $\min\{|p_c^{conn}|\}/(4 M_{BH}^2) = b_{\mathcal{T}}^2/(4 M_{BH}^2)$ decreases with increasing mass. Nonetheless, taking a closer look, the quantitative result is different. In fact, comparing the functional expressions in \cite{AshtekarQuantumExtensionof}, \cite{CorichiLoopquantizationof} and Eq. \eqref{eq:btransition2}, we have
\begin{align*}
&\min\{|p_c^{conn}|\}_{\text{AOS}} = \frac{\gamma}{4} \left(\frac{\gamma \Delta^2}{4 \pi}\right)^{\frac{1}{3}} M_{BH}^{\frac{2}{3}} \;\\
&\min\{|p_c^{conn}|\}_{\text{CS}} = \gamma \sqrt{\Delta} M_{BH} \; \\
&\min\{|p_c^{conn}|\}_{\text{our}} = \begin{cases}
\left(\frac{2 \lambda_1 \mathscr L_o}{m^{\frac{1}{4}}}\right)^{\frac{2}{3}}M_{BH}^{\frac{2}{3}} &\;\text{for}\;\;\beta=\frac{5}{3}\\
&\\
\left(\frac{2 \lambda_1 \mathscr L_o}{m^{\frac{1}{4}}}\right)^{\frac{2}{3}}M_{WH}^{\frac{2}{3}} &\;\text{for}\;\;\beta=\frac{3}{5}
\end{cases}\;,
\end{align*}  
\noindent
from which we see that the AOS and our approach have the same behaviour $\propto M_{BH}^{\frac{2}{3}}$, while in the CS approach the minimal value of $p_c^{(conn)}$ increases linearly with $M_{BH}$. Furthermore, clear differences in the symmetry are visible. While for all depicted masses the AOS solutions are approximately symmetric, neither CS nor ours are. As a consequence, the horizon radius of the black hole and the white hole, which are given by $\sqrt{p_c^{(conn)}}$ at $p_b^{(conn)}=0$, differ for the latter ones. This difference in the horizon radii translates also in differences of the black hole and white hole masses. While in AOS both masses are (in the large mass limit \cite{AshtekarQuantumExtensionof}) approximately equal, i.e. $(M_{WH}/M_{BH})_{AOS} \approx 1$, in CS this relation is $(M_{WH}/M_{BH})_{CS} \sim M_{BH}^2$ \cite{CorichiLoopquantizationof} which corresponds to a mass amplification, while in our model $(M_{WH}/M_{BH}) \sim M_{BH}^{\beta-1}$, where $\beta = \frac{5}{3}, \frac{3}{5}$, respectively corresponding to a mass (de-)amplification smaller than in CS. 

\section{Sketch of quantum theory}\label{quantumtheory}

In this section, we will briefly sketch a possible quantisation of our model from which the effective equations discussed in the main part of the paper are expected to emerge. This hope is founded on similar results in loop quantum cosmology \cite{DienerNumericalSimulationsOf,AshtekarQuantumNatureOfAnalytical} in related models. As we will see, the kernel of the Hamiltonian constraint operator can be found in closed form, which suggests that a complete analytic control of the quantum theory may be possible. 

We recall the effective classical Hamiltonian \eqref{Heff1}:
\be
H_{\text{eff}} = \sqrt{n} \mathcal{H}_{\text{eff}}\quad , \quad \mathcal{H}_{\text{eff}} = 3v_1 \frac{\sin\left(\lambda_1 P_1\right)}{\lambda_1} \frac{\sin\left(\lambda_2 P_2\right)}{\lambda_2} + v_2 \frac{\sin\left(\lambda_2 P_2\right)^2}{\lambda_2^2} - 2 \approx 0 \text{.}
\ee
A comparison with techniques successful for loop quantum cosmology, in particular \cite{AshtekarRobustnessOfKey, Martin-BenitoFurtherImprovementsIn} and also \cite{BodendorferANoteOnTheScalar} for a detailed discussion, suggests to work with a lapse such that $\sqrt{n} = v_2$, corresponding to a density weight 2 Hamiltonian\footnote{E.g., the density weight in $t$-direction of $v_2$, i.e. the scaling with changing $L_0$, is $2$.}. We set $\lambda_1=\lambda_2=2$ for simplicity (we will explain below why)\footnote{A generalisation to the Bohr compactification of the real line can be done by standard techniques \cite{AshtekarMathematicalStructureOf}, which allows to treat arbitrary real non-zero values of $\lambda_1, \lambda_2$. } and choose the ordering 
\begin{align}
H_{\text{eff}} =~& 3 \sqrt{v_1} \left( \frac{\sin\left(2 P_1\right)}{4} \text{sign}({v_1}) +  \text{sign}({v_1}) \frac{\sin\left(2 P_1\right)}{4} \right) \sqrt{v_1}  \notag\\
& ~~~~~~~~~ \times \sqrt{v_2}  \left( \frac{\sin\left(2 P_2\right)}{4}  \text{sign}({v_2}) + \text{sign}({v_2}) \frac{\sin\left(2 P_2\right)}{4} \right) \sqrt{v_2} \notag\\
& + \left(  \sqrt{v_2}  \left( \frac{\sin\left(2 P_2\right)}{4}  \text{sign}({v_2}) + \text{sign}({v_2}) \frac{\sin\left(2 P_2\right)}{4} \right) \sqrt{v_2}  \right)^2 -2 v_2 \approx 0 \text{.} \label{eq:QuantumOrdering}
\end{align}
A basis of the Hilbert space is given by the volume eigenstates $\ket{v_1, v_2}$ on which the operators corresponding to $v_1$ and $\exp(- i n P_1)$, $n \in \mathbb Z$, act as 
\be
\hat v_1\ket{v_1, v_2} = v_1 \ket{v_1, v_2} , ~~~~ \widehat{e^{- i n P_1}} \ket{v_1, v_2} =  \ket{v_1+n, v_2} \text{,}
\ee
and similarly for $v_2$ and $\exp(- i n P_2)$. Using the decomposition $\ket{\chi} = \sum_{v_1, v_2 \in \mathbb Z} \tilde \chi(v_1, v_2) \ket{v_1, v_2}$, elements of the Hilbert space can be written as functions
\be
\tilde \chi(v_1, v_2), ~~~ v_1, v_2 \in \mathbb Z
\ee 
which are square integrable w.r.t. to the scalar product 
\be
\braket{\chi_1}{\chi_2} = \sum_{v_1, v_2 \in \mathbb Z} \overline{\tilde \chi_1(v_1, v_2)} \tilde \chi_2(v_1, v_2) \text{.}
\ee 

The choice $\lambda_1=\lambda_2=2$ leads to four dynamically selected subsectors $v_i \in 2 \mathbb  Z + c_i$, $c_i \in \{0,1\}$. We choose the subsector containing $\ket{0,0}$. The ordering \eqref{eq:QuantumOrdering} then ensures that zero volume states $\ket{0, v_2}$ or $\ket{v_1, 0}$ are annihilated by the Hamiltonian constraint operator and that non-zero volumes are not mapped to zero volumes. This leads to a dynamical decoupling of the zero volume sector. Also, positive and negative volume sectors are not mapped into each other.  

Again, following standard techniques \cite{AshtekarRobustnessOfKey, BodendorferANoteOnTheScalar}, we rescale our wave functions as
\be
\tilde \chi(v_1, v_2) = \sqrt{|v_1 v_2|} \tilde \psi(v_1, v_2) ~~ \text{for}  ~~ v_1, v_2 \neq 0 \text{,} ~~
\ee
leading to the scalar product 
\be
\braket{\psi_1}{\psi_2} = \sum_{v_1, v_2 \in \mathbb Z} \overline{\tilde \psi(v_1, v_2)} |v_1 v_2 |  \tilde \psi(v_1, v_2) \text{.}
\ee
The Fourier transform 
\be
\psi(P_1, P_2) = \sum_{v_1, v_2 \in 2  \mathbb Z} \tilde \psi(v_1, v_2) e^{-i (P_1 v_1+P_2 v_2)}, ~~  ~ \tilde \psi(v_1, v_2) = \frac{ 1}{\pi^2} \int_{0}^{\pi} dP_1 \int_{0}^{\pi}  dP_2 \, e^{i  (v_1 P_1 + v_2 P_2) } {\psi}(P_1, P_2) 
\ee
for $P_1, P_2 \in [0, \pi]$ yields the Hamiltonian constraint operator  
\be
\hat H = - 3  \frac{\sin\left(2 P_1\right)}{2} {  \partial_{P_1}}  \frac{\sin\left(2 P_2\right)}{2} { \partial_{P_2}} - \left(   \frac{\sin \left(2 P_2\right)}{2}   {\partial_{P_2}} \right)^2  + 2 i \partial_{P_2}  \text{.} 
\ee
We now perform the standard variable transform \cite{AshtekarRobustnessOfKey} $x_i=\log(\tan(P_i/2))$, leading to the Hamiltonian constraint operator  
\be
\hat H = - 3 \tanh(x_1)   \partial_{x_1} \tanh(x_2)   \partial_{x_2}  - \left(   \tanh(x_2)   \partial_{x_2} \right)^2 + 2 i \cosh(x_2)   \partial_{x_2}  \text{.} 
\ee
For this transform, it was important that the dynamically selected sublattice has support only on even $v_1, v_2$, leading to $P_1, P_2 \in [0, \pi]$.
Further transforming $y_i=\log(\sinh(x_i))$ leads to 
\be
\hat H = \Bigl( - 3    \partial_{y_1}   -     \partial_{y_2} + 4 i \cosh(y_2)  \Bigr)   \partial_{y_2}   \text{.} 
\ee
We note that for the standard choice of branch cut of the logarithm, $y_i$ is real only for $x_i>0$. One may restrict to having volume-symmetric states, see e.g. \cite{AshtekarRobustnessOfKey}, leading to symmetric functions in $x_i$ so that we can restrict to $x_i>0$ and avoid complex $y_i$. This would restrict us to a either the interior or exterior of the black hole ($a >0$ or $a<0$). Otherwise, one may consider both interior and exterior at the same time at the cost of having to complexify the phase space.

The general solution to the equation $\hat H \ket{\psi} = 0$ reads
\be
\psi_{\text{phys}}(y_1, y_2) =  g(y_1)+\int^{y_2} dy_2' \, e^{4 i \sinh(y_2')} f \left(y_2' - \frac{1}{3} y_1 \right) \text{,}
\ee
where $f$, $g$ are arbitrary functions. 
A complete construction of the quantum theory still requires the physical inner product, observables, and preferably semiclassical states peaked on classical phase space points. An example of an observable would be the operator version of
\be
O_1 = \frac{\sqrt{2}}{3} \left( \lambda_1^2 \lambda_2^2 F_Q \bar{F}_Q \right)^{3/8} = | \sqrt{v_1} \sin(\lambda_1 P_1) \sqrt{v_1}| =  \frac{2}{3} \left(2  \lambda_1^3 \lambda_2^3  \right)^{ \frac{1}{4} } \left (M_{\text{BH}} M_{\text{WH}} \right)^{ \frac{3}{8} } \label{eq:ObsQuantum1}
\ee 
where we left $\lambda_1$ and $\lambda_2$ generic for pedagogical reasons. In the indicated ordering, \eqref{eq:ObsQuantum1} is symmetric and commutes with the Hamiltonian constraint operator. For the mass amplification choice $\beta = 5/3$, \eqref{eq:ObsQuantum1} measures the black hole mass. Another candidate observable is 
\be
O_2 = \frac{F_Q}{\bar F_Q} = \cot \left( \frac{\lambda_1 P_1}{2} \right)^{\frac{2}{3}} \tan \left( \frac{\lambda_2 P_2}{2} \right)^{2}  = \frac{ M_{\text{BH}}} { M_{\text{WH}}}
\ee
We note that $O_1$ and $O_2$ do not commute, as can already be seen classically. Therefore, due to the Heisenberg uncertainty relations, both masses cannot be specified with arbitrary precision at the same time in the quantum theory. 
Further details will be reported elsewhere. 

\section{Conclusions and outlook}\label{sec:conclusions}

In this paper, we presented a novel approach to effective polymer models of spherically symmetric static spacetimes, as inspired by LQG. As in numerous previous attempts of effective models \cite{CorichiLoopquantizationof,ModestoSemiclassicalLoopQuantum,BoehmerLoopquantumdynamics,ChiouPhenomenologicalloopquantum,ChiouPhenomenologicaldynamicsof,JoeKantowski-Sachsspacetimein,BenAchourPolymerSchwarzschildBlack,OlmedoFromblackholesto,AshtekarQuantumTransfigurarationof,AshtekarQuantumExtensionof}, this situation provides an effective loop quantisation of black holes. In contrast to the connection variables commonly used in the LQG community, our starting point is based on variables tailored to implement physically sensible dynamics with constant polymerisation scales as in \cite{AshtekarRobustnessOfKey}. The momenta of these new variables can be interpreted on-shell as proportional to the square root of the Kretschmann scalar and the angular components of the extrinsic curvature (inverse radii of the 2-spheres), which allows a polymerisation with constant polymerisation scales $\lambda_1$, $\lambda_2$.
As discussed, for suitably chosen initial conditions corresponding to a certain mass (de)-amplification, our model provides a unique, i.e. mass independent, Kretschmann curvature scale $\mathcal{K}_{crit}$ at which quantum effects become relevant. Furthermore, curvature invariants are bounded from above for all black hole masses and diverge in the limit of $\lambda_1$, $\lambda_2$ going to zero. This guarantees that quantum effects are only relevant inside the black hole and close to the vicinity of the classical singularity, while leaving general relativity as an excellent approximation for the exterior region. As discussed, only for Planck sized black holes quantum effects become relevant already at the horizon, which can be expected.

The dependence of the physical qualitative features such as onset on quantum effects on the initial conditions, specifically the mass-(in)dependence of $D$ ($DC^2$ on the white hole side), is a clear shortcoming of the model. The origin is that only for a subset of initial conditions with a certain mass (de-)amplification, the on-shell expression for $P_1$ is directly related to the Kretschmann scalar while the on-shell expression for $P_2$ is the inverse radius of the 2-sphere. One would expect that such a shortcoming is not present in a full quantum gravity treatment that should feature the onset of quantum effects at a fixed scale, so that our model may only approximate such a theory for a limited set of initial conditions. For this price to pay, we however obtain a relatively simple model where the quantum theory can be explicitly constructed and appears to be analytically solvable. As such, it is already of interest as a toy model to study the validity of the effective dynamics.

As said, the appearance of $D$ in \eqref{interpretation} is undesirable as it influences the onset of quantum effects via the initial conditions and one would like to eliminate it somehow. As defining different variables which remedy this would probably complicate the Hamiltonian too much, one can try to adopt the strategy advocated in \cite{OlmedoFromblackholesto,AshtekarQuantumExtensionof} to make $\lambda_1$ and $\lambda_2$ dependent on the initial conditions to obtain an onset of quantum effects at a fixed scale. We note that there is a difference in doing so in the Hamiltonian or the equations of motion \cite{BodendorferANoteOnTheHamiltonian}, so that both possibilities should be studied. Preliminary investigations show that one can achieve this way a range of new physically viable black to white hole transitions, including a symmetric bounce as in \cite{AshtekarQuantumExtensionof}. However, we so far did not find a set of equations yielding physically viable spacetimes for all black and white hole masses.

As explicitly shown in the classical setting, the effective equations of motion in the new variables can be solved in the interior of the black hole as well as in the exterior at the same time. This holds true also for the polymerised model which allows us to explicitly construct the Penrose diagram of the full resulting effective quantum extended spacetime. In accordance with other effective models of loop quantum black holes, the classical singularity is resolved by a quantum bounce. Specifically, the classical singularity is replaced by a regular space-like three-dimensional hyper-surface called \textit{transition surface}. This transition surface separates the interior region into a trapped and an anti-trapped region with black hole- and white hole-type horizons respectively at the past and future boundary, showing that the effective solutions provide a black hole to white hole transition. As our approach allows us to describe also the exterior regions, we found two asymptotic Schwarzschild regions with different black hole masses, which we interpret as black hole exterior and white hole exterior, respectively. Moreover, in order to get quantum effects at reasonable scales and an upper bound for the curvature invariants, we found only two compatible possibilities which correspond to a black hole mass amplification and de-amplification, thus providing a mass dependent non-symmetric bounce. Nevertheless, the de-amplification solution is in perfect agreement with the amplification solution as the amplification is exactly the inverse of the de-amplification, thus yielding, as shown by studying the Penrose diagram, an infinite tower of black holes and white holes where the mass is oscillating between the amplified and the de-amplified value. Both solutions can be mapped into each other by identifying black hole with white hole and vice versa, which simply shows time-reversibility and furthermore avoids the problem of unbounded amplification or de-amplification as noted e.g. in \cite{ChiouPhenomenologicalloopquantum}. Such behaviour may be interesting for phenomenological studies as e.g. in \cite{BarrauFastRadioBursts}.

A key difference between our analysis and previous ones seems to be that we discuss a second Dirac observable next to the black hole mass that can be identified as the white hole mass and only exists in the quantum theory. To the best of our knowledge, the existence of such a Dirac observable has not been noted so far. Since the effective equations used in previous approaches are structurally similar to ours, it would be of great interest to revisit those calculations and check for the existence of such an observable. This may change the conclusions of previous works and potentially demand a similar restriction of the initial conditions to achieve physical viability. We note in particular, that the classically most natural choice $D=2/3 \Leftrightarrow m=1$, leading to $b(r) = r$ (that one would adopt without considering the quantum theory and the ensuing second Dirac observable) is precisely in the regime of initial conditions we restrict to.

Further work should include a relation of this model to full LQG, as it is only inspired but not derived from it (see section \ref{Polymerisation} for details).
Another interesting research direction would be to follow previous investigations in the holographic cosmological setting \cite{BodendorferHolographicSignaturesOf,BodendorferHolographicSignaturesofII} and study the effect of resolved black hole singularities for a hypothetic dual gauge theory. In turn, gauge theory computations may help to constrain the polymerisation scheme if a holographic duality is assumed. For this purpose, a generalisation of our work to higher dimensions and the inclusion of a negative cosmological constant would be necessary.

\section*{Acknowledgements}

The authors were supported by an International Junior Research Group grant of the Elite Network of Bavaria. We thank Jibril Ben Achour for suggesting to add \eqref{eq:ExpansionNullRays} and the surrounding discussion.


\end{document}